\documentclass[]{article}
\usepackage{epsf}

\newcommand{\reseteqnum}{\setcounter{equation}{0}}

\newcommand{\bbone}{{\mathchoice {\rm 1\mskip-4mu l} {\rm 1\mskip-4mu l}
{\rm 1\mskip-4.5mu l} {\rm 1\mskip-5mu l}}}
\def\mib#1{\mbox{\boldmath $#1$}}

\newcommand{\rvac}{\vert 0 \rangle}
\newcommand{\rvacp}{\vert + \rangle}
\newcommand{\rvacm}{\vert - \rangle}
\newcommand{\lvac}{\langle 0 }
\newcommand{\lvacp}{\langle + }
\newcommand{\lvacm}{\langle - }

\newcommand{\rVac}{\vert v \rangle}
\newcommand{\rVacp}{\vert v+ \rangle}
\newcommand{\rVacm}{\vert v- \rangle}

\newcommand{\lVacp}{\langle v+}
\newcommand{\lVacm}{\langle v-}

\newcommand{\rN}{\vert n  \rangle}
\newcommand{\lN}{\langle n  }
\newcommand{\rM}{\vert m  \rangle}
\newcommand{\lM}{\langle m  }

\title{
\hfill
\parbox{4cm}{\normalsize KUNS-1446\\HE(TH)~97/09}\\
\vspace{1cm}
Gauge Freedom in Chiral Gauge Theory \\
with Vacuum Overlap -- Four-dimensional case}
\author{Yoshio Kikukawa\thanks{e-mail address:
kikukawa@physics.rutgers.edu, kikukawa@gauge.scphys.kyoto-u.ac.jp}
\\
{\normalsize\em Department of Physics and Astronomy,
                Rutgers University\thanks{On leave of absence from:
         Department of Physics, Kyoto University, Kyoto 606-01, Japan}
}\\
{\normalsize\em Piscataway, NJ 08855-0845}
}

\begin{document}
\maketitle

\begin{abstract}
Dynamical nature of the gauge degree of freedom and its effect
to fermion spectrum are studied for four-dimensional nonabelian
chiral gauge theory in the vacuum overlap formulation.
The covariant gauge fixing term and the Faddeev-Popov determinant
are introduced by hand as a weight for the gauge average.
At $\beta=\infty$, as noticed by Hata some time ago, the model is
renormalizable at one-loop and the gauge fixing term turns into an
asymptotically free self-coupling of the gauge freedom, even in the presence
of the gauge symmetry breaking of the complex phase of chiral determinant.
The severe infrared divergence occurs in the perturbation expansion and it
prevents local order parameters from emerging. The Foerster-Nielsen-Ninomiya
mechanism works in this perturbative framework: the small explicit
gauge symmetry breaking term does not affect the above infrared structure.
Based on these dynamical features, which is quite similar to the
two-dimensional nonlinear sigma model, we assume that the global gauge
symmetry does not break spontaneously and the gauge freedom acquires mass
dynamically. The asymptotic freedom allows us to tame the gauge fluctuation
by approaching the critical point of the gauge freedom without spoiling its
disordered nature. Then it is argued that the disordered gauge freedom does
not necessarily cause the massless chiral state in the (waveguide) boundary
correlation function and that the entire fermion spectrum can be chiral.

\end{abstract}

\newpage
\section{Introduction}
\reseteqnum

It has been one of the most important issues to clarify
the dynamical behavior of the gauge freedom and its effect to
the fermion spectrum for various proposals of lattice
chiral gauge theory\cite{
wilson-yukawa-model,
aoki,kashiwa-funakubo,
wilson-yukawa-model-analysis,
eichten-preskill,
eichten-preskill-analysis,
staggered-fermion,
staggered-fermion-analysis,
original-kaplan,
original-kaplan-analysis,
original-waveguide,
waveguide-analysis-golterman,
waveguide-analysis-general,
waveguide-four-fermi,
waveguide-majorana-yukawa,
original-overlap,overlap}.
In this article, we discuss this issue in the context of
the vacuum overlap formulation\cite{original-overlap}.

In the vacuum overlap formulation of a generic chiral gauge theory,
gauge symmetry is explicitly broken by the complex phase of
fermion determinant. In order to restore the gauge invariance,
gauge average ---the integration along gauge orbit--- is invoked.
Then, what is required for the dynamical nature of the
gauge freedom at $\beta=\infty$ (pure gauge limit) is that
the global gauge symmetry is not broken spontaneously and
all the bosonic field of the gauge freedom could be heavy compared
to a typical mass scale of the theory so as to decouple from physical
spectrum\cite{gauge-symmetry-restoration,wilson-yukawa-model,
aoki,kashiwa-funakubo,staggered-fermion,original-overlap}.
For this mechanism of the gauge symmetry restoration to work,
the gauge symmetry breaking must be ``small'' so that
it does not spoil the disordered nature of the gauge freedom and
it keeps the correlation length of the gauge freedom in the order
of the lattice spacing.

In the previous paper on the two-dimensional nonabelian gauge
theory\cite{kikukawa-gauge-two-dimensions},
we have discussed that a certain gauge symmetry breaking term
turns out to be the asymptotically free self-coupling of the gauge
freedom and it can be made ``large'' without spoiling the disordered
nature of the gauge freedom.
Using this technique to deform the original theory, we have
argued that the disordered gauge degree of freedom does not
necessarily cause the massless chiral state in the (waveguide)
boundary correlation function\cite{waveguide-analysis-golterman}.
We have also argued that the decoupling
of the gauge freedom can occur as the self-coupling is removed,
provided that the IR fixed point due to the Wess-Zumino-Witten term is
absent by anomaly cancellation.

Our objective in this paper is to extend the above argument to
nonabelian chiral gauge theory in {\it four-dimensions}.
For this purpose, we reminds us the following facts
noticed by Hata\cite{pure-gauge-model} some time ago:
{\it at $\beta=\infty$, the covariant gauge fixing term turns into
an asymptotically free self-coupling of the gauge freedom in terms
of gauge parameter and the severe IR divergence,
which occurs in the perturbation (spinwave) expansion, prevents local
order parameters from emerging}.
These dynamical properties are quite similar to the two-dimensional
nonlinear sigma model and they suggest that the gauge freedom
acquires mass dynamically and the global gauge symmetry does not
break spontaneously for the entire region of the gauge parameter.
This is the very nature of the gauge freedom
which is required for the dynamical restoration of the gauge symmetry
in the lattice chiral gauge theory.

With this dynamical picture in mind, we first introduce the covariant
gauge fixing term and the Faddeev-Popov determinant by hand
into the lattice chiral gauge theory defined by the vacuum overlap
{\it as a weight for the gauge average}.
Then we examine the pure gauge model at $\beta=\infty$ by the perturbation
expansion
{\it in the presence of the gauge symmetry breaking terms of the complex
phase of chiral determinants}. We will find at one-loop
that the model is actually renormalizable and the self-coupling
is asymptotically free. We will see that
the IR structure remains same and it prevents local order
parameters from emerging. We will also see that
this IR structure is not affected by the small perturbation
of the gauge symmetry breaking term linear in the link variables, in
accord with the Foerster-Nielsen-Ninomiya mechanism.
Based on these observations, we will assume that
the gauge freedom acquires mass dynamically and the global gauge
symmetry does not break spontaneously in the pure gauge limit.

With this assumption, we can make an argument just as in the case
of the two-dimensional nonabelian theory. Namely,
the asymptotic freedom allows us to tame the gauge fluctuation by
approaching the critical point of the gauge freedom without spoiling
its disordered nature. There we will show by the spinwave
approximation that the spectrum in the {\it invariant boundary
correlation function} has mass gap of the order of the lattice cutoff
and it survives the quantum correction due to the gauge fluctuation.
There is no symmetry against the spectrum mass gap.
Since the overlap correlation function does not depend on the gauge
freedom and does show the chiral spectrum\cite{original-overlap},
the above fact means that the entire fermion spectrum is chiral.

We will not discuss here how the decoupling of the gauge freedom could
occur as the self-coupling is removed, because
we do not know yet much about the non-perturbative dynamics
of the pure gauge model. We will leave this issue for future study.
The lattice formulation of the four-dimensional pure gauge model and
the possibility to examine it by the Monte Carlo simulation will be
discussed elsewhere\cite{kikukawa-lattice-pgm}.

A few comments are in order. The covariant gauge fixing term
and the Faddeev-Popov determinant (Faddeev-Popov ghost fields) are
introduced in the gauge fixing approach for lattice chiral gauge
theory of the Rome group\cite{rome-approach}.
There {\it the restoration of the BRST invariance} is attempted
non-perturbatively by adding and tuning gauge non-invariant counter terms.
Since the chiral gauge symmetry is explicitly broken by the lattice
fermion regularization, the addition of the gauge fixing term does not
mean the gauge fixing. The integration over the compact gauge link
variables means that we are still invoking the gauge average.
What is meant by the gauge fixing approach is {\it to decouple
the gauge degree of freedom kinematically with the help of the
BRST invariance} which could be restored by the fine tuning of
the gauge non-invariant counter terms\footnote{
It is not yet clear how this non-perturbative restoration
of the BRST invariance could overcome the difficulty of the
manifest BRST invariant formulation of lattice gauge theory due
to Gribov copy noticed by H.~Neuberger\cite{lattice_brs_gribov}. }.
In this approach,
the global gauge symmetry does not need to be realized
linearly (symmetrically) and
the gauge degree of freedom does not need to be disordered.

In this spirit of the gauge fixing approach,
the dynamics of the gauge degree of freedom
has been examined by Shamir and Golterman
in the model with the covariant-type gauge fixing
term\cite{renormalizable-gauge-model, covariant-type-gauge-fixing-action}.
One of the points on which the authors put a stress,
among others, is that {\it the continuum limit should be taken inside
the broken phase, because of the species doubling problem
in the symmetric phase}. In the broken phase, both the fermions
(including species doublers) and the gauge boson acquire masses
proportional to the vacuum expectation value of the gauge freedom
(Higgs field). Accordingly, the main concern in
the analysis is to seek the critical points inside the broken phase
at which the gauge boson becomes massless while keeping the species
doublers heavy. The authors identified the boundary between the two broken
phases, the ferromagnetic phase(FM) and the helicoidal-ferromagnetic
phase(FMD), as such critical points. The latter is characterized by the
{\it vectorial local order parameter which breaks spontaneously
the translation and rotation invariance}. Tuning the bare
parameters towards this boundary is regarded as the first step
to restore the BRST invariance.

Although we also introduce the covariant gauge fixing term and
the Faddeev-Popov determinant as a weight for the gauge average,
we do not try to recover the BRST invariance.
Rather, we seek the possibility
to keep the gauge symmetry breaking as small as possible and
to restore the gauge symmetry at low energy by the mechanism
of Foerster-Nielsen-Ninomiya
in the case of the compact nonabelian gauge group.
This requires that the global gauge symmetry is not broken
spontaneously and that the gauge degree of freedom is kept
disordered and heavy compared to the physical scale.
(For this very requirement, we admit certain fine tuning
of the gauge non-invariant counter terms.)
In the overlap formulation, the gauge symmetry breaking is reduced
into only the parity odd (imaginary) terms of the chiral determinant.
A possible dynamical but perturbative picture for the gauge
symmetry restoration for the nonabelian theory is given by Hata,
as described above. In this setting, we try to show that
{\it the disordered nature of the gauge freedom does not contradict
with the chiral fermion spectrum} in the overlap formulation.
We invoke the spinwave approximation
in order to discuss {\it the symmetric phase dynamics},
which validity has been argued in several interesting contexts
of the two-dimensional quantum field theories.
In order to investigate this dynamical possibility fully, we need
the non-perturbative study of the model which could incorporate
the effect of the complex action.
We hope that
the perturbative analysis presented here
will give us a physical and dynamical picture how the
lattice chiral gauge theory could emerge through the gauge average,
and will motivate us to do such non-perturbative study of the dynamics
of the gauge symmetry restoration.

This paper is organized as follows. In
section~\ref{sec:4D-SUN-pure-gauge-limit-with-covariant-gauge-fixing},
we define with
vacuum overlap a generic four-dimensional $SU(N)$ chiral gauge theory.
Then, we clarify its structure in the pure gauge limit and
introduce the covariant gauge fixing term and the Faddeev-Popov
determinant as a weight for the gauge average.  In
section~\ref{sec:pure-gauge-dynamics-with-covriant-gauge-fixing-term},
we discuss the dynamics of the nonabelian pure gauge model in the
presence of the gauge symmetry breaking terms in the complex phase of
the fermion determinant.
In section~\ref{sec:perturbative-FNN-mechamism},
we give a perturbative analysis of the Foerster-Nielsen-Ninomiya
mechanism in the context of the pure gauge model.
In section~\ref{sec:disordered-gauge-freedom-chiral-fermion},
we introduce the boundary correlation functions and examine them
near the critical point.
In section~\ref{sc:conclusion-discussion},
we summarize and discuss our result.

\section{Pure gauge limit with covariant gauge fixing term}
\reseteqnum
\label{sec:4D-SUN-pure-gauge-limit-with-covariant-gauge-fixing}

\subsection{Four-dimensional $SU(N)$ Chiral Gauge Theory}

Let us consider the four-dimensional $SU(N)$ chiral gauge
theory with left-handed Weyl fermions in an anomaly free
representation\footnote{
Our convention for the $SU(N)$ group generators is as follows:
\begin{eqnarray}
       \left[ T^a , T^b \right] &= & i f^{abc} T^c , \\
{\rm Tr}\left( T^a T^b \right)[r] &=& \frac{1}{2} \delta^{ab} \, k[r] , \\
{\rm Tr}\left( T^a \left\{ T^b, T^c \right\} \right)[r]
&=& \frac{1}{2}\, k[r]\, d^{abc}[r]
= \frac{1}{2} d^{abc}[{\rm fundamental}] \, A[r] , \\
\sum_a T^a T^a &=& C_2[r] \, \bbone[r] , \quad
 C_2[r]= \frac{N^2-1}{2 {\rm dim}[r] } k[r] .
\end{eqnarray}
They are normalized by the fundamental representation so that
$k[r]=1$ and $A[r]=1$ for it.
}:
\begin{eqnarray}
\sum_{rep.}
{\rm Tr}\left( T^a \left\{ T^b,T^c \right) \right\}[r]
&=& \sum_{rep.} A[r]  \times \frac{1}{2}\,
    d^{abc}[{\rm fundamental \ rep. }] \nonumber\\
&=& 0 .
\end{eqnarray}

The partition function of the chiral gauge theory
is given by the following formula in the vacuum overlap
formulation\cite{original-overlap}\footnote{
We may start from the partition function with the
manifest local gauge invariance, by introducing the explicit
integration over the gauge degree of freedom.
We will discuss about this alternative gauge invariant formulation
and the various gauge fixings in the discussion
in relation to the Wilson-Yukawa model.
}.
\begin{equation}
    Z= \int [dU] \exp \left(-\beta S_G \right)
      \prod_{rep.} \left(
        \frac{\lvacp \rVacp}{|\lvacp\rVacp|}
               \lVacp \rVacm
        \frac{ \lVacm \rvacm}{| \lVacm \rvacm|}
              \right) .
\end{equation}
In this formula,
$\rVacp$ and $\rVacm$ are the vacua of the second-quantized
Hamiltonians of the five-dimensional Wilson fermion
with positive and negative bare masses, respectively.
\begin{equation}
\label{eq:overlap-hamiltonian}
  \hat H_{\pm} = \hat a^\dagger_{n\alpha}{}^i
  H_{\pm \, n\alpha,m\beta}{}_i^j \hat a_{m\beta}{}_j ,
\end{equation}

\begin{eqnarray}
H_{\pm\, n\alpha,m\beta}{}_i^j
&=&
\left( \begin{array}{cc}
B_{nm}{}_i^j \pm m_0 \delta_{nm} \delta_i^j
& C_{nm}{}_i^j \\
C^\dagger_{nm}{}_i^j
& -B_{nm}{}_i^j \mp m_0 \delta_{nm} \delta_i^j
\end{array} \right) ,
\\
B_{nm}{}_i^j
&=& \frac{1}{2}\sum_\mu \left( 2 \delta_{n,m}\delta_i^j
-\delta_{n+\hat\mu,m}U_{n\mu}{}_i^j
-\delta_{n,m+\hat\mu}U^\dagger_{m\mu}{}_i^j
\right) , \\
C_{nm}{}_i^j &=& \frac{1}{2} \sum_\mu \sigma_\mu
\left(
\delta_{n+\hat\mu,m}U_{n\mu}{}_i^j
 -\delta_{n,m+\hat\mu}U^\dagger_{m\mu}{}_i^j
\right).
\end{eqnarray}
$\rvacp$ and $\rvacm$ are corresponding free vacua.
The Wigner-Brillouin phase convention is explicitly implemented
by the overlaps of vacua with the same signature of mass.
$\prod_{rep.}$ stands for the product over all Weyl fermion multiplets
in the anomaly free representation. $S_G$ is the gauge action.

In the vanishing gauge coupling limit $\beta =\infty$,
the gauge link variable is given in the pure gauge form:
\begin{equation}
  U_{n\mu}= g_n g_{n+\mu}^\dagger ,\quad  g^{}_n \in SU(N) .
\end{equation}
Then the model describes the gauge degree of freedom
coupled to fermion through gauge non-invariant piece of complex
phase of chiral determinants.
\begin{eqnarray}
    Z&=& \int [dg]
      \prod_{rep.} \left(
   \frac{\lvacp \vert \hat G \rvacp}{|\lvacp \vert \hat G \rvacp|}
   \lvacp \rvacm
   \frac{\lvacm \vert \hat G^\dagger \rvacm}
        {|\lvacm \vert \hat G^\dagger \rvacm|}
              \right) .
\end{eqnarray}
$\hat G$ is the operator of the gauge transformation given by:
\begin{equation}
\hat G
= \exp \left( \hat a_n^{\dagger i} \{\log g_n\}_i{}^j \hat a_{n j} \right) .
\end{equation}

As we can see from this pure gauge limit of the original theory,
the gauge average is invoked without any weight for the gauge freedom
except for the complex phase of the chiral determinant. This way of
the gauge average well keep the disordered nature of the gauge freedom
and keep its correlation length within the order of the lattice spacing.

\subsection{Covariant gauge fixing term as a weight of gauge average}

In order to examine the dynamical effect of the gauge average closely,
it is desirable to have control over the fluctuation of the gauge
freedom\cite{kikukawa-gauge-two-dimensions,2d-wess-zumino-by-overlap}.
For this purpose, we add to the original theory by hand
the covariant gauge fixing term and the Faddeev-Popov
determinant (ghost field) {\it as a weight for the gauge average}:
\begin{eqnarray}
&& \exp\left(
      -\sum_{n,a} \frac{1}{2\alpha}
\left( \sum_\mu \bar{\nabla}_\mu \hat{A}^a_{n\mu} \right)^2
\right)
\det\left(\hat{M}^{ab}_{nm}\left[U_{n\mu} \right]\right)
\nonumber\\
&&= \int [d c^a d \bar c^a]
\exp\left(
      -\sum_{n,a} \frac{1}{2\alpha}
\left( \sum_\mu \bar{\nabla}_\mu \hat{A}^a_{n\mu} \right)^2
- \sum_{nm,ab} \bar c_n^a \hat{M}^{ab}_{nm} c_m^b
\right) , \nonumber\\
\end{eqnarray}
where
\begin{equation}
\label{eq:lattice-vector-potential}
\hat{A}_{n\mu}=
 \frac{1}{2 i} \left( U_{n\mu}-U^\dagger_{n\mu} \right)
-\frac{1}{N} \bbone \,
{\rm Tr} \,
    \frac{1}{2 i} \left( U_{n\mu}-U^\dagger_{n\mu} \right) ,
\end{equation}
and
\begin{eqnarray}
\label{eq:lattice-Faddeev-Popov-operator}
\hat{M}^{ab}_{nm}\left[ U_{n\mu}\right]&=&
\sum_\mu \left[
 \left\{ \hat{E}_{ab}^{-1}(U_{n\mu}) \delta_{nm}
        -\hat{E}_{ba}^{-1}(U_{n\mu}) \delta_{n+\hat\mu,m} \right\}
\right. \nonumber\\
&&\qquad \left.
-\left\{ \hat{E}_{ab}^{-1}(U_{n-\hat\mu,\mu}) \delta_{n-\hat\mu,m}
        -\hat{E}_{ba}^{-1}(U_{n-\hat\mu,\mu}) \delta_{n,m} \right\}
\right] , \nonumber\\
\\
\hat{E}_{ab}^{-1}(U_{n,\mu})
&=&{\rm Tr} \left( T^aT^b U_{n\mu}  + T^bT^a U_{n\mu}^\dagger  \right) .
\end{eqnarray}
This covariant gauge fixing term\cite{lattice-covariant-gauge-fixing}
is formally based on the lattice Landau gauge
\begin{equation}
  \bar \nabla_\mu \hat A_{n\mu} = 0 ,
\end{equation}
which minimizes locally the ``functional norm'' of the link
variables on the given gauge orbit,
\begin{equation}
  \label{eq:minimization-function}
  F[ {}^g\!U_{n\mu} ] = 1- \frac{1}{NDV}
\sum_{n\mu} {\rm Tr} \, {}^g\!U_{n\mu} .
\end{equation}
The structure of the Faddeev-Popov operator
Eq.~(\ref{eq:lattice-Faddeev-Popov-operator}) has been examined
extensively in \cite{lattice-Faddeev-Popov-operator}.

It should be noted that
since the gauge symmetry is explicitly
broken in the vacuum overlap formulation, the addition of the
gauge fixing term does not mean the gauge fixing.
The integration over the compact gauge link variables means that
we are still invoking the gauge average to restore the gauge symmetry.
The role of the gauge fixing term is to modify the weight for the
gauge degree of freedom in the gauge average.

Accordingly, the BRST invariance does not hold in this model.
The BRST transformation can be defined in this model
as follows\cite{lattice-brs}:
\begin{eqnarray}
  \label{eq:BRST-transformaion_lattice_Landau-gauge}
  {\mib \delta}_B \hat{A}_{n\mu}^a &=& \left\{
   \hat{E}^{-1}_{ab}(U_{n\mu}) \, c_n^b
-  \hat{E}^{-1}_{ba}(U_{n\mu}) \, c_{n+\hat\mu}^b
                 \right\} \\
{\mib \delta}_B c_n^a &=& -\frac{1}{2} f_{abc} \, c_n^b \, c_n^c , \\
{\mib \delta}_B \bar{c}_n^a &=& - \frac{1}{\alpha}
            \sum_\mu \bar{\nabla}_\mu \hat{A}^a_{n\mu} .
\end{eqnarray}
The nilpotency holds on the gauge potential,
$\hat{A}_{n\mu}^a$:
\begin{equation}
  \label{eq:nilpotency}
  {\mib\delta}_B^2 \hat{A}_{n\mu}^a = 0 .
\end{equation}
Then the gauge fixing term and the Faddeev-Popov ghost action
are BRST invariant. The functional integral measure of the gauge
field and ghost and anti-ghost fields are also BRST invariant.
Only the complex phases of the chiral determinants are BRST
non-invariant, as they should be.
If the BRST invariance would exist on the lattice, the partition
function must vanish as shown by Neuberger\cite{lattice_brs_gribov}.
The failure of the manifestly BRST invariant formulation of
the lattice gauge theory is due to the following reason.
On a finite lattice and for a compact gauge group,
there are even number of solutions for the gauge fixing condition,
namely, the Gribov copies. Because of the compactness of the gauge
configuration space, half of them have the positive Faddeev-Popov
determinants and half of them have the negative ones. They cancel
each other to result in the vanishing partition function.
In the present case, however, the gauge symmetry breaking terms
in the complex phase of the chiral determinant distinguish
these Gribov copies\cite{kikukawa-aoyama-gribov} and resolve the
degeneracy.  They are irrelevant in the naive continuum limit, but
they actually prevent the lattice Gribov copy (at $\beta=\infty$) from
being a classical solution of the equation of motion, as we will see
in the following section
\ref{subsec:puge-gauge-dynamics-classical-solution-Gribov-copy}.

\subsection{Lattice $SU(N)$ pure gauge model with chiral fermions}
In the pure gauge limit, the model reduces to
the $SU(N)$ pure gauge model\cite{pure-gauge-model} in the lattice
regularization which also couples to anomaly-free chiral fermions
through the gauge non-invariant piece of complex phase of chiral
determinant:
\begin{eqnarray}
\label{eq:Chiral-pure-gauge-partition-function-main}
    Z&=& \int [dg]
\exp\left(
      -\sum_{n,a} \frac{1}{2\alpha}
\left( \sum_\mu \bar{\nabla}_\mu \hat{A}^a_{n\mu} \right)^2
\right)
\det\left(\hat{M}^{ab}_{nm}\left[g_n g^\dagger_{n+\hat \mu}
  \right]\right) \times
\nonumber\\
&& \qquad\qquad
      \prod_{rep.} \left(
   \frac{\lvacp \vert \hat G \rvacp}{|\lvacp \vert \hat G \rvacp|}
   \lvacp \rvacm
   \frac{\lvacm \vert \hat G^\dagger \rvacm}
        {|\lvacm \vert \hat G^\dagger \rvacm|}
              \right)
\nonumber\\
&\equiv& \int d\mu[g;\alpha] ,
\end{eqnarray}
where
\begin{equation}
\label{eq:pure-gauge-vector-potential}
\hat{A}_{n\mu}=
 \frac{1}{2 i} \left(
g_n g^\dagger_{n+\hat\mu}- g_{n+\hat\mu}g^\dagger_n
               \right)
-\frac{1}{N} \bbone \,
{\rm Tr} \,
    \frac{1}{2 i}
\left(
g_n g^\dagger_{n+\hat\mu}-g_{n+\hat\mu}g^\dagger_n
\right) .
\end{equation}
Including the imaginary action, the functional integral measure of
the gauge freedom is denoted by $d\mu[g;\alpha]$.

This model is invariant under two global $SU(N)$ transformations
acting on the field of the gauge freedom.
The first one is the global remnant of the gauge transformation:
\begin{equation}
\label{eq:sun-global-gauge-transformation}
 SU(N)_{C}: \qquad   g_n{}_i^j \longrightarrow g_0{}_i^k g_n{}_k^j .
\end{equation}
The second one comes from the arbitrariness of choice of pure gauge
variable $g_n$:
\begin{equation}
\label{eq:sun-hidden-global-transformation}
 SU(N)_{H}: \qquad
g_n{}_i^j \longrightarrow g_n{}_i^k h^\dagger{}_k^j .
\end{equation}
They defines the chiral transformation of
$G= SU(N)_L \times SU(N)_R = SU(N)_{C} \times SU(N)_{H}$
and the model is symmetric under this chiral transformation.

We refer the imaginary part of the action of the gauge freedom
induced from the fermion determinant
as the Wess-Zumino-Witten term,
although the actual Wess-Zumino-Witten terms are canceled
among the fermions.
We denote it by $\Delta\Gamma_{WZW}$:
\begin{equation}
  e^{i \Delta\Gamma_{WZW}[g]} \equiv
   \prod_{rep.} \left(
   \frac{\lvacp \vert \hat G \rvacp}
  {|\lvacp \vert \hat G \rvacp|}
   \frac{\lvacm \vert \hat G^\dagger \rvacm}
        {|\lvacm \vert \hat G^\dagger \rvacm|}
              \right) .
\end{equation}
The explicit formula of the Wess-Zumino-Witten term is given by
\begin{eqnarray}
&& i\Delta\Gamma_{WZW}[g] \nonumber\\
&&\qquad
= \sum_{rep.} \left\{
\, i {\rm Im} {\rm Tr} \, {\rm Ln} \left[
\sum_{m}
v^\dagger_+(p,s) e^{-i p m}
\left( g_m{}_i^j \right)  e^{i q m } v_+(q,s^\prime)
\right]
\right.\nonumber\\
&&\left. \quad\qquad\qquad +
i {\rm Im} {\rm Tr} \, {\rm Ln} \left[
\sum_{m}
v^\dagger_-(p,s) e^{-i p m}
\left( g_m^\dagger{}_i^j \right)  e^{i q m }  v_-(q,s^\prime)
\right] \right\} .
\nonumber\\
\end{eqnarray}
Here $v_+(p,s)$ and $v_-(p,s)$ are negative-energy eigenvectors of
the free Hamiltonians with positive and negative masses,
respectively.

Note also that
even in the limit $\alpha \rightarrow \infty$ where
the covariant gauge fixing term vanishes,
the modified model does not completely reduce to the original model:
the ghost and anti-ghost fields remain to couple to the gauge
freedom.

\section{Pure gauge dynamics with covariant gauge fixing term}
\reseteqnum
\label{sec:pure-gauge-dynamics-with-covriant-gauge-fixing-term}

In this section,
following the argument given by Hata in the continuum limit,
we examine the dynamical effect of the gauge
degree of freedom by the perturbation expansion
(spinwave approximation) in terms of $\lambda$
($\lambda^2 \equiv \alpha$). We first invoke the background field
method\cite{lattice-background-field-method-2D-nonlinear-sigma-model}
in order to examine the quantum effect of the gauge freedom and
the renormalization due to it, including the contributions from the
the Wess-Zumino-Witten term.
Then we examine the realization of the global gauge symmetry
at $\beta=\infty$.

\subsection{Classical solution for the equation of motion and Gribov copy}
\label{subsec:puge-gauge-dynamics-classical-solution-Gribov-copy}

We consider the dynamical degree of freedom of the gauge freedom, the
ghost and anti-ghost fields as the fluctuations from the classical
configurations which satisfies the equation of motion:
\begin{eqnarray}
g_n & \rightarrow&  \exp(i \lambda \pi_n ) \, g_n  , \\
c_n^a & \rightarrow & c_n^a + \xi_n^a , \\
\bar c_n^a & \rightarrow & \bar c_n^a + \bar \xi_n^a .
\end{eqnarray}
By substituting the above decompositions of classical
and quantum components into the action, it is expanded up to the
quadratic terms of the fluctuations $\pi$, $\xi$ and $\bar \xi$.
And then by performing the Gaussian functional integration
of the fluctuations, we evaluate the quantum correction
to the classical action at one-loop.

The linear terms in the fluctuation, $\pi$, $\xi$ and
$\bar \xi$ give the equations of motion for the classical
configurations $g$, $c$ and $\bar c$.
For the gauge freedom, it is given as
\begin{eqnarray}
\label{eq:classical-equation-of-motion-g}
&-&\frac{1}{\alpha} \sum_{n\mu} \bar{\nabla}_\mu \hat A_{n\mu}^a
  \hat M_{nm}^{ac}\left[g_n g^\dagger_{n+\hat \mu}\right] \nonumber\\
&+& \sum_{n\mu} \left[
\left\{
 \bar c_n^a F_{nm}^{abc}
            \left[g_n g^\dagger_{n+\hat \mu}\right] c_n^b
-\bar c_n^a F_{nm}^{bac}
            \left[g_n g^\dagger_{n+\hat \mu}\right] c_{n+\hat\mu}^b
\right\} \right.\nonumber\\
&&\left.  \quad
-\left\{
 \bar c_n^a F_{n-\hat\mu,m}^{abc}
  \left[g_{n-\hat\mu} g^\dagger_{n}\right]  c_{n-\hat\mu}^b
-\bar c_n^a F_{nm}^{bac}
  \left[g_{n-\hat\mu} g^\dagger_{n}\right]  c_n^b
\right\}
\right] \nonumber\\
&+& \frac{i}{2} \sum_{rep.}
\left\{
 {\rm Tr} \left( S_+^v[g^\dagger](n,m) T^c g_m \right)
+{\rm Tr} \left( S_+^v[g](n,m) g_m^\dagger T^c \right)  \right.\nonumber\\
&&\left.  \qquad
-{\rm Tr} \left( S_-^v[g^\dagger](n,m) T^c g_m \right)
-{\rm Tr} \left( S_-^v[g](n,m) g_m^\dagger T^c \right) \right\} = 0 ,
\end{eqnarray}
where
\begin{eqnarray}
F_{nm}^{abc}
            \left[g_n g^\dagger_{n+\hat \mu}\right]
&=& i{\rm Tr}\left( T^a T^b T^c g_n g_{n+\hat\mu}^\dagger
                 -T^c T^b T^a g_{n+\hat\mu} g_n^\dagger \right)
               \delta_{nm} \nonumber\\
&-& i{\rm Tr}\left( T^c T^a T^b g_n g_{n+\hat\mu}^\dagger
                 -T^b T^a T^c g_{n+\hat\mu} g_n^\dagger \right)
               \delta_{n+\hat\mu,m} ,
\end{eqnarray}
and
\begin{eqnarray}
\label{eq:negative-energy-projection-gauge}
&& S_\pm ^v[g](n,m){}_i^j \nonumber\\
&&\qquad \equiv
\int \frac{d^4 p}{(2\pi)^4} \frac{d^4 q}{(2\pi)^4}
\, e^{i p n } e^{-i q m} \times
\nonumber\\
&&\qquad\qquad
v_\pm(p,s)
\left[ v^\dagger_\pm (q,s^\prime)
e^{-i q r} \left( g_r^\dagger{}_j^i \right) e^{i p r }
v_\pm (p,s) \right]^{-1}_{(p,s,i)(q,s^\prime,j)}
v^\dagger_\pm (q,s^\prime)  .
\nonumber\\
\end{eqnarray}
Note that the last term in the l.h.s. of the equation of motion is
imaginary, which comes from the imaginary part of the action,
$i \Delta\Gamma_{WZW}[g]$. Note that this term depends non-locally
on the field variable of the gauge freedom. This non-locality is
an inevitable feature expected for any gauge non-invariant
regularization of fermion. We should note that this classical
equation of motion carries the information about the anomaly
cancellation. In anomaly-free theories, this imaginary term
is irrelevant in the continuum limit which can be taken for the
slowly varying solution. Otherwise, it contains the finite term
which corresponds to the variation of the Wess-Zumino-Witten term
in the continuum limit.

The simplest and trivial solution for the equation of motion is given by
\begin{equation}
  g_n = \bbone, \quad c_n^a = 0, \quad \bar c_n^a = 0 .
\end{equation}
It solves the imaginary part simply because the matrix to be inverted in
Eq.~(\ref{eq:negative-energy-projection-gauge}) become simple
products of the unit matrixes both in the group space and in
the momentum and spinor spaces:
\begin{equation}
\left[ v^\dagger_\pm (q,s^\prime)
e^{-i q r} \left( g_r^\dagger{}_j^i \right) e^{i p r }
v_\pm (p,s) \right]
= (2\pi)^4 \delta^4(q-p) \delta_{ss^\prime} \delta{}_j^i .
\end{equation}
Then the trace over the group indices vanishes.
Moreover, the dependence on the signatures of the masses $\pm m_0$
drop off in each contribution and they can
cancel among the contributions of the positive and negative masses.
(This reason applies to the case of the U(1) gauge theory.)

It is interesting to note the fact that
the lattice Gribov copies (at $\beta=\infty$) in the usual sense,
which solve the real part of the equation of motion, does not
necessarily solve the imaginary part of the equation. To see this,
let us consider the simple example of the lattice Gribov
copy\cite{lattice-gribov-copies,recent-review-on-chiral-fermion,
renormalizable-gauge-model}
given as
\begin{equation}
g_n = \left\{
    \begin{array}{cc}
       \bbone & (n\not = n_0) \\
       {\rm diag}(-1,-1, 1, \ldots, 1) & (n = n_0)
    \end{array} \right.
 \quad c_n^a = 0, \quad \bar c_n^a = 0 ,
\end{equation}
In this case, the group space and the momentum and spinor spaces are
mixed in the matrix to be inverted. Then the trace over the group
indices does not vanish in general. The dependence on the signatures
of the masses $\pm m_0$ does not drop off either and the cancellation
hardly occurs. Therefore, this example of the Gribov copy does not solve
the equation of motion. (The rigorous proof is beyond the scope
of this paper. It is not difficult to check numerically
on finite lattice that the imaginary part does not vanish for the
given lattice Gribov copy.)
The lattice Gribov copies which do not solve the imaginary part of the
equation of motion are no more the stationary point of the
total action and are suppressed by the quantum fluctuation around
them in the gauge average. This is the way how the irrelevant gauge
symmetry breaking terms in the complex phase of chiral determinant
can resolve the degeneracy due to the lattice Gribov copies.
Note that we do not claim here that all the Gribov copies should
be suppressed. Especially, not for the continuum Gribov copies.
Rather we point out a possible mechanism to suppress the lattice
Gribov copies in the case of the lattice chiral gauge theory.

\subsection{Gaussian gauge fluctuation and Infrared singularity}
\label{subsec:gaussian-gauge-fluctuation-IR-singularity}

Let us assume that a classical solution for the equations of motion
of $g$, $c$ and $\bar c$ is given.
We parameterize the classical solution
$g_n$ as\footnote{
The relation to the pure gauge vector potential $\hat A_{n\mu}$ of
Eq.~(\ref{eq:pure-gauge-vector-potential}) is given by
\begin{equation}
  \hat A_{n\mu}^a = 2 {\rm Tr} \left\{ T^a \sin A_{n\mu} \right\} .
\end{equation}
}
\begin{equation}
\label{eq:pure-gauge-link-variable-of-classical solution}
  g_n g^\dagger_{n+\hat\mu} = \exp \left( i A_{n\mu} \right) .
\end{equation}
For the technical reason to extract the local operators
of the background fields, we also assume that this solution is
sufficiently slowly varying in the lattice unit:
\begin{equation}
  \nabla_\nu A_{n\mu} << A_{n\mu} .
\end{equation}
Then we expand the action in terms of the fluctuations
$\pi$, $\xi$, $\bar \xi$ up to quadratic terms as
\begin{eqnarray}
\label{eq:action-of-gaussian-gauge-fluctuation}
  S[\exp(i\pi)g,c+\xi,\bar c+\bar \xi ]
&=& S[g,c,\bar c] + S_0[\pi^2 / \xi,\bar\xi] \nonumber \\
&+& S_1[\pi^2 / \xi,\bar\xi; \sin A]
       + S_{1c}[\pi,\xi / \bar\xi ;g ,c / \bar c] \nonumber\\
&+& S_2[\pi^2 / \xi,\bar\xi; \sin^2 A ,\cos A -1 ]
       + S_{2c}[\pi^2 ; g ,c,\bar c]        \nonumber\\
&+& i \, \Delta S_{WZW}[\pi^2;g] .
\end{eqnarray}
The explicit formula of these terms are given in the appendix
\ref{appendix:action-of-gaussian-gauge-fluctuation}.

$S_0$ is the free action of the fluctuations $\pi$, $\xi$ and
$\bar \xi$.
\begin{equation}
  S_0[\pi^2 / \xi,\bar\xi] =
-\sum_{n,a} \frac{1}{2}\left(\nabla^2 \pi_n^a \right)^2
+ \sum_{n,a} \bar \xi_n^a \nabla^2 \xi_n^a .
\end{equation}
This gives a massless dipole propagator for
$\pi$\cite{pure-gauge-model}:
\begin{equation}
\label{eq:massless-dipole-propagator-of-pi}
  \langle \pi_n^a \pi_m^b \rangle
= \delta^{ab} \int \frac{d^4 p}{(2\pi)^4} e^{ip(n-m)}
    \frac{1}{\left( \sum_\mu 4 \sin^2 \frac{p_\mu}{2} \right)^2 }
\equiv \delta^{ab} G(n-m) .
\end{equation}
This is infrared(IR) divergent in four-dimensions and we need a
certain IR regularization in the perturbation expansion. For this,
we add the mass term for $\pi$,
\begin{equation}
 - \sum_{na} \frac{1}{2}\,  \mu_0^4 \pi_n^a \pi_n^a .
\end{equation}
(We denote the dimensionless mass parameter with subscript ``0'' as
$\mu_0$ and the dimensional one without it: $\mu a = \mu_0$.)
As for the gauge fixing sector, we may invoke the dimensional
regularization, which preserves the BRST invariance of this sector.

$i \Delta S_{ WZW}$ is the contribution from the
Wess-Zumino-Witten term. According to the decomposition of
the classical solution $g_n$ and the quantum fluctuation $\pi_n$,
we can write the Wess-Zumino-Witten term as
\begin{eqnarray}
\label{eq:null-Wess-Zumino-Witten-term-background-fluctuation}
e^{i \Delta S_{WZW}[\pi;g]}
&=&
e^{i \Delta\Gamma_{WZW}[\exp(i\lambda\pi)g]-i \Delta\Gamma_{WZW}[g]}
\nonumber\\
&=&
   \prod_{rep.} \left(
   \frac{\lvacp \vert \hat \Pi \hat G \rvacp}
  {|\lvacp \vert \hat \Pi \hat G \rvacp|}
   \frac{\lvacm \vert \hat G^\dagger \hat \Pi^\dagger \rvacm}
        {|\lvacm \vert \hat G^\dagger \hat \Pi^\dagger \rvacm|}
\left/
   \frac{\lvacp \vert \hat G \rvacp}
  {|\lvacp \vert  \hat G \rvacp|}
   \frac{\lvacm \vert \hat G^\dagger  \rvacm}
        {|\lvacm \vert \hat G^\dagger  \rvacm|}
\right.
              \right) \nonumber\\
&=&
   \prod_{rep.} \left(
   \frac{\lvacp \vert \hat \Pi \rVacp}
  {|\lvacp \vert \hat \Pi \rVacp|}
   \frac{\lVacm \vert \hat \Pi^\dagger \rvacm}
        {|\lVacm \vert \hat \Pi^\dagger \rvacm|}
\left/
   \frac{\lvacp \rVacp}
  {|\lvacp \rVacp|}
   \frac{\lVacm \rvacm}
        {|\lVacm \rvacm|}
\right.
              \right) . \nonumber\\
\end{eqnarray}
$\hat \Pi$ is the operator of the gauge transformation due to
$\pi$ given by:
\begin{equation}
\hat \Pi
=
\exp \left( \hat a_n^{\dagger i}
            \{i \lambda \pi_n\}_i{}^j \hat a_{n j} \right).
\end{equation}
$\rVacp$ and $\rVacm$ in this case are given by
the vacua of the second-quantized Hamiltonians with
the pure gauge link variable
Eq.~(\ref{eq:pure-gauge-link-variable-of-classical solution}),
which consist of the classical solutions $g_n$. We can evaluate
these vacua in the expansion in terms of $A_{n\mu}$ using the
Hamiltonian perturbation
theory\cite{original-overlap,
randjbar-daemi-strathdee,yamada}.
Then $i \Delta S_{WZW}$ is obtained
in the form
\begin{equation}
  i \, \Delta S_{WZW}[\pi;g]
  = \sum_{k=1}^\infty i \, \Delta S_{k \, WZW}[\pi; A^k] .
\end{equation}
This result is further expanded in terms of $\pi$. The linear term
in $\pi$ gives the contribution to the equation of motion.
(Note that the formula for this contribution given in
Eq.~(\ref{eq:classical-equation-of-motion-g}) is evaluated
directly from the first expression in
Eq.~(\ref{eq:null-Wess-Zumino-Witten-term-background-fluctuation}).)
The quadratic term in $\pi$ is what we need for the one-loop
calculation by the background field method. We can see that every
vertexes of $A_\mu$'s and $\pi$'s obtained in this expansion of $i \Delta
S_{WZW}$ are {\it local in the unit of the lattice spacing}. This is because
they can be regarded as the loop effect of the (five-dimensional) Wilson
fermion with the mass $\pm m_0$ of the order of the lattice cutoff.

We can show that the leading and the next-to-leading terms
vanish identically in anomaly-free theories.
The leading term has a generic form as
\begin{eqnarray}
\label{eq:gaussian-fluctuation-induced-vertex1-WZW}
i \Delta S_{1\, WZW}[\pi^2;A]
&=& \sum_{rep.} \sum_{n,l_1,l_2}
{\rm Tr}\left\{ A_{n\mu} \pi_{l_1} \pi_{l_2} \right\} \times
\nonumber\\
&&\quad
\int \frac{d^4 k_1}{(2\pi)^4} \frac{d^4 k_2}{(2\pi)^4}  \,
     e^{ik_1(l_1-m)+ik_2 (l_2-m)} \,
     \Gamma_{1\, WZW\,\mu}(k_1,k_2) . \nonumber\\
\end{eqnarray}
Since this imaginary term must be odd under the charge
conjugation\cite{original-overlap}:
\begin{eqnarray}
 A_{n\mu} &\longrightarrow& - A_{n\mu}^\ast , \\
 \pi_{n} &\longrightarrow& - \pi_{n}^\ast   ,
\end{eqnarray}
it turns out to be proportional to the factor
\begin{eqnarray}
\sum_{rep.}
{\rm Tr}\left\{
    A_{n\mu} \pi_{l_1} \pi_{l_2}
+   \pi_{l_2}^\ast \pi_{l_1}^\ast A_{n\mu}^\ast
\right\}
&=&
\sum_{rep.}
{\rm Tr}\left\{
    A_{n\mu} \left\{ \pi_{l_1} , \pi_{l_2} \right\}
        \right\} \nonumber \\
&=& \sum_{rep.} d^{abc} A_{n\mu}^a \pi_{l_1}^b \pi_{l_2}^c ,
\end{eqnarray}
Therefore, it vanishes in anomaly-free theories.
\begin{equation}
\label{eq:property-WZW-term-first-order}
i \, \Delta S_{1 \, WZW}[\pi^2; A] = 0 .
\end{equation}

The next-to-leading term $i \Delta S_{2\, WZW}$
has the following generic form.
\begin{eqnarray}
\label{eq:gaussian-fluctuation-induced-vertex2-WZW}
&& i \Delta S_{2\, WZW}[\pi^2;A^2] \nonumber\\
&&= \sum_{n,m,l_1,l_2}
{\rm Tr}\left\{ A_{n\mu} A_{m\nu} \pi_{l_1} \pi_{l_2} \right\} \times
\nonumber\\
&&\quad
\int \frac{d^4 p}{(2\pi)^4}
     \frac{d^4 k_1}{(2\pi)^4} \frac{d^4 k_2}{(2\pi)^4}  \,
     e^{ip(m-n)+ik_1(l_1-n)+ik_2 (l_2-n)} \,
     \Gamma_{2\, WZW\, \mu\nu}(p,k_1,k_2) .  \nonumber\\
\end{eqnarray}
This term is also odd under the charge conjugation and it turns out
to be proportional to the factor
\begin{eqnarray}
&&
\sum_{rep.}
{\rm Tr}\left\{ A_{n\mu} A_{m\nu} \pi_{l_1} \pi_{l_2}
   -\pi_{l_2}^\ast  \pi_{l_1}^\ast  A_{m\nu}^\ast  A_{n\mu}^\ast  \right\}
\nonumber\\
&&=\sum_{rep.}
{\rm Tr}\left\{ T^a T^b T^c T^d - T^b T^a T^d T^c \right\}
A_{n\mu}^a A_{m\nu}^b \pi_{l_1}^c \pi_{l_2}^d    \nonumber\\
&&=\sum_{rep} \frac{1}{4} \, k[r] \,
              \left( i f^{abe}d^{cde}[r]+i d^{abe}[r] f^{cde} \right) \,
              A_{n\mu}^a A_{m\nu}^b \pi_{l_1}^c \pi_{l_2}^d ,
\end{eqnarray}
where we have used the identity
\begin{equation}
  T^a T^b = \frac{1}{2} \, i f^{abe} T^e
           + \frac{1}{2} \,  d^{abe}[r] T^e
           + \frac{1}{2} \, k[r] \, \delta^{ab}  \,
              \frac{1}{{\rm dim}[r]}\bbone .
\end{equation}
Therefore, the next-to-leading term also vanishes for anomaly free theories.
\begin{equation}
\label{eq:property-WZW-term-second-order}
i \Delta S_{2\, WZW}[\pi^2;A^2]= 0 .
\end{equation}
As we will see in the next subsection,
the fact that the two leading terms of $i \Delta S_{WZW}$
vanish identically\footnote{
The third order term does not seem to vanish identically.
This term turns out to be proportional to the factor
\begin{eqnarray}
&&
\sum_{rep.}
{\rm Tr}\left\{ A_{n\mu} A_{m\nu} A_{p\lambda} \pi_{l_1} \pi_{l_2}
 +\pi_{l_2}^\ast  \pi_{l_1}^\ast
A_{p\lambda}^\ast  A_{m\nu}^\ast  A_{n\mu}^\ast  \right\}
\nonumber\\
&&=\sum_{rep.}
{\rm Tr}\left\{ T^a T^b T^c T^d T^e + T^e T^d T^c T^b T^a \right\}
A_{n\mu}^a A_{m\nu}^b A_{p\lambda}^c \pi_{l_1}^d \pi_{l_2}^e    \nonumber\\
&&=\sum_{rep} \left(
\frac{i}{8} k[r] \left(i f^{abf} d^{cgf}+ i d^{abf} f^{cgf} \right)
                 f^{deg}
              \right.
\nonumber\\
&&            \left.
\qquad
+\frac{1}{8}k[r]
           \left( 2\frac{k[r]}{{\rm dim}[r]}\delta^{ab}\delta^{cg}
                 - f^{abf} f^{cgf} + d^{abf} d^{cgf} \right)
                 d^{deg}
+\frac{k[r]^2}{4 {\rm dim}[r]} d^{abc} \delta^{de}
\right) \times \nonumber\\
&& \qquad\qquad\qquad
A_{n\mu}^a A_{m\nu}^b A_{p\lambda}^c \pi_{l_1}^d \pi_{l_2}^e . \nonumber
\end{eqnarray}
We can see that the terms which are not
linear in the anomaly coefficient $k[r]\, d^{abc}[r]$
appear in this order.
} is sufficient for the one-loop renormalizability
of the pure gauge model. The structure of these two terms are discussed
in some detail in the
appendix \ref{appendix:contribution-WZW-action}.

\subsection{Quantum effect at one-loop and asymptotic freedom}
\label{subsec:quantum-effect-one-loop-asymptotoic-free}

The quantum correction to the classical action is obtained
by performing the Gaussian functional integration with $S_0$,
\begin{equation}
\exp \left(\Delta S[g,c,\bar c] \right) =
\int[d\pi d\xi d\bar\xi] \, e^{ S_0 } \,
 \sum_{l=0}^\infty
\frac{1}{l !} \left( S_1+S_{1c}+S_2+S_{2c}+iS_{WZW} \right)^l ,
\end{equation}
which can be obtained in the expansion in terms of $A_{\mu}$,
$c$ and $\bar c$,
\begin{equation}
\Delta S[g,c,\bar c] = \sum_{j=1} \Delta S_j [A^j]
                      +\sum_{k,l=1} \Delta S_{k,l} [A^k,(c,\bar c)^l] .
\end{equation}
{}From each one-loop contribution, the local operators
which can be written in terms of the pure gauge vector potential
$A_\mu$ and the ghost and anti-ghost fields, $c$ and $\bar c$,
should be extracted.
The possible local operators of dimensions less than five
can be listed up based on the global $SU(N)$ gauge symmetry, the
hyper-cubic symmetry, the CP invariance, the ghost number conservation
and the anti-ghost shift symmetry\cite{rome-approach}.
The complete list for the $SU(N)$ gauge group is given in the
appendix \ref{appendix:all-local-operator-dim4}.
They are denoted by ${\cal O}_k\,(k=0,\cdots,13)$. Among them,
only ${\cal O}_0$ is the BRST invariant combination.
Using these operators, the quantum correction can be expressed
formally in the form:
\begin{equation}
\Delta S[g,c,\bar c] = \sum_k c_k {\cal O}_k[A,c,\bar c] + \cdots .
\end{equation}

Up to the fourth order in $A_{\mu}$, $c$ and $\bar c$, we obtain
\begin{eqnarray}
\label{eq:one-loop-effective-action-A1}
\Delta S_1 [A]  &=&   \langle S_1 [A] \rangle_0  , \\
\label{eq:one-loop-effective-action-A2}
\Delta S_2 [A^2] &=& \langle S_2[A^2] \rangle_0
                    +\frac{1}{2!} \langle S_1[A] ^2 \rangle_0 , \\
\label{eq:one-loop-effective-action-c2}
\Delta S_{0,1}[(c,\bar c)]
&=&
\langle S_{2c}[(c,\bar c)] \rangle_0
+ \frac{1}{2} \langle S_{1c}[c /\bar c] ^2  \rangle_0  , \\
\label{eq:one-loop-effective-action-Ac2}
\Delta S_{1,1}[A,(c,\bar c)]
&=&
\langle S_{2c}[A,(c,\bar c)] \rangle_0
   + \langle S_{2c}[(c,\bar c)] \, S_1[A] \rangle_0 \nonumber\\
&& + \langle S_{1c}[c /\bar c] \, S_{1c}[A, c /\bar c] \rangle_0 ,
   \nonumber\\
\\
\label{eq:one-loop-effective-action-A3}
\Delta S_3 [A^3] &=&  \langle S_1 [A^3] \rangle_0
                    + \langle S_1[A] S_2[A^2]  \rangle_0
                    + \frac{1}{3!} \langle S_1[A]^3 \rangle_0 \nonumber\\
&& + \langle  iS_{3 \, WZW}[A^3] \rangle_0 , \\
\label{eq:one-loop-effective-action-A4}
\Delta S_4 [A^4] &=&  \langle S_2 [A^4] \rangle_0
                    + \langle S_1[A] S_1[A^3]  \rangle_0
                    +\frac{1}{2!} \langle S_2[A^2]^2 \rangle_0
                    \nonumber\\
&& +\frac{1}{2} \langle S_1[A]^2 S_2[A^2] \rangle_0
   +\frac{1}{4!} \langle S_1[A]^4 \rangle_0 \nonumber\\
&& + \langle  iS_{4 \, WZW}[A^4] \rangle_0
   + \langle  iS_{3 \, WZW}[A^2] S_1[A] \rangle_0  . \nonumber\\
\end{eqnarray}
Here $\langle \, \rangle_0$ stands for the connected part of
the expectation value evaluated with the Gaussian weight $S_0$.
We have taken account of the fact that $\bar c$ is always associated
with the derivative $\nabla_\mu$ because of the invariance of
the original action under the shift symmetry:
\begin{equation}
 \bar c_n \longrightarrow \bar c_n + \bar \epsilon .
\end{equation}
We have also taken account of the fact that
$i \Delta S_{1 \, WZW}$ and $i \Delta S_{2 \, WZW}$ are
identically zero.

We can see in the above expressions that
all the contributions
from the Wess-Zumino-Witten term
are imaginary at this order.
To be real, it needs at least two imaginary induced vertexes.
However, the imaginary induced vertexes quadratic in $\pi$,
$i\Delta S_{k\, WZW}[\pi^2;A^k]$'s,
start from the third order in the pure gauge vector potential $A_\mu$.
Then it must have the dimensions more than five.
On the other hand,
the local operators of dimensions less than five, ${\cal O}_k$'s,
preserve parity and charge conjugation invariance separately and
are real.
Therefore, there occurs no contribution due to
the Wess-Zumino-Witten term
to these local operators of dimensions less than five at the one-loop.
All contributions come from the gauge fixing sector
and they respect the BRST invariance.

The BRST invariant operator of dimensions less than five, which is
given by ${\cal O}_0$ in the list of the
appendix \ref{appendix:all-local-operator-dim4},
is nothing but the classical action:
\begin{equation}
\label{eq:O-0}
{\cal O}_0 = \left[
- \frac{1}{2\alpha} \sum_{n}
  \left( \bar \nabla_\mu A_{n\mu}^a \right)^2
- \sum_{n}\bar c_n^a \bar \nabla_\mu
  \left( - \nabla_\mu c_n^a
         + \frac{1}{2} f^{abc} (c_n^b+c_{n+\hat\mu}^b) A_\mu^c \right)
\right] . \nonumber\\
\end{equation}
Then it is enough to evaluate the terms up to the order
${\cal O}(A^2)$. They are given by
Eqs.~(\ref{eq:one-loop-effective-action-A1}),
(\ref{eq:one-loop-effective-action-A2}),
(\ref{eq:one-loop-effective-action-c2}) and
(\ref{eq:one-loop-effective-action-Ac2}), and are evaluated as follows:
\begin{eqnarray}
\label{eq:one-loop-correction-to-classical-action}
\Delta S_1 + \Delta S_2 + \Delta S_{1c} + \Delta S_{2c}
&=& - \lambda^2 \left[ N \bar G
                      + \left( \frac{N^2-1}{N} \right) [G(0)-G(1)] \right]
    \, {\cal O}_0 . \nonumber\\
\end{eqnarray}
where
\begin{eqnarray}
\label{eq:one-loop-correction-to-classical-action-divergent-part}
  \bar G &=& \frac{1}{2} \int \frac{d^4 k}{(2\pi)^4}
           \frac{ \left( \sum_\mu 4 \sin^2 \frac{k_\mu}{2} \right)
                   \left( \sum_\mu \sin^2 k_\mu \right) }
           {\left[
              \left(\sum_\mu 4 \sin^2 \frac{k_\mu}{2}\right)^2+\mu_0^4
            \right]^2 } \nonumber\\
&\simeq& - \frac{1}{16\pi^2} \, \ln (\mu_0) + \bar C \qquad (\mu_0 << 1).
\end{eqnarray}
The calculation is described in some detail in
the appendix \ref{appendix:detail-one-loop-calculation}.
This logarithmic divergence is renormalizable into $\lambda$:
\begin{equation}
  \frac{1}{\lambda^2} + \frac{N}{16\pi^2} \ln (a \mu) =
  \frac{1}{\lambda_R^2} .
\end{equation}
This proves the one-loop renormalizability of the pure gauge model
of the lattice $SU(N)$ chiral gauge theory defined by the vacuum overlap.
This one-loop renormalizability can be regarded as an indication of the
``smallness'' of the gauge symmetry breaking which is caused
by the gauge non-invariant definition of the complex phase of chiral
determinant\cite{original-overlap}.

The beta function of $\lambda$ can be evaluated as
\begin{equation}
\beta(\lambda)
=  - a \frac{\partial \lambda}{\partial a}
= - \frac{N}{32\pi^2} \lambda^3 ,
\end{equation}
which means that $\lambda$ is asymptotically free, as noticed by
Hata\cite{pure-gauge-model}.

\subsection{IR singularity and unbroken global gauge symmetry}
\label{subsec:IR-singularity-unbroken-global-gauge-symmetry}

In this subsection, we examine the realization of the global
gauge symmetry at $\beta=0$ in the pure gauge model within
the framework of the perturbation theory.
As we have seen in Eq.~(\ref{eq:massless-dipole-propagator-of-pi})
in the
subsection \ref{subsec:gaussian-gauge-fluctuation-IR-singularity},
the gauge fluctuation shows severe IR singularity.
The local order parameter
\begin{equation}
  \left\langle g_n{}_i^j \right\rangle
\end{equation}
suffers from the IR divergence and cannot be regarded to be
the physical observable.
This situation is very analogous to the two-dimensional
chiral $SU(N)$ nonlinear sigma model, as noticed by
Hata\cite{pure-gauge-model}.
This analogy suggests us that
we would examine the chiral $SU(N)$ invariant two-point
function of the gauge freedom, in which the IR divergence is
expected to cancel by the similar mechanism in the two-dimensional
case\cite{IR-finiteness-2D-nonlinear-sigma-model}.
Based on the one-loop renormalizability of the pure gauge model,
which we have observed in the previous section,
we can examine it also by the renormalization group method.

The invariant two-point correlation function of $g$
is defined by
\begin{eqnarray}
G_g(n)&\equiv&  \left\langle
                \frac{1}{N}{\rm Tr } \left\{ g_n g^\dagger_0 \right\}
                \right\rangle  .
\end{eqnarray}
In the perturbation expansion in terms of $\lambda$, it is given
at the leading order as
\begin{equation}
\label{eq:invariant-two-point-correlation-function-leading}
G_g(n)  = 1 + \lambda^2 \left(\frac{N^2-1}{2N}\right)
              \left[ G(n)-G(0) \right]  ,
\end{equation}
where
\begin{eqnarray}
\left[ G(n)-G(0) \right]
&\equiv&
\int \frac{d^4 p}{(2\pi)^4}
    \frac{e^{ip n } - 1}
         {\left( \sum_\mu 4 \sin^2 \frac{p_\mu}{2} \right)^2}
\nonumber\\
&\simeq& -\frac{1}{8\pi^2} \ln \left( \vert n \vert \right) ,
\quad ( |n| >> 1).
\end{eqnarray}
Note that the IR divergence associated to the propagator of $\pi$
is actually canceled in this invariant correlation function.
We assume this cancellation of the IR divergence holds at higher
orders as in the two-dimensional nonlinear sigma
model\cite{pure-gauge-model,IR-finiteness-2D-nonlinear-sigma-model}.
Then $G(n)$ is a function of the dimensionless parameters
$|x|/a \, ( n=a x )$ and $\lambda$.

This two-point function should satisfy the renormalization group equation,
\begin{equation}
 \left[  -a \frac{\partial}{\partial a}
        + \beta(\lambda) \frac{\partial}{\partial \lambda}
        - 2 \gamma_g(\lambda) \right] G_g(|x|/a,\lambda) = 0.
\end{equation}
Using the leading order result
Eq.~(\ref{eq:invariant-two-point-correlation-function-leading}),
the anomalous dimension of $g$ can be evaluated as
\begin{equation}
2 \gamma_g = -\left(\frac{N^2-1}{2N}\right) \frac{1}{8\pi^2} \lambda^2 .
\end{equation}

In order to see the long-distance behavior of the two-point function,
we will solve the renormalization group equation and improve the
two-point function by the renormalization group method.
We first introduce the effective coupling constant $\lambda_x$
at the distance $|x|$ by the equation
\begin{equation}
- x_\mu \frac{\partial}{\partial x_\mu} \lambda_x = \beta(\lambda_x),
\quad  \lambda_a=\lambda .
\end{equation}
$\lambda_a$ is identified with the bare coupling constant $\lambda$.
Then the solution of the renormalization group equation can be written
in the form
\begin{equation}
  G_g(|x|/a,\lambda_a)=
\exp\left( -\int^{\lambda_x}_{\lambda_a} d\lambda
            \frac{2\gamma_g(\lambda) }{\beta(\lambda)} \right)
  G_g(1,\lambda_x) ,
\end{equation}
Using the leading order result
Eq.~(\ref{eq:invariant-two-point-correlation-function-leading})
and the one-loop $\beta$ and $\gamma_g$, we obtain
\begin{eqnarray}
G_g(|x|/a,\lambda) &= &
\left( \frac{\displaystyle a}{\displaystyle |x|}
   \right)^{ \left(\frac{N^2-1}{2N}\right)
                              \frac{\lambda_x^2}{8\pi^2}} \times
\nonumber\\
&&\qquad\qquad
\left(1+\lambda_x^2 \left(\frac{N^2-1}{2N}\right)
              \left[ G(1)-G(0) \right] \right)  .
\end{eqnarray}
\begin{equation}
  \frac{1}{\lambda_x^2}=\frac{1}{\lambda_a^2}
                       -\frac{N}{16\pi^2} \ln ( |x|/a ) .
\end{equation}

We should note that the validity of this improved two-point
function is restricted in the region of the distance
\begin{equation}
 a \le |x| < \frac{1}{\Lambda_\alpha} ,
\end{equation}
where $\Lambda_\alpha$ stands for the renormalization group invariant
mass scale of the pure gauge model:
\begin{equation}
\Lambda_\alpha = \frac{1}{a}
             \exp\left( - \frac{16\pi^2}{N} \frac{1}{\lambda^2} \right) .
\end{equation}
In spite of the restricted validity of the two-point function,
its power behavior strongly suggests that the two-point function
vanishes at large distance and therefore that {\it the global gauge
symmetry does not break spontaneously at $\beta=\infty$}.
Furthermore, the asymptotic freedom is suggesting the
dynamical generation of the mass of the gauge freedom by the
dimensional transmutation:
\begin{equation}
\label{eq:dynamical-mass-of-gauge-degree-of-freedom}
  m
  = c \, \Lambda_\alpha .
\end{equation}

{}From the point of view of the
Foerster-Nielsen-Ninomiya
mechanism\cite{gauge-symmetry-restoration},
we see that the gauge symmetry breaking due to
the covariant gauge fixing term and the Faddeev-Popov ghost action
is very special; it gives the asymptotically free self-coupling of
the gauge degree of freedom and it can be ``large'' without
spoiling the disordered nature of the gauge freedom by virtue
of the severe IR singularity which occurs when it is ``large''.
Moreover, the coupling could control the mass scale of the gauge freedom
by the relation
Eq.~(\ref{eq:dynamical-mass-of-gauge-degree-of-freedom}),
if it would be generated dynamically.
Note that the situation is quite similar to that of
the two-dimensional nonabelian chiral gauge
theory\cite{kikukawa-gauge-two-dimensions}.

\section{Perturbative Foerster-Nielsen-Ninomiya mechanism}
\reseteqnum
\label{sec:perturbative-FNN-mechamism}

\subsection{Effect of explicit gauge symmetry breaking}

The one-loop renormalizability
and the symmetric realization of the global gauge symmetry
due to the asymptotic freedom and the IR singularity are
remarkable properties of the pure gauge model.
However, since the model breaks gauge (BRST) invariance explicitly
by the Wess-Zumino-Witten term, there is no reason based on symmetry
to expect that the renormalizability also holds at higher orders.
Rather, we expect the gauge non-invariant divergent and
finite terms, which can be written in terms of the local operators
of the dimension less than five,  ${\cal O}_k\, (k=1,\cdots, 13)$,
at some higher, but finite orders.

In this section,
in order to examine the effect of such gauge breaking terms,
we will consider the situation in which we have {\it the explicit gauge
breaking term} corresponding to
${\cal O}_1 = - \sum_{n\mu} \frac{1}{2} \, A_{n\mu}^a A_{n\mu}^a $:
\begin{eqnarray}
S_B
&=& K \sum_{n\mu} {\rm Tr}( U_{n\mu} + U_{n\mu}^\dagger) \\
&\longrightarrow &
\label{eq:gauge-breaking-perturbation-pure-gauge-limit}
K  \sum_{n\mu} {\rm Tr}(g_n g_{n+\mu}^\dagger+g_{n+\mu}g_n^\dagger)
\simeq  -  \sum_{n\mu} \frac{1}{2} \, K A_{n\mu}^a A_{n\mu}^a ,
\end{eqnarray}
and examine how this gauge symmetry breaking term would spoil
the disordered nature of the gauge freedom and causes the system
to fall into the broken phase, or how ``small'' such gauge symmetry
breaking terms must be in order to keep the system in the symmetric
phase. Note that this analysis is nothing but the analysis of
the Foerster-Nielsen-Ninomiya
mechanism\cite{gauge-symmetry-restoration}
in the context of the pure gauge model.
Here we can examine this dynamical issue
{\it within the framework of the perturbation theory}.

{}From the point of view of the (continuum limit)
perturbation theory of the nonabelian gauge theory,
it is well-known that the introduction of the mass term
of the vector boson does not affect the renormalizability
of the theory, although it breaks BRST invariance explicitly.
It is also known that based on this fact,
the mass term can be used as an IR regulator for the massless
vector boson\footnote{A proof of the renormalizability of the
nonabelian gauge theory with the mass term of the gauge field
based on the Ward-Takahashi (Slavnov-Taylor) identity
can be found, for example, in
\cite{renormalizability-with-IR-regularizaiton-mass-of-gauge-field}.}.
We will see, as expected, that this is also true in the pure gauge model.
Moreover, we will find a close relation between
the prescription of the IR regularization by the mass term and
the Foerster-Nielsen-Ninomiya mechanism.

\subsection{Renormalizability with gauge symmetry breaking}

Let us first check the renormalizability of the pure
gauge model in the presence of the explicit gauge symmetry breaking
term. We will repeat the one-loop analysis by the background field
method, including the gauge breaking term
Eq.~(\ref{eq:gauge-breaking-perturbation-pure-gauge-limit}).
It is expanded up to the quadratic terms in the quantum gauge
fluctuation $\pi$ as
\begin{eqnarray}
\label{eq:explicit-gauge-break-in-gaussian-gauge-fluctuation}
  S_B[ \exp(i\lambda \pi) g ]
&=& S_B[ g ] + S_{B\, 0}[\pi^2] \nonumber\\
&+& S_{B\, 1}[\pi^2, \sin A ] + S_{B\, 2}[\pi^2, \cos A -1 ]
+ {\cal O}(\pi^3) .
\end{eqnarray}
The explicit formula of these terms are given in the appendix
\ref{appendix:effect-gauge-breaking-term}.
The linear term is assumed to be included into the equation of motion.
$S_{B\, 0}[\pi^2]$, which is given as
\begin{equation}
S_{B\, 0}[\pi^2]=  -\frac{1}{2} K \lambda^2  \sum_{n\mu}
   \nabla_{\mu}\pi_n^a \nabla_{\mu}\pi_n^a ,
\end{equation}
gives a correction to the kinetic term of $\pi$. It modifies
the propagator of $\pi$ as follows:
\begin{eqnarray}
\label{eq:massless-dipole-propagator-of-pi-with-gauge-breaking}
  \langle \pi_n^a \pi_m^b \rangle
&=& \delta^{ab} \int \frac{d^4 p}{(2\pi)^4} e^{ip(n-m)}
    \frac{1}{\left( \sum_\mu 4 \sin^2 \frac{p_\mu}{2} \right)^2
             + M_0^2
             \left( \sum_\mu 4 \sin^2 \frac{p_\mu}{2} \right) }
\nonumber\\
&\equiv& \delta^{ab} G_B(n-m) .
\end{eqnarray}
Here we have set
\begin{equation}
  M_0^2 = K \lambda^2 .
\end{equation}
(We denote the dimensionless mass parameter with subscript ``0'' as
$M_0$ and the dimensional one without it: $M a = M_0$.)
This correction does not change the ultraviolet(UV) behavior of
the propagator, but does change the IR structure substantially:
it is no more IR divergent in four-dimensions.

The quantum correction to the classical action are now divided into
two classes: the first class consists of the contributions which does not
have the gauge-breaking vertexes from $S_B$ and are given by the same
expressions as
Eqs.~(\ref{eq:one-loop-effective-action-A1}),
     (\ref{eq:one-loop-effective-action-A2}),
     (\ref{eq:one-loop-effective-action-c2}),
     (\ref{eq:one-loop-effective-action-Ac2}),
     (\ref{eq:one-loop-effective-action-A3}) and
     (\ref{eq:one-loop-effective-action-A4}).
But these terms should be evaluated with the propagator given by
Eq.~(\ref{eq:massless-dipole-propagator-of-pi-with-gauge-breaking})
rather than  Eq.~(\ref{eq:massless-dipole-propagator-of-pi}).
Taking account of the fact that $M_0^2$ should be counted to have
mass dimension two, they are evaluated as
\begin{eqnarray}
\label{eq:one-loop-correction-to-classical-action-with-explicit-breaking}
&&\Delta S_1 + \Delta S_2 + \Delta S_{1c} + \Delta S_{2c} \nonumber\\
&&\qquad
= - \lambda^2 \left[ N \bar G_B
                   + \left( \frac{N^2-1}{N} [G_B(0)-G_B(1)] \right) \right]
    \, {\cal O}_0  \nonumber\\
&&\qquad \qquad
 + M_0^2
   \left[ N \tilde G_B
            - \left( \frac{N^2-1}{2N}\right)  \nabla^2 G_B(0) \right] \,
  {\cal O}_1 .
\end{eqnarray}
where
\begin{eqnarray}
\label{eq:one-loop-correction-to-classical-action-divergent-part-with-breaking}
  \bar G_B &=& \frac{1}{2} \int \frac{d^4 k}{(2\pi)^4}
           \frac{ \left( \sum_\mu \sin^2 k_\mu \right) }
           {\left[
             \sum_\mu 4 \sin^2 \frac{k_\mu}{2} + M_0^2
            \right]^3 } \nonumber\\
&\simeq& - \frac{1}{16\pi^2} \, \ln (M_0) + \bar C_B  \qquad (M_0 << 1) ,
\end{eqnarray}
and
\begin{eqnarray}
\label{eq:one-loop-correction-to-explicit-breaking-divergent-part}
  \tilde G_B &=& \frac{3}{4} \int \frac{d^4 k}{(2\pi)^4}
           \frac{
                   \left( \sum_\mu \sin^2 k_\mu \right) }
           {\left[
          \left( \sum_\mu 4 \sin^2 \frac{k_\mu}{2} \right)^2
          \left(\sum_\mu 4 \sin^2 \frac{k_\mu}{2}+ M_0^2 \right)
            \right] } \nonumber\\
&\simeq& - \frac{3}{32\pi^2} \, \ln (M_0) + \tilde C_B
\qquad (M_0 << 1) .
\end{eqnarray}

The second class consists of the additional contribution due to the
gauge-breaking vertexes.
They are given as follows up to the second order in $A_\mu$:
\begin{eqnarray}
\label{eq:one-loop-effective-action-A1-gauge-breaking}
\Delta S_{B\, 1}[A]   &=&  \langle S_{B\, 1}[A] \rangle_0 , \\
\label{eq:one-loop-effective-action-A2-gauge-breaking}
\Delta S_{B\, 2}[A^2] &=&  \langle S_{B\, 2}[A^2] \rangle_0
                          + \frac{1}{2} \langle S_{B\, 1}[A]^2 \rangle_0
                          + \langle S_{B\, 1}[A] S_1[A] \rangle_0 .
\end{eqnarray}
They are evaluated as follows:
\begin{eqnarray}
\Delta S_{B\, 1} + \Delta S_{B\, 2}
&=&  - M_0^2
\left(\frac{N^2-1}{2N}\right)
\left[ G_B(0)-G_B(1) \right] \, {\cal O}_1 .
\end{eqnarray}

{}From these results, we can see that the pure gauge model with
the explicit gauge symmetry breaking
Eq.~(\ref{eq:gauge-breaking-perturbation-pure-gauge-limit}) is
actually renormalizable at one-loop. The renormalization of $\lambda$ is
same as before and it is asymptotically free. $M_0^2=K \lambda^2$ is
multiplitically renormalized.
\begin{eqnarray}
  \frac{1}{\lambda^2} + \frac{N}{16\pi^2} \ln (a \mu )
  &=&   \frac{1}{\lambda_R^2} , \\
\label{eq:gauge-breaking-term-renormalization}
  \frac{M^2}{\lambda^2}
  \left( 1 - \lambda^2 \frac{5 N }{32 \pi^2} \ln (a \mu ) \right)
  &=& \frac{M_R^2}{\lambda_R^2} .
\end{eqnarray}
$\mu$ should be understood as a certain renormalization point.

\subsection{Restoration of chiral $SU(N)$ global symmetry}

Let us next examine the realization of the chiral
$SU(N)$ global symmetry.
Since the IR singularity of the pure gauge model is now
``regularized'' by the introduction of the explicit gauge symmetry
breaking term, we may examine the local order parameter
\begin{equation}
  \left\langle g_n{}_i^j \right\rangle .
\end{equation}
At the leading non-trivial order in the perturbation expansion, it
is evaluated as
\begin{eqnarray}
\left\langle g_n{}_i^j \right\rangle
&=& \delta_i^j \left( 1 - \frac{\lambda^2}{2!}
                          \left(\frac{N^2-1}{2N}\right) G_B(0)
               \right) . \nonumber
\end{eqnarray}
For a generic $K$ and a small $\lambda$, it does not vanish and
the chiral $SU(N)$ symmetry breaks down to its vector subgroup
$SU(N)_V$.

However, the second term is potentially IR(UV) divergent and
it becomes large as $K$ decreases:
\begin{eqnarray}
  G_B(0) &=& -\frac{1}{16\pi^2} \ln K \lambda^2  + C_B
                  \qquad ( 0 < K << 1 ) .
\end{eqnarray}
It could be large so that it cancels the first term.
As a crude approximation,
the critical point at which the local order parameter vanishes
can be estimated from this equation as
\begin{equation}
K_c \lambda^2
   \simeq \exp \left\{ - 32\pi^2 \left(\frac{2N}{N^2-1}\right)
                      \left[ \frac{1}{\lambda^2} -
                             \left(\frac{N^2-1}{4N}\right) C_B \right]
               \right\} .
\end{equation}
{}From this relation, we can see that,
if $K$ is sufficiently small so that $0 \le K < K_c$,
the chiral $SU(N)$ global symmetry can be restored.

We may examine the invariant two-point function as before.
At the leading order, it reads
\begin{equation}
\label{eq:invariant-two-point-correlation-function-leading-with-breaking}
G_{g B}(n)  = 1 + \lambda^2 \left(\frac{N^2-1}{2N}\right)
              \left[ G_B(n) - G_B(0) \right] .
\end{equation}
{}From the long distance behavior of $G_B(n)$ for $|n| >> 1$, we
should be able to extract the realization of the chiral $SU(N)$.
But in this case, since we have introduced the
parameter $aM=M_0$ which has mass dimension,
the long distance limit can depend on the relative scale to
this parameter.

In fact, the long distance behavior which
corresponds to the spontaneous breakdown of the chiral $SU(N)$
appears in the limit:
\begin{equation}
  |n| \,\, >> \,\, \frac{1}{M_0} \, \simeq \, 1 .
\end{equation}
In this limit, we can see that $G_B(n)$ falls off
in power and exponentially for large $|n|$,
by decomposing it into the propagators of the massless scalar
and the massive ghost:
\begin{eqnarray}
G_B(n)
&=&
\int \frac{d^4 p}{(2\pi)^4} \, e^{ip n } \,
    \frac{1}{M_0^2}
    \left( \frac{1}{\sum_\mu 4 \sin^2 \frac{p_\mu}{2} }
          -\frac{1}{\sum_\mu 4 \sin^2 \frac{p_\mu}{2} + M_0^2} \right)
\nonumber\\
&\simeq&
          A \frac{1}{M_0^2 |n|^2} - B e^{- c M_0 |n| }
     \qquad ( M_0 |n| > 1) .
\end{eqnarray}
Then the invariant two-point function does not vanish at large distance:
\begin{equation}
G_B(n) \longrightarrow
\frac{1}{N} {\rm Tr}\left\{
\left\langle g \right\rangle \left\langle g \right\rangle
\right\}
\not = 0 ,
 \quad ( |n| \longrightarrow \infty ),
\end{equation}
indicating the spontaneous breakdown of the chiral $SU(N)$ global symmetry.
Thus the ``large'' gauge symmetry breaking term causes the system to fall
into the broken phase. Here ``large'' means that
\begin{equation}
 M_0 = K \lambda^2 \simeq {\cal  O}(1) , \qquad {\rm or} \qquad
 M \simeq {\cal   O}\left(\frac{1}{a}\right) .
\end{equation}

On the other hand, we may consider another long distance limit such that
\begin{equation}
  \frac{1}{M_0} > |n| \,\, >> \,\,  1 .
\end{equation}
Then the behavior of $G_B(n)$ is logarithmic,
\begin{eqnarray}
G_B(n)
&\simeq&
  -\frac{1}{8\pi^2} \ln \left( M_0 |n| \right) ,
\qquad ( 0 < M_0 |n| < 1 ) .
\end{eqnarray}
This means that the invariant two-point function decreases
at the long distance by the power law. Since the perturbative
calculation is subject to the restriction of the validity
for the long distance due to the asymptotic freedom,
this second limit is a possible alternative as long as $M_0$ is
``small'' so that
\begin{equation}
1 \, \, << |n| \,\, < \,\,
 \exp\left( \frac{16\pi^2}{N} \frac{1}{\lambda^2} \right)
\,\, < \,\, \frac{1}{M_0} .
\end{equation}
This relation is giving the criterion about how ``small'' the gauge
symmetry breaking term $M_0$ must be for the chiral $SU(N)$ global
symmetry to be restored at long distance.
Note that this condition is consistent with the previous
crude estimation of the critical point.

Based on the renormalizability, the two-point function can be examined by
the renormalization group equation, which in this case reads
\begin{eqnarray}
 \left[  -a \frac{\partial}{\partial a}
        + \beta(\lambda) \frac{\partial}{\partial \lambda}
        - \gamma_M(\lambda) M^2 \frac{\partial}{\partial M^2 }
        - 2 \gamma_g(\lambda) \right] G_g(|x|/a,\lambda, a M ) = 0.
\nonumber\\
\end{eqnarray}
Here we assume the mass independent
renormalization scheme\cite{mass-independent-renormalization-procedure}
for a technical convenience.
{}From Eq.~(\ref{eq:gauge-breaking-term-renormalization}), we obtain
the anomalous dimension of $M^2$ as
\begin{equation}
\gamma_M (\lambda) = a \frac{\partial \ln M^2 }{\partial a}
                   = \frac{5N}{64\pi^2} \lambda^2 .
\end{equation}
Using the leading order result
Eq.~(\ref
{eq:invariant-two-point-correlation-function-leading-with-breaking}),
we also obtain the anomalous dimension of $g$ as before
\begin{equation}
2 \gamma_g = - \left(\frac{N^2-1}{2N}\right) \frac{1}{8\pi^2} \lambda^2 .
\end{equation}
We introduce the effective coupling constant $\lambda_x$ and the
effective mass $M^2_x$ at the distance $|x|$ by the equation
\begin{eqnarray}
- x_\mu \frac{\partial}{\partial x_\mu} \lambda_x
&=& \beta(\lambda_x); \quad  \lambda_a=\lambda , \\
 x_\mu \frac{\partial}{\partial x_\mu} \ln M^2_x
&=& \gamma_M(\lambda_x) ; \quad M^2_a=M^2 .
\end{eqnarray}
$\lambda_a$ and $M^2_a$ are identified with the bare coupling
constants $\lambda$ and $M^2$.
Then the solution of the renormalization group equation can be written
in the form
\begin{equation}
  G_g(|x|/a,\lambda,aM)=
\exp\left( - \int^{\lambda_x}_{\lambda_a} d\lambda
             \frac{2\gamma_g(\lambda) }{\beta(\lambda)} \right)
  G_g(1,\lambda_x, a M_x ) ,
\end{equation}
Using the leading order result
Eq.~(\ref{eq:invariant-two-point-correlation-function-leading}),
and the one-loop $\beta$, $\gamma_g$ and $\gamma_M$, we obtain
\begin{eqnarray}
G_g(|x|/a,\lambda, aM) &=&
\left( \frac{\displaystyle a}{\displaystyle |x|}
\right)^{\left(\frac{N^2-1}{2N}\right)
         \frac{\lambda_x^2}{8\pi^2}} \times \nonumber\\
&&\qquad\qquad
\left( 1 + \lambda_x^2 \left(\frac{N^2-1}{2N} \right)
                     \left[ G_B(1)-G_B(0) \right](aM_x)  \right) .
\nonumber\\
\end{eqnarray}
\begin{equation}
 M_x^2 = M_a^2
\left( \frac{\displaystyle |x|}{\displaystyle a}
\right)^{\frac{5N}{64\pi^2}\lambda_x^2} .
\end{equation}
{}From this expression, we can see that
the invariant two-point function has the power behavior
same as before, as long as the following relation among the scales is
satisfied:
\begin{equation}
  a \,\, << \,\, |x| \,\, < \,\, \frac{1}{\Lambda_\alpha} \,\,
                   < \frac{1}{M_{\Lambda_\alpha}} .
\end{equation}
This is the renormalized version of the criterion for the restoration
of the chiral $SU(N)$ global symmetry.
{}From the point of view of the prescription of the IR regularization,
this condition is just telling that
the IR regularization mass $M$ must be smaller than the scale of
the theory $\Lambda_\alpha$ so that its effect should be irrelevant
physically.

\subsection{Phase diagram of the pure gauge model}

This criterion for the restoration of the chiral $SU(N)$ global
symmetry can be regarded to define
the phase boundary
between the symmetric phase and the broken phase
in the coupling space $( \frac{1}{\alpha}, K )$
for the region of the small $\lambda$ and $K$:
\begin{equation}
  K_c(\lambda)  \, \simeq  \, \frac{1}{\lambda^2}
       \exp\left( -\frac{16\pi^2}{N} \frac{1}{\lambda^2} \right) .
\end{equation}
We can see that,
although it is tiny for a small $\lambda$,
a finite region of $K$ can exist, where the chiral $SU(N)$ global
symmetry is restored.
Namely, if $K$ is sufficiently small so that
$0 \le K < K_c(\lambda)$, the chiral $SU(N)$ would not be broken
spontaneously. As $\lambda$ becomes large, this symmetric region
is expected to become larger, because the bound becomes milder.

\begin{figure}
\epsfxsize=9cm
\centerline{\epsffile{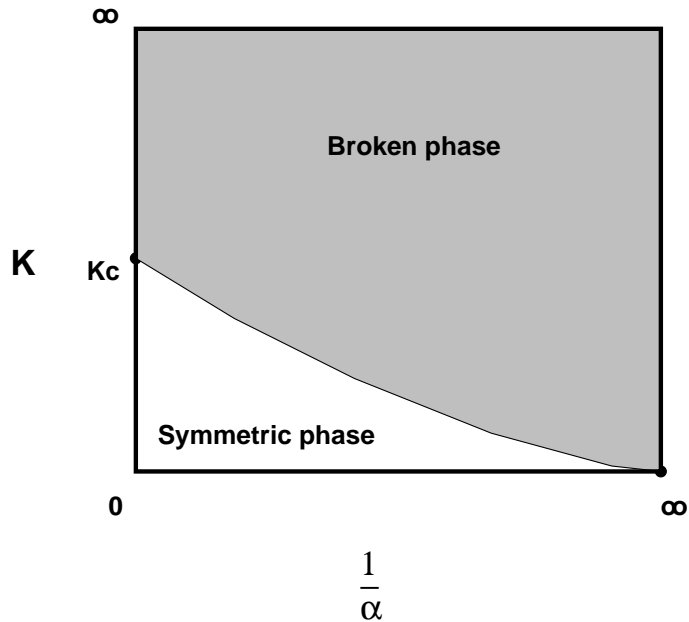}}
\caption{Possible phase structure for the pure gauge model of
anomaly-free $SU(N)$ non-abelian chiral gauge theory.}
\label{fig:phase-diagram-1}
\end{figure}

If no phase transition would occur in the course of removing
the covariant gauge fixing term along the limit:
\begin{equation}
\lambda \rightarrow \infty, \qquad K=0 ,
\end{equation}
then it is naturally expected that
the critical line $K_c(\lambda)$ finally reaches
the critical point of the usual four-dimensional $SU(N)$
non-linear sigma model (coupled to the fermion and ghost fields),
$K_c(\infty)$.
In the case of the anomaly-free theory, this assumption seems plausible.
Then the expected phase diagram in the coupling space
of $(1/\alpha,K)$ is given by Fig.~\ref{fig:phase-diagram-1}.
We should note that the existence of the symmetric phase in the region of
the small $\lambda$ can be possible here due to the IR singularity
and the non-abelian nature of the pure gauge dynamics, namely
the asymptotic freedom.

By the mean-field method, the phase structure has been examined
in \cite{renormalizable-gauge-model}
for the similar model with the higher derivative
coupling term which is induced from the covariant-type
gauge fixing action proposed in \cite{covariant-type-gauge-fixing-action}.
According to this result, the paramagnetic (PM) phase terminates at
a finite value of $1/\alpha$ on the line $K=0$.
Beyond this point, the critical line which separates
the ferromagnetic phase(FM) and the helicoidal-ferromagnetic
phase(FMD) lies along the $K=0$ line to the limit
$1/\alpha = \infty$.

The point we make here is the possibility that
this transition point among PM, FM and FMD phases
and the phase boundary line of FM and FMD phases
can be pushed away to the limit $1/\alpha = \infty$,
due to the non-abelian nature of the gauge freedom.
By virtue of the asymptotic freedom, the mass of the gauge freedom
can be generated and the paramagnetic(PM) phase can persist
up to the UV fixed point of $\alpha$ at $1/\alpha = \infty$.
Our analysis of the effect of the gauge symmetry breaking term based on
the perturbation theory is suggesting this possibility,
supporting the picture of the dynamical restoration of
the chiral $SU(N)$ symmetry given by Hata\cite{pure-gauge-model}.
This dynamical restoration is protected from the ``small''
gauge symmetry breaking by the Foerster-Nielsen-Ninomiya mechanism.

On the other hand, if an IR fixed point of $\lambda$ could occur,
the disordered phase which is found in the perturbation theory,
may not necessarily be connected to the disordered phase
of the usual four-dimensional non-linear sigma model
along the line
\begin{equation}
   0  \le   K   \le K_c(\infty), \qquad \lambda=\infty .
\end{equation}
In the case of the two-dimensional nonabelian chiral gauge
theory, the quantum effect of anomaly (Wess-Zumino-Witten term)
produces such an IR fixed point.
It is quite interesting if it could happen in four-dimensions,
because it would be possible to distinguish the
anomalous theory from the anomaly-free theory by the dynamical
nature of the gauge degree of freedom at $\beta=\infty$.
To explore this possibility, we need to examine the effects of the higher
order quantum corrections in both anomalous and anomaly-free
theories,
which is beyond the scope of this article.

In order to examine the phase structure of the pure gauge model fully,
we need the analysis beyond the perturbation theory.
We will leave this issue for future study.
The lattice formulation of the four-dimensional pure gauge model and
the possibility to examine it by the Monte Carlo simulation will be
discussed elsewhere\cite{kikukawa-lattice-pgm}.

\subsection{Induced gauge symmetry breaking and counter term}

Next we consider the effect of the induced gauge symmetry breaking
terms from the Wess-Zumino-Witten term.
Since there is no reason based on the symmetry, we cannot escape them.
Here we consider the breaking term of dimension two, ${\cal O}_1$,
as before.  Let us assume the situation in which
we get additive contributions of the induced gauge breaking
in the coupling $K$ and it is written effectively in the form
\begin{equation}
  K_{\rm eff} = K + K_{\rm ind}(\lambda) .
\end{equation}
By the consideration based on the charge conjugation and the anomaly
cancellation, the induced term is expected to appear first at the
three-loop order\footnote{The author owes this point to the discussion
with A.~Yamada.}:
\begin{equation}
  K_{\rm ind}(\lambda) =  \sum_{l\ge 3} \Delta_l \lambda^{2 l} .
\end{equation}
Since $K_c(\lambda)$ has the essential singularity in $\lambda$,
any polynomial of $\lambda$ cannot
satisfy the criterion of the symmetry restoration for small
$\lambda$. This means that any perturbatively induced term
can cause the spontaneous break down of the chiral $SU(N)$.
Therefore we are forced to introduce the counter term as a part of
the explicit breaking term so that $K_{\rm eff}$ satisfies
\begin{equation}
  0 \, < \, K_{\rm eff}
    \, < \,  K_c(\lambda) ,
\end{equation}
where
\begin{equation}
K_{\rm eff}= K + K_{\rm ind}(\lambda)+ \delta K_{\rm count}  .
\end{equation}
This fine tuning can be achieved,
since the finite region of $K$ for the symmetry restoration
can exist for the nonabelian gauge theory, as we have argued in the
previous subsection. As $\lambda$ becomes larger, we can expect
that this fine tuning becomes milder.
With this counter terms, the IR structure of the pure gauge model
can be maintained and all the properties of the pure gauge model
hold true.

\section{Disordered gauge freedom and chiral fermion}
\reseteqnum
\label{sec:disordered-gauge-freedom-chiral-fermion}

Based on the dynamical features observed in the previous section,
we assume that the global gauge symmetry does not break spontaneously
and the gauge freedom acquires mass dynamically in the pure gauge
limit for the entire region of the gauge parameter.
Under this assumption, we examine the spectrum of the fermion
correlation functions in the vacuum overlap formulation.
Our strategy is, with the help of the asymptotic freedom,
to tame the gauge fluctuation by approaching the critical point of
the gauge degree of freedom without spoiling its disordered nature.
Namely, we invoke the perturbation (spinwave) expansion in order
to examine the fermion spectrum in {\it the disordered phase}.

\subsection{Overlap correlation function}

Before discussing the dynamical effect of the disordered gauge freedom
to the fermion spectrum in lattice chiral gauge theory,
let us first summarize what happens in the pure gauge limit of the
lattice QCD. In the limit of vanishing gauge coupling,
the gauge variable is restricted to the pure gauge form
\begin{equation}
 U_{n\mu}= g^{}_n g_{n+\mu}^\dagger,  \quad  g^{}_n \in SU(3)
\qquad(\beta=\infty).
\end{equation}
Then the quark action reads
\begin{equation}
\sum_n \left\{
\frac{1}{2} \sum_{\mu} \left(
 \bar q_n
 (1-\gamma_\mu ) g^{}_n g_{n+\hat\mu}^\dagger
 q_{n+\hat\mu}
+\bar q_{n+\hat\mu}
 (1+\gamma_\mu ) g^{}_{n+\hat\mu} g_n^\dagger
 q_n
\right)
- m_0 \, \bar q^{}_n q^{}_n
\right\}.
\end{equation}

Since the action still possess the local gauge invariance,
we cannot identify directly the ``colored'' quark
even in the pure gauge limit through
the gauge non-invariant correlation function of the quark:
\begin{equation}
\label{eq:gauge-non-invariant-quark-correlation-function}
\left\langle   q^{}_n{}_i   \bar   q^{}_m{}^j \right\rangle .
\end{equation}
This is because
it must vanish according to the Elitzur's
theorem\cite{Elitzur-theorem}.

On the other hand, by the change of variable,
\begin{equation}
  q^{}_n{}_i \longrightarrow
  \psi_n{}_i \equiv \left( g_n^\dagger \right)_i^j q^{}_n{}_j,
\end{equation}
the action turns into that of the free Wilson fermion.
The functional integral measure of the quark field is invariant
under the change of variable. Therefore the gauge freedom decouples
completely from the fermion. This action has a symmetry under the global
transformation,
\begin{equation}
  \psi_n{}_i \rightarrow h_i^j \psi_n{}_j \quad h \in SU(3) ,
\end{equation}
This is not the global gauge symmetry but comes from the arbitrariness
of choice of the pure gauge variable.  We refer this as $SU(3)_{H}$.
$\psi_n$ belongs to the fundamental representation of $SU(3)_{H}$.

The fermion spectrum is measurable by the gauge invariant
correlation function
\begin{equation}
\label{eq:QCD_gauge-invariant_corr}
  \langle \psi_n{}_i \bar \psi_m{}^j \rangle ,
\end{equation}
which is given by the correlation function of the free Wilson fermion.
By virtue of the Wilson term, the masses of the species doubles
are lifted to cutoff scale. Since $\psi_n$  belongs to the fundamental
representation of the $SU(3)_{H}$, the correct number of
light {\it Dirac} fermions emerge just same as the quark in the
fundamental representation of $SU(3)_{Color}$.

In the lattice chiral gauge theory based on the vacuum overlap,
the correlation functions of fermion are
defined by putting creation and annihilation operators in the
overlap of vacua with different masses.
We refer the correlation function in this definition as
{\it overlap correlation function}.
The two-point overlap correlation function of the fermion
in the representation $r$ is defined as follows:
\begin{eqnarray}
\label{eq:covariant-overlap-correlation-function}
\langle a_n{}_i a^\dagger_m{}^j  \rangle_r
&\equiv& \frac{1}{Z}
\int [dU][d\bar c d c]
      \exp \left(-\beta S_G
           -\frac{1}{2\alpha} \left(\bar \nabla \hat A \right)^2
           -\bar c M[ U ] c \right)
\times \nonumber\\
&& \qquad
      \prod_{rep.} \left(
        \frac{ \lvacp \rVacp }
             {|\lvacp \rVacp|}
               \lVacp
               \rVacm
        \frac{ \lVacm \rvacm }
             {|\lVacm \rvacm|}
              \right)  \times
\nonumber\\
&& \qquad
         \lVacp
         \vert
         \left\{ \hat a_n{}_i \hat a^\dagger_m{}^j
                -\frac{1}{2} \delta_{nm} \delta_i^j \right\}
         \rVacm_r \Big/
         \lVacp \rVacm_r .
\end{eqnarray}
This correlation function does not transform covariantly
under the local gauge transformation
because of the explicit gauge symmetry breaking.
Only under the global gauge transformation
Eq.~(\ref{eq:sun-global-gauge-transformation}), it does.

In the pure gauge limit, it turns out to be
\begin{eqnarray}
\label{eq:covariant-overlap-correlation-function-pure-gauge}
\langle a_n{}_i a^\dagger_m{}^j  \rangle_r
&=&
\frac{1}{Z}
\int d\mu[g;\alpha] \times
\nonumber\\
&& \qquad
\lvacp \vert \hat G^\dagger
            \left\{ \hat a_n{}_i \hat a^\dagger_m{}^j
                    -\frac{1}{2} \delta_{nm} \delta_i^j \right\}
      \hat G \rvacm_r \Big/
      \lvacp \rvacm_r .
\nonumber\\
\end{eqnarray}
By noting the transformation property of the creation and annihilation
operators under the action of $\hat G$,
we can see that this correlation function has the structure
of the product of the free fermion two-point correlator and
the correlation function of the gauge degree of freedom:
\begin{equation}
  \left\langle \left( g_n{}_i^k \right)_r \,
               \left( g_m^\dagger{}_l^j \right)_r
  \right\rangle \times
\lvacp \vert
       \left\{ \hat a_n{}_k \hat a^\dagger_m{}^l
              -\frac{1}{2} \delta_{nm} \delta_k^l \right\}
      \rvacm_r \Big/
      \lvacp \rvacm_r  .
\end{equation}

If the gauge degree of freedom would be disordered and would have a very
short correlation length of the order of the lattice spacing,
it is hardly expected that we could find a physical fermion
in this correlation function, even when the free fermion
two-point correlator could have a pole of physical particle.
In fact, according to our perturbative analysis with the covariant gauge
fixing term given in the previous section,
the non-invariant correlation functions suffer from the IR divergence
in general and they cannot be regarded as the physical observables.
We may think of the invariant counterpart of the above correlation
function by taking the trace over the group indexes.
However, since the invariant correlation function of the gauge
degree of freedom shows the power behavior
\begin{equation}
\left\langle
    \frac{1}{N}{\rm Tr } \left\{ g_n g^\dagger_0 \right\}
\right\rangle
=
\left( \frac{\displaystyle 1}{\displaystyle |n|}
   \right)^{\left(\frac{N^2-1}{2N}\right)
                              \frac{\lambda^2}{8\pi^2}} ,
\end{equation}
its combolution with the free fermion correlator cannot describe
a physical free particle\cite{screening-of-quark-continuum-limit}.
We may regard this screening phenomenon
as the counterpart of the Elitzur's theorem in the Lattice QCD.
The theorem, as noticed in \cite{original-overlap}, does not seem
possible to be established in the lattice chiral gauge theory
defined by the vacuum overlap because of the nonlocal nature
of the fermion determinant, even when it is formulated
with the manifest local gauge invariance by the explicit
integration over the gauge degree of freedom.
But, due to the disordered gauge degree of freedom,
the screening of the ``color charge'' can take place effectively.
This correspondence is quite analogous to the mapping of the
dynamical features between the two-dimensional nonlinear sigma model
and the four-dimensional Yang-Mills theory.

On the other hand, just in the same manner as the lattice QCD
at $\beta=\infty$ as we reviewed above,
we can consider the change of variables
\begin{equation}
  a_n{}_i \longrightarrow b_n{}_i = g_n^\dagger{}_i^j a_n{}_j .
\end{equation}
And we may define another fermion two-point correlation function by
\begin{eqnarray}
\label{eq:H-charged-fermion-correlation-function}
\langle b_n{}_i b^\dagger_m{}^j  \rangle_r
&\equiv&
\frac{1}{Z}
\int d\mu[g;\alpha] \times
\nonumber\\
&& \qquad
\lvacp \vert \hat G^\dagger
            \left\{ \hat b_n{}_i \hat b^\dagger_m{}^j
                    -\frac{1}{2} \delta_{nm} \delta_i^j \right\}
      \hat G \rvacm_r \Big/
      \lvacp \rvacm_r .
\nonumber\\
&=&
\lvacp \vert
       \left\{ \hat a_n{}_k \hat a^\dagger_m{}^l
              -\frac{1}{2} \delta_{nm} \delta_k^l \right\}
      \rvacm_r \Big/
      \lvacp \rvacm_r  .
\nonumber\\
\end{eqnarray}
This correlation function is invariant under the global
gauge transformation $SU(N)_C$.
It transforms covariantly under the global transformation
$SU(N)_H$ given by Eq.~(\ref{eq:sun-hidden-global-transformation}).
It was shown in \cite{original-overlap} that free {\it Weyl}
fermions appear from this correlation function, which degeneracy
is just same as that of the continuum {\it Weyl} fermion in the
representation $r$. The appearance of poles of species doublers are
suppressed by the appearance of zeros at the vanishing momenta
for doublers.\footnote{This has been pointed out to be a
possible way out of the Nielsen-Ninomiya
theorem.\cite{NN-theorem,NN-Shamir-theorem}
However,
the examples are known in which they cause ghost states
which contribute to the vacuum polarization with wrong signature
and lead to the wrong normalization\cite{rebbis,zeros-ghosts}.
In the case of the vacuum overlap formulation,
it has been shown that the perturbative calculation gives
the correct normalization of the vacuum
polarization\cite{saoki-levien,
randjbar-daemi-strathdee,
kikukawa_lattice-vacuum-pol}.
This result is naturally understood from the point of view
of the infinite number of the Pauli-Villars
fields\cite{frolov-slavnov}.
Still it is desirable to clarify this point in relation to
the Ward identity\cite{zeros-ghosts}.
}
Note also that the gauge degree of freedom completely decouples
from this correlation function. This is a remarkable property
of the overlap correlation function. It makes it easy and clear
to identify the fermion spectrum in the pure gauge limit.

By these observations, we can see that
the way the Weyl fermion emerges in the pure gauge limit in the
overlap correlation function is quite analogous to the way
the Dirac fermion does in the pure gauge limit of the lattice QCD,
although the mechanism to suppress the species doublers is
quite different.

\subsection{Boundary fermion correlation function}

However, through the analysis of the waveguide
model\cite{original-waveguide}, it has been claimed
that the required disordered nature of the gauge freedom
causes the vector-like spectrum of
fermion\cite{waveguide-analysis-golterman}.
In this argument, the fermion correlation functions at
the waveguide boundaries were examined.
One may think of
the counter parts of these correlation functions in the overlap
formulation by putting creation and annihilation operators in
the overlap of vacua with the same signature of mass.
Let us refer this kind of correlation function as
{\it boundary correlation function}.
We should note that the boundary correlation functions
are no more the observables in the sense defined in the overlap
formulation\cite{original-overlap}; they cannot be expressed by
the overlap of two vacua with their phases fixed by the Wigner-Brillouin
phase convention. But, they are still relevant because they can probe
the auxiliary fermionic system for the definition of
the complex phase of chiral determinant and therefore the
anomaly (the Wess-Zumino-Witten term).
If massless chiral states could actually appear in the boundary
correlation functions, we would have difficulty defining the
complex phase.

As for the overlap of the vacuum of the Hamiltonian with negative mass,
there are three possible definitions of boundary correlation function
in the representation $r$:
\begin{eqnarray}
\langle \phi_n{}_i \phi^\dagger_m{}^j  \rangle_{-r}
&\equiv&\frac{1}{Z} \int d\mu[g;\alpha]
\frac{\lvacm \vert \hat G^\dagger
  \left\{ \hat a_n{}_i \hat a^\dagger_m{}^j
         -\frac{1}{2} \delta_{nm} \delta_i^j \right\}
      \rvacm_r}
     {\lvacm \vert \hat G^\dagger \rvacm_r } ,
\\
\langle \varphi_n{}_i \varphi^\dagger_m{}^j  \rangle_{-r}
&\equiv&\frac{1}{Z} \int d\mu[g;\alpha] \,
\frac{\lvacm \vert
  \left\{ \hat a_n{}_i \hat a^\dagger_m{}^j
         -\frac{1}{2} \delta_{nm} \delta_i^j \right\}
      \hat G^\dagger \rvacm_r}
     {\lvacm \vert \hat G^\dagger \rvacm_r } ,
\\
\label{eq:boundary-correlation-function-gauge-mixed}
\langle \varphi_n{}_i \phi^\dagger_m{}^j  \rangle_{-r}
&\equiv&\frac{1}{Z}\int d\mu[g;\alpha] \,
\frac{\lvacm \vert
  \left\{ \hat a_n{}_i \hat G^\dagger \hat a^\dagger_m{}^j
         -\frac{1}{2} \delta_{nm} \left(g^\dagger_m\right)_i^j \right\}
      \rvacm_r}
     {\lvacm \vert \hat G^\dagger \rvacm_r } .
\nonumber\\
\end{eqnarray}

The transformation properties of these correlation functions
under the chiral $SU(N)$ can be read as follows:
\begin{equation}
\langle \phi_n{}_i \phi^\dagger_m{}^j  \rangle_{-r}
\longrightarrow
(g_0{}_i^s)
\langle \phi_n{}_s \phi^\dagger_m{}^t  \rangle_{-r}
(g_0^\dagger{}_t^j)   ,
\end{equation}
\begin{equation}
\langle \varphi_n{}_i \varphi^\dagger_m{}^j  \rangle_{-r}
\longrightarrow
(h_i^s)
\langle \varphi_n{}_s \varphi^\dagger_m{}^t  \rangle_{-r}
(h^\dagger{}_t^j)   ,
\end{equation}
\begin{equation}
\langle \varphi_n{}_i \phi^\dagger_m{}^j  \rangle_{-r}
\longrightarrow
(h_i^s)
\langle \varphi_n{}_s \phi^\dagger_m{}^t  \rangle_{-r}
(g_0^\dagger{}_t^j)   .
\end{equation}

Note that contrary to the case of the overlap correlation functions,
the gauge freedom never decouple from these boundary correlation
functions.
Then, due to the severe IR singularity,
only chiral $SU(N)$ invariant quantities can be used as
observables\cite{IR-finiteness-2D-nonlinear-sigma-model}.
The first and second correlation functions can be made invariant under
the chiral $SU(N)$ by taking the trace over the group indices.
The third one cannot be made invariant and we should discard
it from observables.

We will examine these two invariant correlation functions associated to
the overlap of the vacuum of the Hamiltonian with negative mass in
the following.
As for the positive mass case,
there are similarly three possible definitions of boundary correlation
function in the representation $r$. They could be examined
in a similar manner as the boundary correlation function associated
with the negative mass.

\subsubsection{Expression of boundary correlation functions}
The invariant boundary correlation functions associated
to the overlap of the vacuum of the Hamiltonian with negative mass
are evaluated as follows:
\begin{eqnarray}
\langle \phi_n{}_i \phi^\dagger_m{}^i  \rangle_{-r}
&\equiv&\frac{1}{Z} \int d\mu[g;\alpha]
\frac{\lvacm \vert \hat G^\dagger
  \left\{ \hat a_n{}_i \hat a^\dagger_m{}^i
         -\frac{1}{2} \delta_{nm} \delta_i^i \right\}
      \rvacm_r}
     {\lvacm \vert \hat G^\dagger \rvacm_r }
\nonumber\\
&=&\frac{1}{Z} \int d\mu[g;\alpha]
\left[
\frac{1}{2} \delta_{nm} \delta_i^i
- S_-^v[g](n;m){}_i^o \left( g^\dagger_m{}_o^i \right)
\right] ,
\\
\langle \varphi_n{}_i \varphi^\dagger_m{}^i  \rangle_{-r}
&\equiv&\frac{1}{Z} \int d\mu[g;\alpha] \,
\frac{\lvacm \vert
  \left\{ \hat a_n{}_i \hat a^\dagger_m{}^i
         -\frac{1}{2} \delta_{nm} \delta_i^i \right\}
      \hat G^\dagger \rvacm_r}
     {\lvacm \vert \hat G^\dagger \rvacm_r }
\nonumber\\
&=&\frac{1}{Z} \int d\mu[g;\alpha]
\left[
\frac{1}{2} \delta_{nm} \delta_i^i
- \left( g^\dagger_m{}_i^o \right) S_-^v[g](n;m){}_o^i
\right] ,
\end{eqnarray}
where
\begin{eqnarray}
S_-^v[g](n,m){}_i^j
&\equiv&
\int \frac{d^4 p}{(2\pi)^4} \frac{d^4 q}{(2\pi)^4} \times
\nonumber\\
&&
e^{i p n } v_-(p)
\left[ v^\dagger_-(q)
e^{-i q r} \left( g_r^\dagger{}_j^i \right) e^{i p r }
v_-(p) \right]^{-1}_{(p,i)(q,j)}
v^\dagger_-(q) e^{-i q m} .
\nonumber\\
\end{eqnarray}
We refer the reader to \cite{kikukawa-gauge-two-dimensions}
for the detail of the calculation.

In the perturbation theory in $\lambda$, $S_+^v[g](n,m)$ can
be expanded as follows:
\begin{eqnarray}
S_-^v[g](n,m){}_s^o
&=&
S^v_-( n-m ) \left( \bbone{}_s^o \right)
-
\sum_r S^v_-( n-r ) \left( (-i\lambda) \pi_r{}_s^o \right) S^v_-( r-m )
\nonumber\\
&&
-\sum_r S^v_-( n-r )
\left( \frac{(-i\lambda)^2}{2!}\pi_r^2{}_s^o \right) S^v_-( r-m )
\nonumber\\
&&
+ \sum_{r,l}
S^v_-( n-r ) \left( (-i\lambda) \pi_r{}_s^u \right)
S^v_-( r-l ) \left( (-i\lambda) \pi_l{}_u^o \right)
S^v_-( l-m )
\nonumber\\
&&
+{\cal O}(\lambda^3) .
\end{eqnarray}
Then we obtain at the one-loop order
\begin{eqnarray}
  \langle \phi_n{}_i \phi^\dagger_m{}^i  \rangle_{-r}
&=&
\frac{1}{2} \delta_{nm} \delta_i^i - S^v_-( n-m ) \delta_i^i
\nonumber\\
&-&
\lambda^2  \sum_{r}
S^v_-( n-r )
\left[
\langle \pi_r{}_i^o \pi_m{}_o^i \rangle^\prime S^v_-( r-m )
\right]
\nonumber\\
&+&
\lambda^2  \sum_{r,l}
S^v_-( n-r )
\left[
\langle \pi_r{}_i^o \pi_l{}_o^i \rangle^\prime S^v_-( r-l )
\right] S^v_-( l-m ) +{\cal O}(\lambda^4) ,
\nonumber\\
\label{eq:boundary-correlation-function-colored}
&&\\
  \langle \varphi_n{}_i \varphi^\dagger_m{}^i  \rangle_{-r}
&=&
\frac{1}{2} \delta_{nm} \delta_i^i - S^v_-( n-m ) \delta_i^i
\nonumber\\
&-&
\lambda^2  \sum_{r}
\left[
\langle \pi_n{}_i^o \pi_r{}_o^i \rangle^\prime S^v_-( n-r )
\right]
S^v_-( r-m )
\nonumber\\
&+&
\lambda^2  \sum_{r,l}
S^v_-( n-r )
\left[
\langle \pi_r{}_i^o \pi_l{}_o^i \rangle^\prime S^v_-( r-l )
\right] S^v_-( l-m ) +{\cal O}(\lambda^4) ,
\nonumber\\
\label{eq:boundary-correlation-function-noncolored}
&&\\
\label{eq:gauge-freedom-correlation-function}
\langle \pi_r{}_i^o \pi_l{}_o^i \rangle^\prime
&=& \delta_i^i \int \frac{d^4 p}{(2\pi)^4}
\frac{ e^{i p(r-l) } -1 }
{\left( \sum_\mu 4 \sin^2 \frac{p_\mu}{2} \right)^2 }.
\end{eqnarray}
Note that the IR divergences associated to the correlation
function of the gauge freedom
\begin{equation}
\langle \pi_r{}_i^o \pi_l{}_o^i \rangle
= \delta_i^i \int \frac{d^4 p}{(2\pi)^2}
\frac{ e^{i p(r-l) } }
{\left( \sum_\mu 4 \sin^2 \frac{p_\mu}{2} \right)^2} ,
\end{equation}
cancel among the second and third terms in the r.h.s. of
Eqs.~(\ref{eq:boundary-correlation-function-colored}) and
(\ref{eq:boundary-correlation-function-noncolored}).
To show this fact explicitly, we have used
$\langle \pi_r{}_i^o \pi_l{}_o^i \rangle^\prime $ instead.\footnote{
It is not difficult to see that the infrared divergence remains
in the correlation function of the type of
Eq.~(\ref{eq:boundary-correlation-function-gauge-mixed}). }

\subsubsection{Boundary correlation functions at criticality}

At the critical point, $\lambda=0$, the correlation functions
reduce to the expression:
\begin{eqnarray}
\langle \phi_n{}_i \phi^\dagger_m{}^i  \rangle_{-r}
&=&
\langle \varphi_n{}_i \varphi^\dagger_m{}^i  \rangle_{-r} ,
\\
\langle \varphi_n{}_i \varphi^\dagger_m{}^i  \rangle_{-r}
&=&
\int \frac{d^4 p}{(2\pi)^4} e^{i p (n-m)} \delta_i^i \times
\nonumber\\
&& \quad
\frac{1}{2 \lambda_-}
\left( \begin{array}{cc}
-m_0+B(p) & C(p) \\
C^\dagger(p) & m_0-B(p)
       \end{array}\right) ,
\end{eqnarray}
where
\begin{eqnarray}
\lambda_- &=& {\sqrt{C_\mu^2(p)+(B(p)-m_0)^2}} , \\
C(p) &=& i \sigma_\mu \sin p_\mu  = i \sigma_\mu C_\mu(p) , \\
B(p)    &=& \sum_\mu (1-\cos p_\mu ) .
\end{eqnarray}

This boundary correlation function does not show any pole which
can be interpreted as particle. Rather, it consists of the
continuum spectrum with a mass gap. To see it, we consider the
boundary correlation function without the spinor structure for
simplicity. For the fixed spatial momentum $p_k \, (k=1,2,3)$,
the correlation function can be evaluated as
\begin{eqnarray*}
D(n_4;p_k)
&=& \int\frac{d p_4}{(2\pi)} e^{i p_4 n_4} \frac{1}{2 \lambda_-} \\
&=& \int\frac{d p_4}{(2\pi)}\frac{d \omega}{(2\pi)}
 e^{i p_4 n_4} \frac{1}{\omega^2+ \lambda_-^2} \\
&=& \int \frac{d \omega}{(2\pi)}
\frac{1}{\left( 1-\sum_k(1-\cos p_k)- m_0 \right)}
\frac{ e^{-M(\omega,p_k) |n_4|}}{2 \sinh M(\omega,p_k)} ,
\end{eqnarray*}
where
\begin{equation}
 \cosh M(\omega,p_k) =
1 +
\frac{  \omega^2 + \sum_k \sin^2 (p_k)
      + \left(\sum_k(1-\cos p_k) - m_0\right)^2}
     {2 \left(1-\sum_k(1-\cos p_k)- m_0\right)} .
\end{equation}
The minimum of $M(\omega,p_k)$ can be identified as the mass gap.
For $m_0 = 0.5$, it appears at $\omega=0$ and $p_k=0$.
In this case,  the mass gap $M_B$ is given by
\begin{equation}
\cosh M_B = 1 + \frac{ m_0^2}{2 (1 - m_0)} .
\end{equation}
Since the mass gap is of order of the cutoff,
no light physical particle emerges
in the boundary correlation functions on the critical point.
They have very short-distance
nature.\footnote{In fact, the boundary correlation function can
be derived from the correlation function of the three-dimensional
massive fermion by the reduction to the two-dimensional space-time.
This is why the continuum spectrum with mass gap emerges. }

\subsubsection{Boundary correlation functions off criticality}

Once we know that the boundary correlation functions have
short distance nature at $\lambda=0$,
the quantum correction to such quantities can be evaluated
by the perturbation theory in $\lambda$ rather reliably
by virtue of the asymptotic freedom.

Since $S_-^v$ has the short correlation length of the order of the
lattice spacing as we have shown,
its combolutions in
Eqs.~(\ref{eq:boundary-correlation-function-colored})
and (\ref{eq:boundary-correlation-function-noncolored})
with the correlation function of the gauge freedom
Eq.~(\ref{eq:gauge-freedom-correlation-function})
also have short correlation lengths, even though
Eq.~(\ref{eq:gauge-freedom-correlation-function}) is
logarithmically increasing function of the distance $|n|$.
There is no symmetry against the mass gap.
This result shows that even inside of the symmetric phase
off the critical point, no light particle emerges
in the boundary correlation functions.
Since the overlap correlation function does not depend on the gauge
freedom at all and does show the chiral
spectrum\cite{original-overlap},
as we have seen,
the above fact means that the entire fermion spectrum is chiral.
Thus, {\it the disordered nature of
the gauge freedom does not contradict with the chiral fermion spectrum}.

\section{Summary and Discussion}
\label{sc:conclusion-discussion}
\reseteqnum

In summary,
we have examined the dynamical nature of the gauge degree of freedom
in the lattice pure gauge model.
The model is obtained in the limit $\beta=\infty$
from the four-dimensional $SU(N)$ nonabelian chiral gauge theory
which is defined by the vacuum overlap and
is modified in the weight of the gauge average
by introducing the covariant gauge fixing term and the Faddeev-Popov
determinant.

In the continuum theory, it has been noticed by Hata
that the model has the dynamical features
which is quite similar to the two-dimensional nonlinear sigma model.
Namely, it is renormalizable in the perturbation theory
and the self-coupling of the gauge freedom is asymptotically free.
The severe IR divergence, which occurs in the perturbation expansion,
prevents local order parameters from emerging. These dynamical
features suggest that the global gauge symmetry does not
break spontaneously and the gauge degree of freedom
acquires mass dynamically by the dimensional transmutation.

The different feature in the lattice counterpart, which is obtained
from the lattice chiral gauge theory in consideration,
is that the gauge degree of freedom does not decouple from the
fermion determinant and has coupling through the gauge (BRST)
non-invariant piece of the complex phase of chiral determinant.
What we showed first is that even in the presence of these
gauge (BRST) symmetry breaking terms,
the model is renormalizable at one-loop and the self-coupling
of the gauge freedom is indeed asymptotically free.
The IR divergence also occurs and it prevents local order parameters
from emerging. Therefore, as in the continuum theory,
the gauge degree of freedom is disordered at $\beta=\infty$.
In this analysis, we also found that the
lattice Gribov copies are no more the stationary points of the
total action and should be suppressed in the gauge average.

Because of the lack of the gauge (BRST) symmetry, however,
it is expected that the gauge non-invariant divergent and finite terms
would be induced at some higher but finite orders.
In order to clarify the effect of such terms, we next performed
the perturbative analysis of the Foerster-Nielsen-Ninomiya mechanism.
Namely, we introduced the explicit real gauge symmetry breaking term,
${\cal O}_1$, and examined how this term would spoil the disordered
nature of the gauge freedom or how small the breaking term must be
in order to keep that nature.
Although it is small for small $\lambda$, we found a finite region
of the coupling constant of ${\cal O}_1$ for which the effect of the
breaking term does not alter the power-law long distance behavior
of the invariant two-point function of the gauge freedom.
This means that, even if the gauge breaking term of the type
${\cal O}_1$ is induced, by tuning the counter term, we can keep
the disordered nature of the gauge freedom.  From this analysis,
the phase structure of the pure gauge model was also argued.
We also found that the model remains renormalizable with
${\cal O}_1$ and that the Foerster-Nielsen-Ninomiya mechanism
in this perturbative framework is closely related to the
IR regularization prescription of the continuum perturbation theory
by the mass term for the gauge boson.

Based on these dynamical features, which is quite similar to the
nonlinear sigma model in two-dimensions, and following Hata's scenario
for the dynamical gauge symmetry restoration,
we then assumed that the global gauge symmetry does not break
spontaneously and the gauge freedom acquires mass dynamically in the
pure gauge limit for the entire region of the gauge parameter.

Under this assumption,
the asymptotic freedom allows us to tame the gauge fluctuation by
approaching the critical point of the gauge freedom without
spoiling its disordered nature. There we showed
by the perturbation expansion that
the spectrum in the invariant boundary correlation functions, which
is free from IR divergence, have mass gap of the order of the lattice
cutoff and it survives the quantum correction due to the gauge
fluctuation. There is no symmetry against the spectrum mass gap.
This means that no light state appear in the boundary
correlation functions and they cannot be regarded as physically
relevant observables.

{}From the overlap correlation functions, on the other hand, the gauge
degree of freedom decouples completely.
This is a remarkable property of the overlap correlation function.
It makes it easy and clear to identify the fermion spectrum in
the pure gauge limit.
In fact, the correct number of free Weyl fermions emerge
from this correlation function. The appearance of the
poles of species doublers are suppressed by the appearance
of zeros at the vanishing momenta for doublers.
Therefore the entire fermion spectrum is chiral and consistent
with the continuum target theory.
Thus we argued that the dynamical restoration of the gauge
symmetry due to the disordered gauge degree of freedom
does not contradict with the chiral fermion spectrum
in the vacuum overlap formulation of lattice $SU(N)$ chiral gauge theory.

Finally, let us make a few comments related to
the formulation with the manifest local gauge invariance.
We might start from the partition function with the manifest local
gauge invariance, by introducing the explicit integration over the
gauge degree of freedom:
\begin{eqnarray}
\label{eq:manifest-gauge-invariant-partition-function}
Z_{inv} &=& \int [d\omega][dU][d\bar c d c]
      \exp \left(-\beta S_G
           -\frac{1}{2\alpha} \left(\bar \nabla {}^\omega \hat A \right)^2
           -\bar c M[{}^\omega U ] c \right)
\times \nonumber\\
&& \qquad\qquad
      \prod_{rep.} \left(
        \frac{ \lvacp \vert \hat \Omega \rVacp }
             {|\lvacp \vert \hat \Omega \rVacp|}
               \lVacp
               \rVacm
        \frac{ \lVacm \vert \hat \Omega^\dagger \rvacm }
             {|\lVacm \vert \hat \Omega^\dagger \rvacm|}
              \right)  . \nonumber\\
\end{eqnarray}
Here the field variable $\omega_n$ takes its value
in $SU(N)$ and is assumed to transform under the gauge transformation
as follows:
\begin{equation}
  \omega_n \longrightarrow  \omega_n g_n^\dagger .
\end{equation}
$\hat \Omega$ is defined by
\begin{equation}
\hat \Omega
= \exp \left( \hat a_n^{\dagger i}
       \{\log \omega_n\}_i{}^j \hat a_{n j} \right) .
\end{equation}
As to the gauge link variable in the gauge fixing term and
the Faddeev-Popov ghost action, they are
replaced by
\begin{equation}
  {}^\omega U_{n\mu} = \omega_n U_{n\mu} \omega_{n+\hat\mu}^\dagger ,
\end{equation}
Accordingly, ${}^\omega \hat{A}_{n\mu}$ is defined
from ${}^\omega U_{n\mu}$ by the definition of the lattice
vector potential Eq.~(\ref{eq:lattice-vector-potential}).
Besides the gauge symmetry, there exists a global $SU(N)$
symmetry under the transformation which we refer
as global $SU(N)_{H'}$:
\begin{equation}
  \omega_n \longrightarrow  h' \omega_n , \qquad h' \in SU(N)_{H'} .
\end{equation}
At $\beta =\infty$, the partition function reduces to
\begin{eqnarray}
Z_{inv} &=& \int [d\omega][dg][d\bar c d c]
      \exp \left(-\frac{1}{2\alpha}
            \left(\bar \nabla {}^{\omega g}\hat A \right)^2
           -\bar c M[\omega g] c \right)
\times \nonumber\\
&& \qquad\qquad
      \prod_{rep.} \left(
        \frac{ \lvacp \vert \hat \Omega \hat G \rvacp }
             {|\lvacp \vert \hat \Omega \hat G \rvacp|}
               \lvacp
               \rvacm
      \frac{ \lvacm \vert \hat G^\dagger \hat \Omega^\dagger \rvacm }
           {|\lvacm \vert \hat G^\dagger \hat \Omega^\dagger \rvacm|}
              \right)   \nonumber \\
&=& \int d\mu[\omega g;\alpha] .
\end{eqnarray}
In this limit, there emerges another global $SU(N)$ symmetry,
which we have refered as global $SU(N)_{H}$:
\begin{equation}
  g_n \longrightarrow  g_n h^\dagger , \qquad h \in SU(N)_{H} .
\end{equation}

If we fix the local gauge symmetry so that
\begin{equation}
  \omega_n = 1 , \qquad \hat \Omega = \bbone ,
\end{equation}
the model reduces to the form with which we have discussed
in this article. The symmetry is then reduced to
\begin{equation}
SU(N)_{C} \otimes SU(N)_{H} , \quad  SU(N)_{H'} \sim SU(N)_{C} .
\end{equation}
On the other hand, we can fix the local gauge symmetry so that
\begin{equation}
  g_n = 1, \qquad  \hat G = \bbone .
\end{equation}
In this case, the symmetry is reduced to
\begin{equation}
SU(N)_{H'} \otimes SU(N)_{C}, \quad  SU(N)_{C} \sim SU(N)_{H} .
\end{equation}
We note that, in the context of the Wilson-Yukawa
model\cite{wilson-yukawa-model,aoki,kashiwa-funakubo},
this way of the gauge fixing leads to the formulation in
terms of the ``ordinary'' fermion, with $\omega_n$ to appear
in the Wilson-Yukawa coupling term. There, the ``$C$-charged''
fermion is called ``charged'' and ``$H'$-charged'' fermion is
called ``neutral''. With this correspondence in mind, we next
clarify the correspondence of the fermion correlation functions.

The gauge covariant overlap correlation function in the pure gauge limit,
which corresponds to
Eq.~(\ref{eq:covariant-overlap-correlation-function-pure-gauge}),
is given as follows:
\begin{eqnarray}
\langle a_n{}_i a^\dagger_m{}^j  \rangle_r
&=& \frac{1}{Z} \int d\mu[\omega g ;\alpha] \times \nonumber\\
&&\qquad\qquad
         \lvacp
         \vert \hat G^\dagger
         \left\{ \hat a_n{}_i \hat a^\dagger_m{}^j
                -\frac{1}{2} \delta_{nm} \delta_i^j \right\}
         \hat G
         \rvacm_r \Big/
         \lvacp \rvacm_r . \nonumber\\
\end{eqnarray}
In the second gauge, it reduces to the form
\begin{eqnarray}
\langle a_n{}_i a^\dagger_m{}^j  \rangle_r
&=& \frac{1}{Z} \int d\mu[\omega ;\alpha] \times \nonumber\\
&&\qquad\qquad
         \lvacp
         \vert
         \left\{ \hat a_n{}_i \hat a^\dagger_m{}^j
                -\frac{1}{2} \delta_{nm} \delta_i^j \right\}
         \rvacm_r \Big/
         \lvacp \rvacm_r . \nonumber\\
&=&
         \lvacp
         \vert
         \left\{ \hat a_n{}_i \hat a^\dagger_m{}^j
                -\frac{1}{2} \delta_{nm} \delta_i^j \right\}
         \rvacm_r \Big/
         \lvacp \rvacm_r . \nonumber\\
\end{eqnarray}
Note that this is the ``$C$-charged'' (``charged'')
correlation function and the gauge freedom $\omega$ decouples
completely from it.
(Note also that this correlation function corresponds
to the ``colored'' quark correlation function
Eq.~(\ref{eq:gauge-non-invariant-quark-correlation-function})
when the gauge is fixed so that $g_n=1$ in QCD.)
This correlation function has the same structure as
the ``$H$-charged'' correlation function
Eq.~(\ref{eq:H-charged-fermion-correlation-function})
in the first gauge, in which, as we know,
the correct number of the Weyl fermions appear.
This means that in the vacuum overlap formulation,
the ``charged'' fermions appear in the correct chiral spectrum.
As for the ``neutral'' fermions, we have seen in the
first gauge that the ``C-charged'' fermions cannot be observed
due to the IR singularity and it means that
the ``$H'$-charged''(``neutral'') fermion does not appear
in the physical spectrum.
Therefore, there is no problem here concerning
the ``neutral''-ness of the fermion and the triviality of the
chiral vector coupling\cite{triviality-wilson-yukawa}.
\footnote{We can find in \cite{renormalizable-gauge-model}
(latest revised version) the similar observation.
For the explicit calculation, the author refers to forthcoming
papers.
Note that, in the overlap formulation, once we
understand that any light states do not appear
in the boundary correlation functions,
the spectrum of the ``charged'' fermion can be easily and clearly
read from the overlap correlation function because the gauge degree
of freedom decouples completely from it.}

{}From these observations, we see that the several questions raised to
the Wilson-Yukawa model;
the absence of the ``charged'' fermion,
the vector-like spectrum of the ``neutral'' fermion
and the triviality of its chiral coupling to vector boson,
do not apply to the lattice chiral gauge theory
defined by the vacuum overlap formulation.
As to the fermion number violation,
it has been already demonstrated in several contexts
that the overlap formulation can correctly reproduce the
effect of the instanton and
can describe such physical process of the fermion number
violation\cite{original-overlap,overlap-fermion-number-violation}.

As to the Wilson-Yukawa model for the standard model itself,
the possible answer for these long standing questions
has been described in \cite{renormalizable-gauge-model}(latest revised
version) in relation to the dynamics of the
gauge freedom governed by the covariant-type gauge fixing
term\cite{covariant-type-gauge-fixing-action}.
It may be interesting to reconsider the Eichten-Preskill model
for the nonabelian chiral gauge theory
\cite{eichten-preskill,eichten-preskill-analysis,
Eichten-Preskill-model-revisited}
from the point of view of the nonabelian dynamics of
the gauge degree of freedom discussed in this paper.

\section*{Acknowledgments}
The author would like to thank H. Neuberger and R. Narayanan
for enlightening discussions.
He would like to thank H.~Hata for discussion on
the dynamics of the pure gauge model.
He would also like to thank A.~Yamada for discussion.
The author express his sincere thanks to
High energy physics theory group, Nuclear physics group and the
computer staffs of the department of physics and astronomy
of Rutgers university for their kind hospitality.

\appendix
\vspace{2cm}
\section*{Appendix}

\section{Expansion of the action in gauge fluctuation}
\label{appendix:action-of-gaussian-gauge-fluctuation}
\reseteqnum

In this appendix, we give the explicit formula of
Eq.~(\ref{eq:action-of-gaussian-gauge-fluctuation}):
the action expanded in terms of the fluctuations of the gauge freedom,
ghost and anti-ghost field around a classical configuration.
\begin{eqnarray}
  S[\exp(i\pi)g,c+\xi,\bar c+\bar \xi ]
&=& S[g,c,\bar c] + S_0[\pi^2 / \xi,\bar\xi] \nonumber \\
&+& S_1[\pi^2 / \xi,\bar\xi; \sin A]
       + S_{1c}[\pi,\xi / \bar\xi ;g ,c / \bar c] \nonumber\\
&+& S_2[\pi^2 / \xi,\bar\xi; \sin^2 A ,\cos A -1 ]
       + S_{2c}[\pi^2 ; g ,c,\bar c]        \nonumber\\
&+& i \, \Delta S_{WZW}[\pi^2;g] ,
\end{eqnarray}
where
\begin{eqnarray}
S[g,c,\bar c] &=&
      -\sum_{n,a} \frac{1}{2\alpha}
\left( \bar{\nabla}_\mu \hat{A}^a_{n\mu} \right)^2
- \sum_{nm,ab} \bar c_n^a
  \hat{M}^{ab}_{nm}[g_n g^\dagger_{n+\hat\mu} ] c_m^b \nonumber\\
&+& i\Delta\Gamma_{WZW}[g] ,
\\
\label{eq:vertex-quadratic-in-pi-0-appendix}
S_0[\pi^2 / \xi,\bar\xi] &=&
-\sum_{n,a} \frac{1}{2}\left(\nabla^2 \pi_n^a \right)^s
+ \sum_{n,a} \bar \xi_n^a \nabla^2 \xi_n^a ,
\\
\label{eq:vertex-quadratic-in-pi-A1-appendix}
S_1[\pi^2 / \xi,\bar\xi;A]
&=&
-\sum_{n\mu} \frac{1}{2} f^{abc} \nabla^2 \pi_n ^a
\left( \pi_{n+\hat\mu}^b \hat A_{n\mu}^c
      -\pi_{n-\hat\mu}^b \hat A_{n-\hat\mu,\mu}^c \right)
\nonumber\\
&+&
 \sum_n \frac{1}{2} f^{abc}
   \bar \xi_n^a \bar \nabla_\mu
  \left\{
   \left( \xi_n^b + \xi_{n+\hat\mu}^b \right) \hat A_{n\mu}^c
  \right\}
\nonumber\\
&=&
\sum_{n\mu}
\frac{1}{2} \bar \nabla_\mu
\hat A_{n\mu}^a f^{abc} \pi_n^b \nabla^2 \pi_n^c  \nonumber\\
&-&
\sum_{n\mu}
\frac{1}{2} \nabla^2 \pi_n^a f^{abc}
 \left(\nabla_\mu \pi_n^b \hat A_{n\mu}^c
      +\bar \nabla_\mu \pi_n^b \hat A_{n-\hat\mu,\mu}^c \right)
\nonumber\\
&+&
 \sum_n \frac{1}{2} f^{abc}
   \bar \xi_n^a \bar \nabla_\mu
  \left\{
   \left( \xi_n^b + \xi_{n+\hat\mu}^b \right) \hat A_{n\mu}^c
  \right\} ,
\\
\label{eq:vertex-quadratic-in-pi-A2-appendix}
S_2[\pi^2 / \xi,\bar\xi;A^2] &=&
\sum_n \frac{1}{2} \bar \nabla_\nu \hat A_{n\nu}^a  \times
\nonumber\\
&&
\bar \nabla_\mu \left\{
{\rm Tr}\left[
  \left( \pi_{n+\hat\mu}[ \pi_{n+\hat\mu},T^a]
\right.\right.\right.\nonumber\\
&&\left.\left.\left. \qquad
        +\pi_{n}[ \pi_{n},T^a]
        +[T^a,\pi_{n+\hat\mu} ] \pi_{n+\hat\mu}
\right.\right.\right.\nonumber\\
&&\left.\left.\left. \qquad
        +[T^a,\pi_{n}] \pi_{n}
        +2 \nabla_\mu \pi_n T^a \nabla_\mu \pi_n
  \right) \sin A_{n\mu}
        \right] \right\} \nonumber\\
&-&\sum_n \frac{1}{8}
   f^{abc}
   \left( 2 \pi_n^b \bar \nabla_\nu \hat A_{n\nu}^c
         +\nabla_\nu \pi_n^b \hat A_{n\nu}^c
         +\bar \nabla_\nu \pi_n^b \hat A_{n-\hat\nu,\nu}^c \right) \times
\nonumber\\
&& \qquad
   f^{ade}
   \left( 2 \pi_n^d \bar \nabla_\mu \hat A_{n\mu}^e
         +\nabla_\mu \pi_n^d \hat A_{n\mu}^e
         +\bar \nabla_\mu \pi_n^d \hat A_{n-\hat\mu,\mu}^e \right)
\nonumber\\
&-&\sum_n \nabla^2 \pi_n^a \bar\nabla_\mu \left\{ {\rm Tr}
        \left[ \{ T^a,T^b \}\left(\cos A_{n\mu}-1\right)\right]
        \nabla_\mu \pi_n^b \right\}
\nonumber\\
&+& \sum_n \bar\xi_n^a \bar\nabla_\mu \left\{ {\rm Tr}
\left[ \{ T^a,T^b \}\left(\cos A_{n\mu}-1\right) \right]
\nabla_\mu \xi_n^b \right\}  ,
\\
\label{eq:vertex-quadratic-in-pi-gc1-appendix}
S_{1c}[\pi,\xi / \bar\xi;g,c / \bar c]
&=&\sum_{n\mu} i\, \bar \xi_n^a \bar \nabla_\mu
   \left\{
    {\rm Tr}\left[ \left( T^aT^b\pi_n -\pi_{n+\hat\mu}T^aT^b \right)
                   g_n g^\dagger_{n+\hat\mu}
   \right.   \right. \nonumber\\
&& \left.    \left. \qquad
                  +\left( T^bT^a\pi_{n+\hat\mu}-\pi_n T^bT^a \right)
                   g^{}_{n+\hat\mu} g^\dagger_n \right] c_n^b
   \right.\nonumber\\
&&\qquad
   \left.
   -{\rm Tr}\left[ \left( T^bT^a\pi_n -\pi_{n+\hat\mu}T^bT^a \right)
                   g_n g^\dagger_{n+\hat\mu}
   \right.  \right. \nonumber\\
&& \left.   \left. \qquad
                  +\left( T^aT^b\pi_{n+\hat\mu}-\pi_n T^aT^b \right)
                   g_{n+\hat\mu} g^\dagger_n \right] c_{n+\hat\mu}^b
   \right\} \nonumber\\
&+&\sum_{n\mu} i\, \bar c_n^a \bar \nabla_\mu
   \left\{
    {\rm Tr}\left[ \left( T^aT^b\pi_n -\pi_{n+\hat\mu}T^aT^b \right)
                   g_n g^\dagger_{n+\hat\mu}
   \right.   \right. \nonumber\\
&& \left.    \left. \qquad\qquad
                  +\left( T^bT^a\pi_{n+\hat\mu}-\pi_n T^bT^a \right)
                   g^{}_{n+\hat\mu} g^\dagger_n \right] \xi_n^b
   \right.\nonumber\\
&&\qquad\qquad
   \left.
   -{\rm Tr}\left[ \left( T^bT^a\pi_n -\pi_{n+\hat\mu}T^bT^a \right)
                   g_n g^\dagger_{n+\hat\mu}
   \right.  \right. \nonumber\\
&& \left.   \left. \qquad\qquad
                  +\left( T^aT^b\pi_{n+\hat\mu}-\pi_n T^aT^b \right)
                   g_{n+\hat\mu} g^\dagger_n \right] \xi_{n+\hat\mu}^b
   \right\} ,
\nonumber\\
\\
\label{eq:vertex-quadratic-in-pi-gc2-appendix}
S_{2c}[\pi^2;g,c,\bar c]
&=& \sum_{n\mu} \frac{1}{2} \, \nabla_\mu \bar c_n^a \times \nonumber\\
&&
   \left\{
    {\rm Tr}\left[ \left( T^aT^b\pi_n^2 - 2 \pi_{n+\hat\mu}T^aT^b\pi_n
                         +\pi_{n+\hat\mu}^2 T^aT^b   \right)
                   g_n g^\dagger_{n+\hat\mu}
   \right.   \right. \nonumber\\
&& \left.    \left.
                  +\left( T^bT^a\pi_{n+\hat\mu}^2
                        -2 \pi_n T^bT^a \pi_{n+\hat\mu}
                        +\pi_n^2 T^bT^a \right)
                   g^{}_{n+\hat\mu} g^\dagger_n \right] c_n^b
   \right.\nonumber\\
&& \left.
   -{\rm Tr}\left[ \left( T^bT^a\pi_n^2 - 2 \pi_{n+\hat\mu}T^bT^a\pi_n
                         +\pi_{n+\hat\mu}^2 T^bT^a   \right)
                   g_n g^\dagger_{n+\hat\mu}
  \right. \right. \nonumber\\
&&\left.  \left.
    +\left( T^aT^b\pi_{n+\hat\mu}^2
          -2 \pi_n T^aT^b \pi_{n+\hat\mu}
          +\pi_n^2 T^aT^b \right)
          g^{}_{n+\hat\mu} g^\dagger_n \right]
  \right. \nonumber\\
&& \left.
\qquad\qquad c_{n+\hat\mu}^b
   \right\} .
\end{eqnarray}

\section{Local operators of dimensions less than five}
\label{appendix:all-local-operator-dim4}
\reseteqnum

The possible local operators of dimensions less than five
can be listed up based on the global $SU(N)$ gauge symmetry, the
hyper-cubic symmetry, the CP invariance, the ghost number conservation
and the anti-ghost shift symmetry\cite{rome-approach}.
The complete list for the $SU(N)$ gauge group is given in this appendix.

\begin{eqnarray}
{\cal O}_0 &=&
- \frac{1}{2 \alpha} \sum_{n}
  \left( \bar \nabla_\mu A_{n\mu}^a \right)^2
- \sum_{n}\bar c_n^a \bar \nabla_\mu
  \left( - \nabla_\mu c_n^a
         + \frac{1}{2} f^{abc} (c_n^b+c_{n+\hat\mu}^b) A_\mu^c \right),
\nonumber\\
\\
{\cal O}_1 &=&
-\frac{1}{2} \sum_{n\mu} A_{n\mu}^a A_{n\mu}^a ,
\\
{\cal O}_2 &=&
-\frac{1}{2}
\sum_{n\mu\nu} \bar \nabla_\mu A_{n\nu}^a \bar \nabla_\mu A_{n\nu}^a ,
\\
{\cal O}_3 &=&
-\frac{1}{2}
\sum_{n\mu} \bar \nabla_\mu A_{n\mu}^a \bar \nabla_\mu A_{n\mu}^a ,
\\
{\cal O}_4 &=&
 \sum_{n\mu}\bar c_n^a \bar \nabla_\mu \nabla_\mu c_n^a  ,
\\
{\cal O}_5 &=&
-\frac{1}{2}
 \sum_{n\mu}\bar c_n^a \bar \nabla_\mu
  \left( f^{abc} (c_n^b+c_{n+\hat\mu}^b) A_\mu^c \right) ,
\\
{\cal O}_6 &=&
\sum_{n\mu\nu} f^{abc} \bar \nabla_\mu A_{n\nu}^a A_{n\mu}^b
A_{n\nu}^c ,
\\
{\cal O}_7 &=&
\sum_{n\mu\nu} A_{n\mu}^a A_{n\mu}^a  A_{n\nu}^b A_{n\nu}^b ,
\\
{\cal O}_8 &=&
\sum_{n\mu\nu} A_{n\mu}^a A_{n\mu}^b  A_{n\nu}^a A_{n\nu}^b ,
\\
{\cal O}_9 &=&
\sum_{n\mu\nu} f^{ace}f^{bde}
A_{n\mu}^a A_{n\mu}^b  A_{n\nu}^c A_{n\nu}^d ,
\\
{\cal O}_{10} &=&
\sum_{n\mu\nu} d^{abe}d^{cde}
A_{n\mu}^a A_{n\mu}^b  A_{n\nu}^c A_{n\nu}^d ,
\\
{\cal O}_{11} &=&
\sum_{n\mu\nu} d^{ace}d^{bde}
A_{n\mu}^a A_{n\mu}^b  A_{n\nu}^c A_{n\nu}^d ,
\\
{\cal O}_{12} &=&
\sum_{n\mu} A_{n\mu}^a A_{n\mu}^a  A_{n\mu}^b A_{n\mu}^b ,
\\
{\cal O}_{13} &=&
\sum_{n\mu} d^{abe}d^{cde} A_{n\mu}^a A_{n\mu}^b  A_{n\mu}^c A_{n\mu}^d .
\end{eqnarray}

\section{Contribution of the Wess-Zumino-Witten term}
\label{appendix:contribution-WZW-action}
\reseteqnum

$i \Delta S_{ WZW}$ in
Eq.~(\ref{eq:action-of-gaussian-gauge-fluctuation})
is the contribution from the Wess-Zumino-Witten term.
This is evaluated from
Eq.~(\ref{eq:null-Wess-Zumino-Witten-term-background-fluctuation}),
\begin{eqnarray}
e^{i \Delta S_{WZW}[\pi;g]}
&=&
e^{i \Delta\Gamma_{WZW}[\exp(i\lambda\pi)g]-i \Delta\Gamma_{WZW}[g]}
\nonumber\\
&=&
   \prod_{rep.} \left(
   \frac{\lvacp \vert \hat \Pi \hat G \rvacp}
  {|\lvacp \vert \hat \Pi \hat G \rvacp|}
   \frac{\lvacm \vert \hat G^\dagger \hat \Pi^\dagger \rvacm}
        {|\lvacm \vert \hat G^\dagger \hat \Pi^\dagger \rvacm|}
\left/
   \frac{\lvacp \vert \hat G \rvacp}
  {|\lvacp \vert  \hat G \rvacp|}
   \frac{\lvacm \vert \hat G^\dagger  \rvacm}
        {|\lvacm \vert \hat G^\dagger  \rvacm|}
\right.
              \right) \nonumber\\
&=&
   \prod_{rep.} \left(
   \frac{\lvacp \vert \hat \Pi \rVacp}
  {|\lvacp \vert \hat \Pi \rVacp|}
   \frac{\lVacm \vert \hat \Pi^\dagger \rvacm}
        {|\lVacm \vert \hat \Pi^\dagger \rvacm|}
\left/
   \frac{\lvacp \rVacp}
  {|\lvacp \rVacp|}
   \frac{\lVacm \rvacm}
        {|\lVacm \rvacm|}
\right.
              \right) . \nonumber\\
\end{eqnarray}
$\hat \Pi$ is the operator of the gauge transformation due to
$\pi$ given by:
\begin{equation}
\hat \Pi
=
\exp \left( \hat a_n^{\dagger i}
            \{i \lambda \pi_n\}_i{}^j \hat a_{n j} \right).
\end{equation}
$\rVacp$ and $\rVacm$ are given in this case by
the vacua of the second-quantized Hamiltonians with
the pure gauge link variable which satisfies the classical
equation of motion and is parameterized by the pure gauge vector
potential $A_{n\mu}$ as in
Eq.~(\ref{eq:pure-gauge-link-variable-of-classical solution}),
\begin{equation}
  g_n g^\dagger_{n+\hat\mu} = \exp \left( i A_{n\mu} \right) .
\end{equation}

We can evaluate these vacua in the expansion
in terms of $A_{n\mu}$ using the Hamiltonian perturbation theory.
Then $i \Delta S_{WZW}$ is obtained in the following expansion,
\begin{equation}
  i \, \Delta S_{WZW}[\pi;g]
  = \sum_{k=1}^\infty i \,\Delta S_{k\,WZW}[\pi; A^k] .
\end{equation}
This result is further expanded in terms of $\pi$. The linear term
in $\pi$ gives the contribution to the classical
equation. (Note that the formula of this contribution given in
Eq.~(\ref{eq:classical-equation-of-motion-g}) is evaluated
directly from the first expression of
Eq.~(\ref{eq:null-Wess-Zumino-Witten-term-background-fluctuation}).)
The quadratic term in $\pi$ is what we need for the one-loop
calculation by the background field method.

\subsection{Background vacua}

The second quantized Hamiltonians Eq.~(\ref{eq:overlap-hamiltonian})
with the pure gauge link variable of the classical solution
are divided into the free part and the perturbative part.
(Here after we omit the suffixes of the signature of masses, $\pm$,
as long as it does not introduce any confusions.)
\begin{equation}
 \hat H \left[g_n g^\dagger_{n+\hat\mu}\right]
=\hat H_{0} + \hat V\left[g_n g^\dagger_{n+\hat\mu}\right] ,
\end{equation}
where
\begin{equation}
  \hat V = \hat V_1 + \hat V_2 ,
\end{equation}
\begin{eqnarray}
V_1(n,m){}_i^j
&=& i \gamma_5
       \left[
        \left(\frac{\gamma_\mu-\bbone}{2}\right)
        \delta_{n+\hat\mu,m} \sin A_{n\mu}
       +\left(\frac{\gamma_\mu+\bbone}{2}\right)
        \delta_{n,m+\hat\mu} \sin A_{m\mu}
       \right]
\nonumber\\
&=& \int\frac{d^4 q}{(2\pi)^4}\frac{d^4 p}{(2\pi)^4} \,
     e^{i q n - i p m} \times
\nonumber\\
&&
    \gamma_5
    \left[ i \gamma_\mu \cos \left(\frac{q_\mu+p_\mu}{2}\right)
           + \sin \left(\frac{q_\mu+p_\mu}{2}\right) \right]
    \sin A_\mu (q-p) ,
\nonumber\\
\\
V_2(n,m){}_i^j
&=&  \gamma_5
       \left[
        \left(\frac{\gamma_\mu-\bbone}{2}\right)
        \delta_{n+\hat\mu,m} \left( \cos A_{n\mu} -1 \right)
      \right. \nonumber\\
&&\qquad\qquad \left.
       -\left(\frac{\gamma_\mu+\bbone}{2}\right)
        \delta_{n,m+\hat\mu} \left( \cos A_{m\mu} -1 \right)
       \right]
\nonumber\\
&=& \int\frac{d^4 q}{(2\pi)^4}\frac{d^4 p}{(2\pi)^4} \,
     e^{i q n - i p m} \times
\nonumber\\
&&
\gamma_5
\left[ i \gamma_\mu \sin \left(\frac{q_\mu+p_\mu}{2}\right)
        - \cos \left(\frac{q_\mu+p_\mu}{2}\right) \right]
    \left(\cos A_\mu (q-p) - 1 \right) .
\nonumber\\
\end{eqnarray}

We denote the eigenstates of the free Hamiltonian as
\begin{equation}
  \hat H_0 \rN = E_n \rN, \quad (n=0,1,\ldots) .
\end{equation}
We also denote the one-particle state normalized eigenvectors of
Hamiltonian by $u(p,s)$ and $v(p,s)$ $(s=1,2)$:
\begin{eqnarray}
H_{0\, nm} \left( e^{ip m} u(p,s) \right)
&=& + \lambda(p) \left( e^{i p n} u(p,s) \right) , \\
H_{0, nm} \left( e^{ip m} v(p,s) \right)
&=& - \lambda(p) \left( e^{i p n} v(p,s) \right) ,
\end{eqnarray}
and introduce the projection operators to the positive and
negative eigenvectors by
\begin{eqnarray}
  S^u(p) &=& \sum_s u(p,s) u(p,s)^\dagger ,  \\
  S^v(p) &=& \sum_s v(p,s) v(p,s)^\dagger .
\end{eqnarray}

The ground state is evaluated up to the second order of the
perturbation $\hat V$ as follows:
\begin{eqnarray}
 \rVac &=& \rvac
          \left( 1
                -\frac{1}{2}\sum_{n > 0} \lvac \vert V
                       \rN \frac{1}{(E_0-E_n)^2} \lN \vert V \rvac
           \right) \nonumber\\
&+& \sum_{n > 0} \rN \frac{1}{E_0-E_n} \lN \vert V \rvac  \nonumber\\
&+& \sum_{m,n > 0} \rN \frac{1}{E_0-E_n} \lN \vert V
                      \rM \frac{1}{E_0-E_m} \lM \vert V \rvac
\nonumber\\
&-& \sum_{n > 0} \rN \frac{1}{(E_0-E_n)^2} \lN \vert V \rvac
           \lvac\vert V \rvac  \nonumber\\
&+& {\cal O}(V^3 ).
\end{eqnarray}

Since our perturbation $\hat V$ is bilinear in the fermion operators,
only the two-particle states can contribute as intermediate states.
The fermion number is also conserved by the perturbation and
the total fermion number of the intermediate states always vanishes
because we are considering the vacuum to vacuum transition amplitude.
The intermediate-state projection operator is then evaluated as
\begin{eqnarray}
\sum_{\rm two\ particle}
    \rN \frac{1}{E_0-E_n} \lN \vert
&=&
-\int\frac{d^4 q}{(2\pi)^4}\frac{d^4 p}{(2\pi)^4} \,
    \frac{\displaystyle e^{i q (n-n^\prime) + i p (m-m^\prime)}}
         {\lambda(q)+\lambda(p)}
\nonumber\\
&&\qquad
\left\{ S^u(p)_{\alpha\beta} S^v(q)_{\gamma\delta} \,
        a_{m\alpha}^\dagger a_{n\delta} \rvac \lvac \vert
        a_{n^\prime\gamma}^\dagger a_{m^\prime\beta}
\right. \nonumber\\
&&\qquad
\left. +S^v(p)_{\alpha\beta} S^u(q)_{\gamma\delta} \,
        a_{m\beta} a_{n\gamma}^\dagger \rvac \lvac \vert
        a_{n^\prime\delta} a_{m^\prime\alpha}^\dagger
\right\} . \nonumber\\
\end{eqnarray}

Matrix elements which appear in the above formula of the ground state
perturbation can be evaluated as follows:
\begin{eqnarray}
\lvac \vert \hat V \rvac &=& {\rm Tr}\left\{ V(s,t)S^v(t-s) \right\},
\nonumber\\
\\
\lvac \vert a_{m\alpha} a_{n\beta}^\dagger \hat V \rvac
&=&
+ S^v(m-n)_{\alpha\beta} {\rm Tr} \left\{ V \cdot S^v \right\}
- S^u \cdot V \cdot S^v (m,n)_{\alpha\beta} ,
\nonumber\\
\\
\lvac \vert a_{m\alpha}^\dagger a_{n\beta} \hat V \rvac
&=&
- S^v(n-m)_{\beta\alpha} {\rm Tr} \left\{ V \cdot S^v \right\}
+ S^u \cdot V \cdot S^v (n,m)_{\beta\alpha} ,
\nonumber\\
\end{eqnarray}
\begin{eqnarray}
&& \lvac \vert a_{m_1\alpha_1} a_{n_1\beta_1}^\dagger \hat V
a_{m_2\alpha_2} a_{n_2\beta_2}^\dagger \rvac
\nonumber\\
&&=
+S^u(m_1-n_1)_{\alpha_1\beta_1} S^u(m_2-n_2)_{\alpha_2\beta_2}
{\rm Tr} \left\{ V \cdot S^v \right\}
\nonumber\\
&&+
S^u(m_1-n_2)_{\alpha_1\beta_2} S^v(m_2-n_1)_{\alpha_2\beta_1}
{\rm Tr} \left\{ V \cdot S^v \right\}
\nonumber\\
&&- S^u(m_1-n_1)_{\alpha_1\beta_1}
    S^v \cdot V \cdot S^u(m_2,n_2)_{\alpha_2\beta_2}
\nonumber\\
&&- S^u(m_1-n_2)_{\alpha_1\beta_2}
    S^v \cdot V \cdot S^v(m_2,n_1)_{\alpha_2\beta_1}
\nonumber\\
&&+ S^u \cdot V \cdot S^u(m_1,n_2)_{\alpha_1\beta_2}
    S^v(m_2-n_1)_{\alpha_2\beta_1}
\nonumber\\
&&- S^u \cdot V \cdot S^u(m_1,n_1)_{\alpha_1\beta_1}
    S^u(m_2-n_2)_{\alpha_2\beta_2} ,
\nonumber\\
\\
&&\lvac \vert a_{n_1\beta_1}^\dagger a_{m_1\alpha_1} \hat V
a_{m_2\alpha_2} a_{n_2\beta_2}^\dagger \rvac
\nonumber\\
&&=
+S^v(m_1-n_1)_{\alpha_1\beta_1} S^u(m_2-n_2)_{\alpha_2\beta_2}
{\rm Tr} \left\{ V \cdot S^v \right\}
\nonumber\\
&&
-S^u(m_1-n_2)_{\alpha_1\beta_2} S^v(m_2-n_1)_{\alpha_2\beta_1}
{\rm Tr} \left\{ V \cdot S^v \right\}
\nonumber\\
&&- S^v(m_1-n_1)_{\alpha_1\beta_1}
    S^v \cdot V \cdot S^u(m_2,n_2)_{\alpha_2\beta_2}
\nonumber\\
&&+ S^u(m_1-n_2)_{\alpha_1\beta_2}
    S^v \cdot V \cdot S^v(m_2,n_1)_{\alpha_2\beta_1}
\nonumber\\
&&-S^u \cdot V \cdot S^u(m_1,n_2)_{\alpha_1\beta_2}
    S^v(m_2-n_1)_{\alpha_2\beta_1}
\nonumber\\
&&+ S^u \cdot V \cdot S^v(m_1,n_1)_{\alpha_1\beta_1}
    S^u(m_2-n_2)_{\alpha_2\beta_2} .
\nonumber\\
\end{eqnarray}

Using these results, we obtain
\begin{eqnarray}
&&\sum_{\rm two\ particle}\rN \frac{1}{E_0-E_n} \lN \vert \hat V \rvac
\nonumber\\
&&=
\int\frac{d^4 q}{(2\pi)^4}\frac{d^4 p}{(2\pi)^4} \,
    \frac{2}{\lambda(q)+\lambda(p)}
e^{i q n} \left[ S^u(q) V(q;p) S^v(p) \right]_{\alpha\beta} e^{-ipm}
\times \nonumber\\
&&\qquad
\frac{1}{2} \left\{ a_{m\beta}a_{n\alpha}^\dagger
                  -a_{n\alpha}^\dagger a_{m\beta} \right\} \rvac .
\nonumber\\
\end{eqnarray}
\begin{eqnarray}
&& \sum_{\rm two\ particle}
\rN \frac{1}{E_0-E_n} \lN \vert V
\rM \frac{1}{E_0-E_m} \lM \vert V \rvac
\nonumber\\
&&=
\int\frac{d^4 q}{(2\pi)^4}\frac{d^4 p}{(2\pi)^4}\frac{d^4 k}{(2\pi)^4} \,
    \frac{2}{\lambda(q)+\lambda(k)}\frac{2}{\lambda(k)+\lambda(p)}
\nonumber\\
&& \qquad
e^{i q n}
\left[ S^u(q) V(q;k) \right( S^v(k)-S^u(k) \left) V(k;p) S^v(p)
\right]_{\alpha\beta}
e^{-ip m} \times \nonumber\\
&& \qquad
\frac{1}{2} \left\{ a_{m\beta}a_{n\alpha}^\dagger
                  -a_{n\alpha}^\dagger a_{m\beta} \right\} \rvac .
\nonumber\\
\end{eqnarray}
\begin{eqnarray}
&&\sum_{\rm two\ particle} \rN \frac{1}{(E_0-E_n)^2} \lN \vert V \rvac
           \lvac\vert V \rvac
\nonumber\\
&&=
\int\frac{d^4 q}{(2\pi)^4}\frac{d^4 p}{(2\pi)^4}\frac{d^4 k}{(2\pi)^4} \,
    \frac{4}{\left(\lambda(q)+\lambda(p)\right)^2}
\nonumber\\
&& \qquad
{\rm Tr}\left\{ V(k;k) S^v(k) \right\}
e^{i q n} \left[ S^u(q) V(q;p) S^v(p)\right]_{\alpha\beta} e^{-ip m}
\times \nonumber\\
&&
\frac{1}{2} \left\{ a_{m\beta}a_{n\alpha}^\dagger
                  -a_{n\alpha}^\dagger a_{m\beta} \right\} \rvac .
\nonumber\\
\end{eqnarray}
\begin{eqnarray}
&& \frac{1}{2}\sum_{\rm two\ particle} \lvac \vert V
                       \rN \frac{1}{(E_0-E_n)^2} \lN \vert V \rvac
\nonumber\\
&&=
- \frac{1}{2}
\int\frac{d^4 q}{(2\pi)^4}\frac{d^4 p}{(2\pi)^4} \,
    \frac{4}{\left(\lambda(q)+\lambda(p)\right)^2}
{\rm Tr} \left\{ S^u(q) V(q;p) S^v(p) V(p;q) \right\} . \nonumber\\
\end{eqnarray}

\subsection{Overlap with insertion of $\hat \Pi$}

Given the expression of the background vacuum, its overlap to
the free vacuum with the insertion of the operator of the gauge
fluctuation $\hat \Pi$ is evaluated as
\begin{eqnarray}
\label{eq:appendix-overlap-with-Pi}
&& \lvac \vert \hat \Pi \rVac \nonumber\\
&&= \lvac \vert \hat \Pi \rvac \times \nonumber\\
&&\quad
\left(
1
+ \frac{1}{2}
\int\frac{d^4 q}{(2\pi)^4}\frac{d^4 p}{(2\pi)^4} \,
    \frac{4}{\left(\lambda(q)+\lambda(p)\right)^2}
{\rm Tr} \left\{ S^u(q) V(q;p) S^v(p) V(p;q) \right\}
\right) \nonumber\\
&&
+
\lvac \vert \hat \Pi \,
\frac{1}{2} \left\{ a_{m\beta}a_{n\alpha}^\dagger
                  -a_{n\alpha}^\dagger a_{m\beta} \right\} \rvac
\times \nonumber\\
&&\qquad
\left\{
\int\frac{d^4 q}{(2\pi)^4}\frac{d^4 p}{(2\pi)^4} \,
    \frac{2}{\lambda(q)+\lambda(p)}
e^{i q n} \left[ S^u(q) V(q;p) S^v(p) \right]_{\alpha\beta} e^{-ipm}
\right.
\nonumber\\
&&\qquad
+\int\frac{d^4 q}{(2\pi)^4}\frac{d^4 p}{(2\pi)^4}\frac{d^4 k}{(2\pi)^4} \,
    \frac{2}{\lambda(q)+\lambda(k)}\frac{2}{\lambda(k)+\lambda(p)}
\nonumber\\
&& \quad\qquad\qquad
e^{i q n}
\left[ S^u(q) V(q;k) \right( S^v(k)-S^u(k) \left) V(k;p) S^v(p)
\right]_{\alpha\beta}
e^{-ip m} \nonumber\\
&&\qquad
-\int\frac{d^4 q}{(2\pi)^4}\frac{d^4 p}{(2\pi)^4}\frac{d^4 k}{(2\pi)^4} \,
    \frac{4}{\left(\lambda(q)+\lambda(p)\right)^2}
\nonumber\\
&& \quad\qquad\qquad
\left.
{\rm Tr}\left\{ V(k;k) S^v(k) \right\}
e^{i q n} \left[ S^u(q) V(q;p) S^v(p)\right]_{\alpha\beta} e^{-ip m}
\right\} +{\cal O}(\hat V^3) . \nonumber\\
\end{eqnarray}

The correlation function in the above equation
\begin{equation}
\lvac \vert \hat \Pi
\frac{1}{2} \left\{ a_{m\beta}a_{n\alpha}^\dagger
                  -a_{n\alpha}^\dagger a_{m\beta} \right\} \rvac
/ \lvac \vert \hat \Pi \rvac
\end{equation}
is nothing but the boundary correlation function
calculated in \cite{kikukawa-gauge-two-dimensions} and is given by
\begin{equation}
 \frac{1}{2} \delta_{mn}\delta_i^j - S^v[\exp(-i\lambda\pi)](m,n){}_i^0
                \, \exp(i\lambda\pi){}_o^j ,
\end{equation}
where
\begin{eqnarray}
\label{eq:negative-energy-projection-gauge-appendix}
&&
S^v[\exp(-i\lambda\pi)](n,m){}_i^j \nonumber\\
&&\equiv
\int \frac{d^4 p}{(2\pi)^4} \frac{d^4 q}{(2\pi)^4}
\, e^{i p m } e^{-i q n} \times
\nonumber\\
&&\quad
v(p,s)
\left[ v^\dagger (q,s^\prime)
e^{-i q r} \left(
\exp(i\lambda\pi_r){}_j^i \right) e^{i p r }
v (p,s) \right]^{-1}_{(p,s,i)(q,s^\prime,j)}
v^\dagger (q,s^\prime)  .
\nonumber\\
\end{eqnarray}
It has the following expansion in terms of $\pi$ ($\lambda$).
\begin{eqnarray}
&&\lvac \vert \hat \Pi
\frac{1}{2} \left\{ a_{m\beta}a_{n\alpha}^\dagger
                  -a_{n\alpha}^\dagger a_{m\beta} \right\} \rvac
/ \lvac \vert \hat \Pi \rvac
\nonumber\\
&&= \frac{1}{2} \delta_{mn} - S^v(m-n) \nonumber\\
&&  - \sum_r S^v(m-r) (i\lambda \pi_r ) S^v(r-n)
    - S^v(m-n) (-i\lambda \pi_n )
\nonumber\\
&&  - \sum_r S^v(m-r)
    \frac{1}{2!}(i\lambda \pi_r )^2 S^v(r-n)
\nonumber\\
&&  + \sum_{r,l} S^v(n-r)(i\lambda \pi_r )
               S^v(r-l)(i\lambda \pi_l ) S^v(l-m)
\nonumber\\
&& - \sum_r S^v(m-r) (i\lambda \pi_r ) S^v(r-n) (-i\lambda \pi_n )
 - S^v(m-n) \frac{1}{2!}(-i\lambda \pi_n )^2
\nonumber\\
&& +{\cal O}(\lambda^3) .
\end{eqnarray}

\subsection{Expressions of $i\Delta S_{1\, WZW}[\pi^2;A]$ and
                           $i\Delta S_{2\, WZW}[\pi^2;A^2]$}

By inserting the quadratic terms of the above expansion into
Eq.~(\ref{eq:appendix-overlap-with-Pi}) and extracting the
imaginary part, we finally obtain
\begin{eqnarray}
&& i\Delta S_{1\, WZW}[\pi^2;A] \nonumber\\
&&=
\frac{1}{2} \sum_{rep.} {\rm Tr}
\left[  \bar V_{1+}[A]  \times \phantom{\frac{1}{2}}
\right. \nonumber\\
&&\qquad\qquad
\left.
  \left\{  \frac{1}{2}\left(S^v_+ \pi^2 S^u_+ -S^u_+ \pi^2 S^v_+ \right)
          -\left(S^v_+ \pi S^v_+ \pi S^u_+
                 -S^u_+ \pi S^v_+ \pi S^v_+ \right) \right\} \right]
\nonumber\\
&&-
\frac{1}{2} \sum_{rep.} {\rm Tr}
\left[  \bar V_{1-}[A]  \times \phantom{\frac{1}{2}}
\right. \nonumber\\
&&\qquad\qquad
\left.
  \left\{  \frac{1}{2}\left(S^v_- \pi^2 S^u_- -S^u_- \pi^2 S^v_- \right)
          -\left(S^v_- \pi S^v_- \pi S^u_-
                 -S^u_- \pi S^v_- \pi S^v_- \right) \right\} \right] ,
\nonumber\\
\\
&& i\Delta S_{2\, WZW}[\pi^2;A] \nonumber\\
&&=
\frac{1}{2} \sum_{rep.} {\rm Tr}
\left[ \left( \bar V_{2+}[A^2]
            + \bar V_{1+}[A]( S^v_+ - S^u_+)\bar V_{1+}[A] \right)
       \times \phantom{\frac{1}{2}}
\right. \nonumber\\
&&\qquad\qquad
\left.
  \left\{  \frac{1}{2}\left(S^v_+ \pi^2 S^u_+ -S^u_+ \pi^2 S^v_+ \right)
          -\left(S^v_+ \pi S^v_+ \pi S^u_+
                 -S^u_+ \pi S^v_+ \pi S^v_+ \right) \right\} \right]
\nonumber\\
&&-
\frac{1}{2} \sum_{rep.} {\rm Tr}
\left[ \left(\bar V_{2-}[A^2]
     + \bar V_{1-}[A]( S^v_- - S^u_-)\bar V_{1-}[A] \right)
       \times \phantom{\frac{1}{2}}
\right. \nonumber\\
&&\qquad\qquad
\left.
  \left\{  \frac{1}{2}\left(S^v_- \pi^2 S^u_- -S^u_- \pi^2 S^v_- \right)
          -\left(S^v_- \pi S^v_- \pi S^u_-
                 -S^u_- \pi S^v_- \pi S^v_- \right) \right\} \right] ,
\nonumber\\
\end{eqnarray}
where
\begin{eqnarray}
\bar V_{1\pm} [A]_{nm}{}_i^j
&=& \int \frac{d^4 q}{(2\pi)^4}\frac{d^4 p}{(2\pi)^4} \, e^{i q n -i p m}
     \frac{2}{\lambda_\pm(q)+\lambda_\pm(p)} \times \nonumber\\
&&
     \gamma_5 \left[ i \gamma_\mu \cos \left(\frac{q_\mu+p_\mu}{2}\right)
                     + \sin \left(\frac{q_\mu+p_\mu}{2}\right) \right]
     \, \tilde A_\mu(q-p){}_i^j  ,
\nonumber\\
\\
\bar V_{2\pm}[A^2]_{nm}{}_i^j
&=& \int \frac{d^4 q}{(2\pi)^4}\frac{d^4 p}{(2\pi)^4} \, e^{i q n -i p m}
     \frac{2}{\lambda_\pm(q)+\lambda_\pm(p)} \times \nonumber\\
&&
     \gamma_5 \left[-i \gamma_\mu \sin \left(\frac{q_\mu+p_\mu}{2}\right)
                    +\cos \left(\frac{q_\mu+p_\mu}{2}\right) \right]
     \times
\nonumber\\
&&\quad
\int \frac{d^4 k_1}{(2\pi)^4}\frac{d^4 k_2}{(2\pi)^4}
     (2\pi)^4\delta(q-p-k_1-k_2) \,
     \frac{1}{2!} ( \tilde A_\mu(k_1) \tilde A_\mu(k_2) ){}_i^j  ,
\nonumber\\
\end{eqnarray}
and
\begin{eqnarray}
&& S^u_\pm(n-m)  \nonumber\\
&&=
\int \frac{d^4 p}{(2\pi)^4} \, e^{i p (n-m)}
 \left\{ \frac{1}{2}
        +\frac{\gamma_5
   \left[i \gamma_\mu \sin p_\mu +\sum_\mu(1-\cos p_\mu )\pm m_0\right]
              }{2\lambda_\pm(p)} \right\} , \nonumber\\
\\
&& S^v_\pm(n-m) \nonumber\\
&&=
\int \frac{d^4 p}{(2\pi)^4} \, e^{i p (n-m)}
 \left\{ \frac{1}{2}
        -\frac{\gamma_5
   \left[i \gamma_\mu \sin p_\mu +\sum_\mu(1-\cos p_\mu )\pm m_0\right]
              }{2\lambda_\pm(p)} \right\} , \nonumber\\
\\
&&\lambda_\pm(p)=\sqrt{ \sum_\mu \sin^2 p_\mu
                     +\left(\sum_\mu(1-\cos p_\mu )\pm m_0 \right)^2 } .
\end{eqnarray}

It is not difficult to check that these two terms
change the sign under parity and charge conjugation transformations and
invariant under CP transformation. There is no reason based on
symmetry for these terms to vanish at finite lattice cutoff.

\subsection{Properties of $i\Delta S_{1\, WZW}[\pi^2;A]$ and
                          $i\Delta S_{2\, WZW}[\pi^2;A^2]$}

We first show that the leading term
$i \Delta S_{1\, WZW}$ vanishes identically,
\begin{equation}
i \, \Delta S_{1 \, WZW}[\pi^2; A] = 0 .
\end{equation}
It is explicitly evaluated as follows:
\begin{eqnarray}
\label{eq:gaussian-fluctuation-induced-vertex1-WZW-appendix}
i \Delta S_{1\, WZW}[\pi^2;A]
&=& \sum_{rep.} \sum_{n,l_1,l_2}
{\rm Tr}\left\{ A_{n\mu} \pi_{l_1} \pi_{l_2} \right\} \times
\nonumber\\
&&
\int \frac{d^4 k_1}{(2\pi)^4} \frac{d^4 k_2}{(2\pi)^4}  \,
     e^{ik_1(l_1-m)+ik_2 (l_2-m)} \,
     \Gamma_{1\, WZW\,\mu}(k_1,k_2) , \nonumber\\
\end{eqnarray}
where ($p=-k_1-k_2$)
\begin{eqnarray}
&& \Gamma_{1\, WZW\,\mu}(k_1,k_2) \nonumber\\
&&=
\frac{1}{2} \lambda^2 \, \int \frac{d^4 l}{(2\pi)^4}
\frac{2}{\lambda_\pm(l)+\lambda_\pm(l+p)}  \times \nonumber\\
&&\quad
     {\rm Tr}\left[
     \gamma_5 \left[
                 i\gamma_\mu \cos \left(l_\mu+\frac{p_\mu}{2}\right)
                +            \sin \left(l_\mu+\frac{p_\mu}{2}\right)
              \right]
            \right.
\times     \nonumber\\
&&\qquad
             \left.
\left\{
\left(\frac{1}{2}
   -\frac{\gamma_5
    \left[i \gamma_\rho \sin l_\rho +\sum_\rho(1-\cos l_\rho)\pm m_0\right]
              }{2\lambda_\pm(l)} \right)
\right.
             \right.
\times     \nonumber\\
&&\qquad\quad
             \left.
\left.
\left(\frac{\gamma_5
      \left[ i\gamma_\sigma \sin (l_\sigma-k_{2\sigma})
            +\sum_\sigma(1-\cos (l_\sigma-k_{2\sigma}) )\pm m_0\right]
              }{2\lambda_\pm(l-k_2)} \right)
\right.
             \right.
\times     \nonumber\\
&&\qquad\quad
             \left.
\left.
\left(\frac{1}{2}
     +\frac{\gamma_5
      \left[ i\gamma_\lambda \sin (l_\lambda+p_\lambda)
            +\sum_\lambda(1-\cos (l_\lambda+p_\lambda))\pm m_0\right]
              }{2\lambda_\pm(l+p)} \right)
\right.
             \right.
\nonumber\\
&&\qquad
             \left.
\left.
-\left(\frac{1}{2}
     +\frac{\gamma_5
      \left[i \gamma_\rho \sin l_\rho +\sum_\rho(1-\cos l_\rho)\pm m_0\right]
              }{2\lambda_\pm(l)} \right)
\right.
             \right.
\times     \nonumber\\
&&\qquad\quad
             \left.
\left.
\left(\frac{\gamma_5
      \left[ i\gamma_\sigma \sin (l_\sigma-k_{2\sigma})
            +\sum_\sigma(1-\cos (l_\sigma-k_{2\sigma}))\pm m_0\right]
              }{2\lambda_\pm(l-k_2)} \right)
\right.
             \right.
\times     \nonumber\\
&&\qquad\quad
             \left.
\left.
\left(\frac{1}{2}
     -\frac{\gamma_5
      \left[ i\gamma_\lambda \sin (l_\lambda+p_\lambda)
            +\sum_\lambda(1-\cos (l_\lambda+p_\lambda))\pm m_0\right]
              }{2\lambda_\pm(l+p)} \right)
\right\}
             \right] \nonumber\\
&&=
\frac{1}{2} \lambda^2 \, \int \frac{d^4 l}{(2\pi)^4}
\frac{2}{\lambda_\pm(l)+\lambda_\pm(l+p)} \times \nonumber\\
&&\quad
     {\rm Tr}\left[
     \gamma_5 \left[
                 i\gamma_\mu \cos \left(l_\mu+\frac{p_\mu}{2}\right)
                +            \sin \left(l_\mu+\frac{p_\mu}{2}\right)
              \right]
             \right.
\times     \nonumber\\
&&\qquad
             \left.
\left\{
\left(\frac{\gamma_5
      \left[ i\gamma_\sigma \sin (l_\sigma-k_{2\sigma})
            +\sum_\sigma(1-\cos (l_\sigma-k_{2\sigma}) )\pm m_0\right]
              }{2\lambda_\pm(l-k_2)} \right)
\right.
             \right.
\times     \nonumber\\
&&\qquad\quad
             \left.
\left.
\left(
     \frac{\gamma_5
      \left[ i\gamma_\lambda \sin (l_\lambda+p_\lambda)
            +\sum_\lambda(1-\cos (l_\lambda+p_\lambda))\pm m_0\right]
              }{2\lambda_\pm(l+p)} \right)
\right.
             \right.
\nonumber\\
&&\qquad
             \left.
\left.
-\left(
     \frac{\gamma_5
      \left[i \gamma_\rho \sin l_\rho +\sum_\rho(1-\cos l_\rho)\pm m_0\right]
              }{2\lambda_\pm(l)} \right)
\right.
             \right.
\times     \nonumber\\
&&\qquad\quad
             \left.
\left.
\left(\frac{\gamma_5
      \left[ i\gamma_\sigma \sin (l_\sigma-k_{2\sigma})
            +\sum_\sigma(1-\cos (l_\sigma-k_{2\sigma}))\pm m_0\right]
              }{2\lambda_\pm(l-k_2)} \right)
\right\}
             \right] .  \nonumber\\
\end{eqnarray}
By the charge conjugation transformation, we can see that
this vertex is satisfy the relation
\begin{equation}
\Gamma_{1\, WZW\,\mu}(k_1,k_2) = \Gamma_{1\, WZW\,\mu}(k_2,k_1) .
\end{equation}
Then, this contribution turns out to be proportional to
the anomaly coefficient,
\begin{eqnarray}
i \Delta S_{1\, WZW}[\pi^2;A]
&=& \sum_{rep.} \frac{1}{2} d^{abc}
\sum_{n,l_1,l_2} A_{n\mu}^a \pi_{l_1}^b \pi_{l_2}^c \times
\nonumber\\
&&
\int \frac{d^4 k_1}{(2\pi)^4} \frac{d^4 k_2}{(2\pi)^4}  \,
     e^{ik_1(l_1-m)+ik_2 (l_2-m)} \,
     \Gamma_{1\, WZW\,\mu}(k_1,k_2) ,  \nonumber\\
\end{eqnarray}
and vanishes identically for anomaly free theories.
Even for anomalous case, it is easily seen by the $\gamma_5$
book-keeping that $\Gamma_{1\, WZW\,\mu}$ itself vanishes
because of the vanishing trace in the spinor space.

Next we turn to the property of the next-to-leading term
$i \Delta S_{2\, WZW}$. Its explicit formula is given as
\begin{eqnarray}
i \Delta S_{2\, WZW}[\pi^2;A^2]
&=& \sum_{rep.}\sum_{n,m,l_1,l_2}
{\rm Tr}\left\{ A_{n\mu} A_{m\nu} \pi_{l_1} \pi_{l_2} \right\} \times
\nonumber\\
&&\quad
\int \frac{d^4 p}{(2\pi)^4}
     \frac{d^4 k_1}{(2\pi)^4} \frac{d^4 k_2}{(2\pi)^4}  \,
     e^{ip(m-n)+ik_1(l_1-n)+ik_2 (l_2-n)}  \times \nonumber\\
&&\qquad\qquad\qquad
       \Gamma_{2\, WZW\, \mu\nu}(p,k_1,k_2) ,  \nonumber\\
\end{eqnarray}
where ($q+p+k_1+k_2=0$)
\begin{eqnarray}
&& \Gamma_{2\, WZW\, \mu\nu}(p,k_1,k_2)
  \nonumber\\
&&= -\frac{1}{2} \lambda^2 \,
\int \frac{d^4 l}{(2\pi)^4}
\frac{2}{\lambda_\pm(l+q)+\lambda_\pm(l)}
\frac{2}{\lambda_\pm(l-p)+\lambda_\pm(l)}
\frac{1}{\lambda_\pm(l)}
\times     \nonumber\\
&&\quad
     {\rm Tr}\left[
     \gamma_5 \left[
     i\gamma_\mu \cos \left(l_\mu+\frac{q_\mu}{2}\right)
    +            \sin \left(l_\mu+\frac{q_\mu}{2}\right)
              \right]
             \right.
\times     \nonumber\\
&&\quad
             \left.
    \gamma_5
      \left[ i\gamma_\rho \sin l_\rho
            +\sum_\mu(1-\cos l_\rho) \pm m_0\right] \,
             \right.
\times     \nonumber\\
&&\quad
             \left.
    \gamma_5 \left[
                 i\gamma_\mu \cos \left(l_\nu-\frac{p_\nu}{2}\right)
                +            \sin \left(l_\nu-\frac{p_\nu}{2}\right)
              \right]
             \right.
\times     \nonumber\\
&&\quad
             \left.
\left\{
\left(\frac{\gamma_5
      \left[ i\gamma_\sigma \sin (l_\sigma-p_\sigma-k_{2\sigma})
            + \sum_\sigma(1-\cos(l_\sigma-p_\sigma-k_{2\sigma}))
            \pm m_0\right]
              }{2\lambda_\pm(l-p-k_2)} \right)
             \right.
\right.
\times     \nonumber\\
&&\quad\quad
             \left.
\left.
\left(
     \frac{\gamma_5
      \left[ i\gamma_\lambda \sin (l_\lambda+q_\lambda)
            +\sum_\lambda(1-\cos (l_\lambda+q_\lambda))
            \pm m_0\right]
              }{2\lambda_\pm(l+q)} \right)
             \right.
\right.
\nonumber\\
&&\quad
             \left.
\left.
-\left(
     \frac{\gamma_5
    \left[i \gamma_\sigma \sin(l_\sigma-p_\sigma)
         +\sum_\sigma(1-\cos (l_\sigma-p_\sigma))
            \pm m_0\right]
              }{2\lambda_\pm(l-p)} \right)
             \right.
\right.
\times     \nonumber\\
&&\quad\quad
             \left.
\left.
\left(\frac{\gamma_5
      \left[ i\gamma_\lambda \sin (l_\lambda-p_\lambda-k_{2\lambda})
            +\sum_\lambda(1-\cos (l_\lambda-p_\lambda-k_{2\lambda}))
            \pm m_0\right]
              }{2\lambda_\pm(l-p-k_2)} \right)
\right\}
             \right] .  \nonumber\\
\end{eqnarray}
By the charge conjugation transformation again, we can see that
this vertex is satisfy the relation
\begin{equation}
\Gamma_{2\, WZW\,\mu}(p,k_1,k_2) = \Gamma_{2\, WZW\,\mu}(q,k_2,k_1) .
\end{equation}
Then, this contribution also turns out to be proportional to
the anomaly coefficient,
\begin{eqnarray}
&& i \Delta S_{2\, WZW}[\pi^2;A^2]  \nonumber\\
&&= \sum_{rep.} \frac{1}{8}(-if^{abe}d^{cde}-id^{abe}f^{cde} )
\sum_{n,m,l_1,l_2} A_{n\mu}^a A_{m\nu}^b \pi_{l_1}^c \pi_{l_2}^c \times
\nonumber\\
&&\quad
\int \frac{d^4 p}{(2\pi)^4}
     \frac{d^4 k_1}{(2\pi)^4} \frac{d^4 k_2}{(2\pi)^4}  \,
     e^{ip(m-n)+ik_1(l_1-n)+ik_2 (l_2-n)}  \times \nonumber\\
&&\qquad\qquad\qquad
       \Gamma_{2\, WZW\, \mu\nu}(p,k_1,k_2) ,  \nonumber\\
\end{eqnarray}

On the other hand, the factor of the trace of the spinor space
is evaluated as
\begin{eqnarray}
&& \frac{1}{4 \lambda_\pm(l-p-k_2) \lambda_\pm(l+q) } \times
\nonumber\\
&&
\left\{
+i\epsilon_{\rho\nu\sigma\lambda}
 \sin \left(l_\mu+\frac{q_\mu}{2}\right)
 \sin l_\rho
 \cos \left(l_\nu-\frac{p_\nu}{2}\right)
 \sin (l_\sigma-p_\sigma-k_{2\sigma})
 \sin (l_\lambda+q_\lambda)
\right.
\nonumber\\
&&
\left.
-i\epsilon_{\mu\nu\sigma\lambda}
\cos \left(l_\mu+\frac{q_\mu}{2}\right)
\left(B(l) \pm m_0\right)
 \cos \left(l_\nu-\frac{p_\nu}{2}\right)
 \sin (l_\sigma-p_\sigma-k_{2\sigma})
 \sin (l_\lambda+q_\lambda)
\right.
\nonumber\\
&&
\left.
+i\epsilon_{\mu\rho\sigma\lambda}
 \cos \left(l_\mu+\frac{q_\mu}{2}\right)
 \sin l_\rho
 \sin \left(l_\nu-\frac{p_\nu}{2}\right)
 \sin (l_\sigma-p_\sigma-k_{2\sigma})
 \sin (l_\lambda+q_\lambda)
\right.
\nonumber\\
&&
\left.
-i\epsilon_{\mu\rho\nu\lambda}
 \cos \left(l_\mu+\frac{q_\mu}{2}\right)
 \sin l_\rho
\cos \left(l_\nu-\frac{p_\nu}{2}\right)
\left( B(l-p-k_2)\pm m_0 \right)
 \sin (l_\lambda+q_\lambda)
\right.
\nonumber\\
&&
\left.
+i\epsilon_{\mu\rho\nu\sigma}
 \cos \left(l_\mu+\frac{q_\mu}{2}\right)
 \sin l_\rho
\cos \left(l_\nu-\frac{p_\nu}{2}\right)
 \sin (l_\sigma-p_\sigma-k_{2\sigma})
 \left( B(l+q)\pm m_0 \right)
\right\}
\nonumber\\
&& -\frac{1}{4 \lambda_\pm(l-p) \lambda_\pm(l-p-k_2) } \times
\nonumber\\
&&
\left\{
+i\epsilon_{\rho\nu\sigma\lambda}
 \sin \left(l_\mu+\frac{q_\mu}{2}\right)
 \sin l_\rho
 \cos \left(l_\nu-\frac{p_\nu}{2}\right)
 \sin (l_\sigma-p_\sigma)
 \sin (l_\lambda-k_{2\lambda}-p_\lambda)
\right.
\nonumber\\
&&
\left.
-i\epsilon_{\mu\nu\sigma\lambda}
\cos \left(l_\mu+\frac{q_\mu}{2}\right)
\left(B(l) \pm m_0\right)
 \cos \left(l_\nu-\frac{p_\nu}{2}\right)
 \sin (l_\sigma-p_\sigma)
 \sin (l_\lambda-p_\lambda-k_{2\lambda})
\right.
\nonumber\\
&&
\left.
+i\epsilon_{\mu\rho\sigma\lambda}
 \cos \left(l_\mu+\frac{q_\mu}{2}\right)
 \sin l_\rho
 \sin \left(l_\nu-\frac{p_\nu}{2}\right)
 \sin (l_\sigma-p_\sigma)
 \sin (l_\lambda-p_\lambda-k_{2\lambda})
\right.
\nonumber\\
&&
\left.
-i\epsilon_{\mu\rho\nu\lambda}
 \cos \left(l_\mu+\frac{q_\mu}{2}\right)
 \sin l_\rho
\cos \left(l_\nu-\frac{p_\nu}{2}\right)
\left( B(l-p)\pm m_0 \right)
 \sin (l_\lambda-p_\lambda-k_{2\lambda})
\right.
\nonumber\\
&&
\left.
+i\epsilon_{\mu\rho\nu\sigma}
 \cos \left(l_\mu+\frac{q_\mu}{2}\right)
 \sin l_\rho
 \cos \left(l_\nu-\frac{p_\nu}{2}\right)
 \sin (l_\sigma-p_\sigma)
 \left( B(l-p-k_2)\pm m_0 \right)
\right\} .
\nonumber\\
\end{eqnarray}
{}From this expression, we can see that the trace vanishes if we set
$p=0$ and $q=0 (k_1+k_2=0)$.
Therefore $i \Delta S_{2\, WZW}$ satisfy the following property
\begin{equation}
      \Gamma_{2\, WZW\, \mu\nu}(p,k_1,k_2)= 0, \quad
{\rm at \ } p=0, k_1+k_2=0 .
\end{equation}

In summary, $i \Delta S_{1\, WZW}$ vanishes identically
for anomaly free theories as well as anomalous ones.
$i \Delta S_{2\, WZW}$ vanishes identically for anomaly free
theories. It vanishes at the kinematical limit
$p=0$ and $q=0 (k_1+k_2=0)$ for anomalous theories.
In fact, the latter property of $i \Delta S_{2 \, WZW}$ and
the vanishing $i \Delta S_{1\, WZW}$ are
sufficient to prove the one-loop renormalizability of the pure gauge
model, as we will show in the next section.

\section{One-loop renormalizability}
\label{appendix:one-loop-renormalizability-any-theory}
\reseteqnum

In the text, we have shown that the pure gauge model of the
anomaly-free nonabelian chiral gauge theory defined through
the vacuum overlap is renormalizable at one-loop.
In this appendix, we discuss the one-loop renormalizability of the
pure gauge model from a more general point of view which could
cover anomalous theories, as well.
We will see that even for anomalous theories,
the pure gauge model is one-loop renormalizable.

It is instructive to have a general formula for the quantum
correction to the classical action without taking account of
the properties of $i \Delta S_{WZW}$ specific to anomaly-free
theories. The quantum correction to the classical action can be expressed
up to the fourth order in $A_{n\mu}$, $c$ and $\bar c$ as follows:
\begin{eqnarray}
\label{eq:one-loop-effective-action-A1-WZW}
\Delta S_1 [A]  &=&   \langle S_1 [A] \rangle_0
                     +\langle  i\Delta S_{1 \, WZW}[A] \rangle_0 , \\
\label{eq:one-loop-effective-action-A2-WZW-appendix}
\Delta S_2 [A^2] &=& \langle S_2[A^2] \rangle_0
                    +\frac{1}{2!} \langle S_1[A] ^2 \rangle_0
                    -\frac{1}{2!} \langle \Delta S_{1 \, WZW}[A]^2 \rangle_0
\nonumber\\
&&           +\langle  i\Delta S_{2 \, WZW}[A^2] \rangle_0
             +\langle  i\Delta S_{1 \, WZW}[A] \, S_1[A] \rangle_0 ,\\
\label{eq:one-loop-effective-action-c2-WZW}
\Delta S_{0,1}[(c,\bar c)]
&=&
\langle S_{2c}[(c,\bar c)] \rangle_0
+ \frac{1}{2} \langle S_{1c}[c /\bar c] ^2  \rangle_0  , \\
\label{eq:one-loop-effective-action-Ac2-WZW}
\Delta S_{1,1}[A,(c,\bar c)]
&=&
\langle S_{2c}[A,(c,\bar c)] \rangle_0
   + \langle S_{2c}[(c,\bar c)] \, S_1[A] \rangle_0
  \nonumber\\
&& + \langle S_{1c}[c /\bar c] \, S_{1c}[A, c /\bar c] \rangle_0
  \nonumber\\
&& + \langle i\Delta S_{1 \, WZW}[A] \, S_{2c}[(c,\bar c)] \rangle_0
\\
\label{eq:one-loop-effective-action-A3-WZW}
\Delta S_3 [A^3] &=&  \langle S_1 [A^3] \rangle_0
                    + \langle S_1[A] S_2[A^2]  \rangle_0
                    + \frac{1}{3!} \langle S_1[A]^3 \rangle_0 \nonumber\\
&& - \frac{1}{2} \langle \Delta S_{1 \, WZW}[A]^2 \, S_1[A] \rangle_0
\nonumber\\
&& - \langle  \Delta S_{1 \, WZW}[A] \,
              \Delta S_{2 \, WZW}[A^2] \rangle_0  \nonumber \\
&& + \langle  i\Delta S_{3 \, WZW}[A^3] \rangle_0
   + \langle  i\Delta S_{2 \, WZW}[A^2] S_1[A] \rangle_0  \nonumber \\
&& + \langle  i\Delta S_{1 \, WZW}[A] S_2[A^2] \rangle_0
   - \frac{1}{3!} \langle  i \Delta S_{1 \, WZW}[A]^3 \rangle_0
\nonumber\\
&& + \frac{1}{2} \langle  i \Delta S_{1 \, WZW}[A] S_1[A]^2 \rangle_0,
\nonumber\\
\\
\label{eq:one-loop-effective-action-A4-WZW}
\Delta S_4 [A^4] &=&  \langle S_2 [A^4] \rangle_0
                    + \langle S_1[A] S_1[A^3]  \rangle_0
                    +\frac{1}{2!} \langle S_2[A^2]^2 \rangle_0
\nonumber\\
&&                 +\frac{1}{2} \langle S_1[A]^2 S_2[A^2] \rangle_0
                   +\frac{1}{4!} \langle S_1[A]^4 \rangle_0
\nonumber\\
&& -\frac{1}{2!} \langle \Delta S_{2 \, WZW}[A^2] ^2 \rangle_0
   -\frac{1}{2} \langle \Delta S_{1 \, WZW}[A]^2 S_2[A^2] \rangle_0
   \nonumber\\
&&   - \langle  \Delta S_{3 \, WZW}[A^3] \Delta S_{1 \, WZW}[A] \rangle_0
\nonumber\\
&&   -\frac{1}{2}\langle \Delta S_{1 \, WZW}[A^1]^2 S_1[A]^2 \rangle_0
\nonumber\\
&&     +\frac{1}{4!}\langle \Delta S_{1 \, WZW}[A^1]^4 \rangle_0
\nonumber\\
&& + \langle  i\Delta S_{4 \, WZW}[A^4] \rangle_0
   + \langle  i\Delta S_{3 \, WZW}[A^2] S_1[A] \rangle_0
\nonumber\\
&& + \langle  i\Delta S_{2 \, WZW}[A^2] S_2[A^2] \rangle_0 \nonumber\\
&& + \langle  i\Delta S_{1 \, WZW}[A] S_1[A^3] \rangle_0
   + \langle  i\Delta S_{1 \, WZW}[A] S_1[A] S_2[A^2] \rangle_0 \nonumber\\
&& + \frac{1}{3!} \langle  i\Delta S_{1 \, WZW}[A] S_1[A]^3 \rangle_0
   - \frac{1}{3!} \langle  i\Delta S_{1 \, WZW}[A]^3 S_1[A] \rangle_0
. \nonumber\\
\end{eqnarray}
The local operators of dimensions less than five
which can be written in terms of the pure gauge potential
$A_\mu$ and the ghost and anti-ghost fields, $c$ and $\bar c$,
preserve parity and charge conjugation.
Therefore the imaginary parts of the above expressions do not contribute
to these operators.
Then we can see from the above expansion that
if $i\Delta S_{1 \, WZW}[A]$ vanishes identically,
the nontrivial contribution from the imaginary action
$ i \Delta \Gamma_{WZW}$ to the local operators of dimensions
less than five first appear at the order
${\cal O}(A^4)$ in $\Delta S_4[A^4]$ as
\begin{equation}
   -\frac{1}{2!} \langle \Delta S_{2 \, WZW}[\pi^2;A^2] ^2 \rangle_0  .
\end{equation}
This contribution, however, turns out to vanish by the following reason.
We can first take the local operator limit of the product of
the two pure gauge vector potential $A_\mu$'s in each vertex:
this corresponds to the kinematical limit $p=0$ of
$\Gamma_{2\, WZW}(p,k_1,k_2)$ in
Eq.~(\ref{eq:gaussian-fluctuation-induced-vertex2-WZW}).
By the Gaussian functional integration, the two $\pi$'s in one
vertex are ``Wick-contracted'' to the two $\pi$'s of the other vertex.
The local operator limit of the four $A_\mu$'s
\begin{equation}
\left( A_{n\mu} A_{n\nu} \right) \left( A_{l\lambda} A_{l\sigma} \right)
\longrightarrow
A_{n\mu} A_{n\nu} A_{n\lambda} A_{n\sigma}
+ \left( A_{n\mu} A_{n\nu} \right) (l-n)_\gamma \nabla_\gamma
\left( A_{n\lambda} A_{n\sigma} \right) + \cdots ,
\end{equation}
then corresponds to the kinematical limit of the vanishing
net momentum flow through the two propagators connecting the
two vertexes. This is nothing but the kinematical limit
$k_1+k_2=0$ of the vertex $\Gamma_{2\, WZW}(p=0,k_1,k_2)$,
at which it vanishes. We should note that this limit
is IR finite because we have the mass term for
the IR regularization. The momentum dependent terms should
give the local operators of dimension higher than four.

Therefore, the pure gauge model is renormalizable at one-loop,
both for anomaly-free theories and for anomalous theories.
This means that the quantum and dynamical effect of anomaly
will show up at higher order, unlike the two-dimensional case.

Using the explicit expression, it is also not difficult to see that
the imaginary contribution in
Eq.~(\ref{eq:one-loop-effective-action-A2-appendix}) vanishes:
\begin{equation}
\langle  i\Delta S_{2 \, WZW}[\pi^2;A^2] \rangle_0  = 0 .
\end{equation}

\section{Detail of one-loop calculation}
\label{appendix:detail-one-loop-calculation}
\reseteqnum

In this appendix, we describe in some detail the calculation of
the one-loop contributions,
Eqs.~(\ref{eq:one-loop-effective-action-A1}),
(\ref{eq:one-loop-effective-action-A2}),
(\ref{eq:one-loop-effective-action-c2}) and
(\ref{eq:one-loop-effective-action-Ac2}):
\begin{eqnarray}
\label{eq:one-loop-effective-action-A1-appendix}
\Delta S_1 [A]  &=&   \langle S_1 [A] \rangle_0  , \\
\label{eq:one-loop-effective-action-A2-appendix}
\Delta S_2 [A^2] &=& \langle S_2[A^2] \rangle_0
                    +\frac{1}{2!} \langle S_1[A] ^2 \rangle_0 , \\
\label{eq:one-loop-effective-action-c2-appendix}
\Delta S_{0,1}[(c,\bar c)]
&=&
\langle S_{2c}[(c,\bar c)] \rangle_0
+ \frac{1}{2} \langle S_{1c}[c /\bar c] ^2  \rangle_0  , \\
\label{eq:one-loop-effective-action-Ac2-appendix}
\Delta S_{1,1}[A,(c,\bar c)]
&=&
\langle S_{2c}[A,(c,\bar c)] \rangle_0
   + \langle S_{2c}[(c,\bar c)] \, S_1[A] \rangle_0
\nonumber\\
&& + \langle S_{1c}[c /\bar c] \, S_{1c}[A, c /\bar c] \rangle_0 .
   \nonumber\\
\end{eqnarray}

The propagators of the quantum fluctuations of the gauge freedom
and the ghost fields, $\pi$ and $\xi, \bar \xi$, are given by
\begin{eqnarray}
\label{eq:massless-dipole-propagator-of-pi-appendix}
  \langle \pi_n^a \pi_m^b \rangle
&=& \delta^{ab} \int \frac{d^4 p}{(2\pi)^4} e^{ip(n-m)}
    \frac{1}{\left( \sum_\mu 4 \sin^2 \frac{p_\mu}{2} \right)^2 +\mu^4}
\equiv \delta^{ab} G(n-m) , \nonumber\\
\\
\label{eq:propagator-of-xi-appendix}
  \langle \xi_n^a \bar \xi_m^b \rangle
&=& \delta^{ab} \int \frac{d^4 p}{(2\pi)^4} e^{ip(n-m)}
    \frac{1}{\sum_\mu 4 \sin^2 \frac{p_\mu}{2}}
\equiv \delta^{ab} G_c(n-m) . \nonumber\\
\end{eqnarray}

The contribution,
Eq.~(\ref{eq:one-loop-effective-action-A1-appendix}), is easily seen
to vanish because of the anti-symmetric nature of the group structure
constant in the vertex,
Eq.~(\ref{eq:vertex-quadratic-in-pi-A1-appendix}).
\begin{equation}
\langle S_1 [A] \rangle_0 =0 .
\end{equation}

The first term in
Eq.~(\ref{eq:one-loop-effective-action-A2-appendix}) is evaluated
as follows:
\begin{eqnarray}
\langle S_2[A^2] \rangle_0
&=&
\left\langle
\sum_n \frac{1}{2} \bar \nabla_\nu \hat A_{n\nu}^a \times
\right.
\nonumber\\
&& \left. \quad
\bar \nabla_\mu \left\{
{\rm Tr}\left[
  \left( \pi_{n+\hat\mu}[ \pi_{n+\hat\mu},T^a]
\right.\right.\right.\right.\nonumber\\
&&\left.\left.\left.\left. \qquad\qquad
        +\pi_{n}[ \pi_{n},T^a]
        +[T^a,\pi_{n+\hat\mu} ] \pi_{n+\hat\mu}
\right.\right.\right.\right.\nonumber\\
&&\left.\left.\left.\left. \qquad\qquad
        +[T^a,\pi_{n}] \pi_{n}
        +2 \nabla_\mu \pi_n T^a \nabla_\mu \pi_n
  \right) \sin A_{n\mu}
        \right] \right\} \phantom{\frac{1}{2}} \right\rangle_0 \nonumber\\
&-&
\left\langle
\sum_n \frac{1}{8}
   f^{abc}
   \left( 2 \pi_n^b \bar \nabla_\nu \hat A_{n\nu}^c
         +\nabla_\nu \pi_n^b \hat A_{n\nu}^c
         +\bar \nabla_\nu \pi_n^b \hat A_{n-\hat\nu,\nu}^c \right) \times
\right.\nonumber\\
&& \qquad
\left.   f^{ade}
   \left( 2 \pi_n^d \bar \nabla_\mu \hat A_{n\mu}^e
         +\nabla_\mu \pi_n^d \hat A_{n\mu}^e
         +\bar \nabla_\mu \pi_n^d \hat A_{n-\hat\mu,\mu}^e \right)
\phantom{\frac{1}{8}}
\right\rangle_0
\nonumber\\
&-&
\left\langle
\sum_n \nabla^2 \pi_n^a \bar\nabla_\mu \left\{ {\rm Tr}
        \left[ \{ T^a,T^b \}\left(\cos A_{n\mu}-1\right)\right]
        \nabla_\mu \pi_n^b \right\}
\right\rangle_0
\nonumber\\
&+&
\left\langle
  \sum_n \bar\xi_n^a \bar\nabla_\mu \left\{ {\rm Tr}
\left[ \{ T^a,T^b \}\left(\cos A_{n\mu}-1\right) \right]
\nabla_\mu \xi_n^b \right\}
\right\rangle_0
\nonumber\\
&=&
\sum_n
\frac{1}{2} \bar \nabla_\mu \hat A_{n\mu}^a \bar \nabla_\nu \hat A_{n\nu}^b
\,  \left\{ G(0) f^{acd}f^{bcd}
    \right.
\nonumber\\
&&  \left. \quad
          +4 \left[ G(0)-G(1) \right]
           \left( \frac{1}{4N} \delta^{ab}
                   -\frac{1}{8}f^{acd}f^{bcd}+\frac{1}{8}d^{acd}d^{bcd}
             \right)
    \right\}
\nonumber\\
&-&
\sum_n
\frac{1}{8} \bar \nabla_\mu \hat A_{n\mu}^a \bar \nabla_\nu \hat A_{n\nu}^b
\,  f^{acd}f^{bcd}
\left\{ 4 G(0) - 4 \left[ G(0)-G(1) \right] \right\}
\nonumber\\
&-&
\sum_n
\frac{1}{8} \hat A_{n\mu}^a \hat A_{n\nu}^b
\,  f^{acd}f^{bcd}
\left[ G(\mu-\nu)+G(0)-G(\mu)-G(-\nu) \right]
\nonumber\\
&-&
\sum_n
\frac{1}{8} \hat A_{n-\hat\mu,\mu}^a \hat A_{n-\hat\nu,\nu}^b
\,  f^{acd}f^{bcd} \nonumber\\
&& \qquad
\left[ G(-\mu+\nu)+G(0)-G(-\mu)-G(\nu) \right]
\nonumber\\
&-&
\sum_n
\frac{1}{8} \hat A_{n\mu}^a \hat A_{n-\hat\nu,\nu}^b
\,  f^{acd}f^{bcd}
\left[ G(\mu)+G(\nu)-G(\mu+\nu)-G(0) \right]
\nonumber\\
&-&
\sum_n
\frac{1}{8} \hat A_{n-\hat\mu,\mu}^a \hat A_{n\nu}^b
\,  f^{acd}f^{bcd} \times \nonumber\\
&& \qquad
\left[ G(-\mu)+G(-\nu)-G(-\mu-\nu)-G(0) \right]
\nonumber\\
&-&
\sum_n (\nabla^2)^2 G (0)
        \left(\frac{N^2-1}{N}\right)
        {\rm Tr}\left(\cos A_{n\mu}-1\right)
\nonumber\\
&-&
\sum_n (\nabla^2) G_c(0)
        \left(\frac{N^2-1}{N}\right)
        {\rm Tr}\left(\cos A_{n\mu}-1\right)
\nonumber\\
\end{eqnarray}
Noting the relations,
$f^{acd}f^{bcd}=N\delta^{ab}$
and $d^{acd}d^{bcd}=\frac{N^2-4}{N}\delta^{ab}$
and using the identities
\begin{eqnarray}
(\nabla^2)^2 G (n-m) &=& \delta_{nm} - \mu^4 G(n-m)  ,  \\
\nabla^2 G_c (n-m) &=& - \delta_{nm}  ,
\end{eqnarray}
the above expression is reduced to
\begin{eqnarray}
\label{eq:reduced-formula-of-first-term-in-A2-appendix}
\langle S_2[A^2] \rangle_0
&=&
\sum_n \left(\frac{N^2-1}{2N}\right)\, \left[ G(0)-G(1) \right] \,
 \bar \nabla_\mu \hat A_{n\mu}^a \bar \nabla_\nu \hat A_{n\nu}^a
\nonumber\\
&-&
\sum_n \frac{N}{8} \hat A_{n\mu}^a \hat A_{n\nu}^a \,
\left[ G(\mu-\nu)+G(0)-G(\mu)-G(-\nu) \right]
\nonumber\\
&-&
\sum_n \frac{N}{8} \hat A_{n-\hat\mu,\mu}^a \hat A_{n-\hat\nu,\nu}^a\,
\left[ G(-\mu+\nu)+G(0)-G(-\mu)-G(\nu) \right]
\nonumber\\
&-&
\sum_n \frac{N}{8} \hat A_{n\mu}^a \hat A_{n-\hat\nu,\nu}^a \,
\left[ G(\mu)+G(\nu)-G(\mu+\nu)-G(0) \right]
\nonumber\\
&-&
\sum_n \frac{N}{8} \hat A_{n-\hat\mu,\mu}^a \hat A_{n\nu}^a \,
\left[ G(-\mu)+G(-\nu)-G(-\mu-\nu)-G(0) \right]
\nonumber\\
&+& {\cal O}(\mu^4) .
\nonumber\\
\end{eqnarray}

The explicit formula of the second term in
Eq.~(\ref{eq:one-loop-effective-action-A2-appendix}) is given
as follows:
\begin{eqnarray}
\frac{1}{2!} \langle S_1[A]^2 \rangle_0
&=&
\frac{1}{2!} \left\langle
\left(
\sum_{n\mu}
\frac{1}{2} \bar \nabla_\mu \hat A_{n\mu}^a f^{abc} \pi_n^b \nabla^2 \pi_n^c
\right.\right.
\nonumber\\
&&
\left.\left. \qquad\qquad
-\sum_{n\mu}
\frac{1}{2} \nabla^2 \pi_n^a f^{abc}
 \left(\nabla_\mu \pi_n^b \hat A_{n\mu}^c
      +\bar \nabla_\mu \pi_n^b \hat A_{n-\hat\mu,\mu}^c \right)
\right.\right.
\nonumber\\
&&
\left.\left. \qquad\qquad\qquad
+ \sum_n \frac{1}{2} f^{abc}
   \bar \xi_n^a \bar \nabla_\mu
  \left\{
   \left( \xi_n^b + \xi_{n+\hat\mu}^b \right) \hat A_{n\mu}^c
  \right\}
\right)^2
\right\rangle_0  \nonumber\\
\end{eqnarray}
It can be divided into the four contributions and
they are evaluated separately as follows:
\begin{eqnarray}
&& \frac{1}{2!}
\left\langle
\left(
\sum_{n\mu}
\frac{1}{2}
 \bar \nabla_\mu \hat A_{n\mu}^a f^{abc} \pi_n^b \nabla^2 \pi_n^c
\right)^2
\right\rangle_0
\nonumber\\
&&\qquad
= \frac{N}{8}\sum_{nm} \bar \nabla_\mu \hat A_{n\mu}^a
                     \bar \nabla_\nu \hat A_{m\nu}^a \times
\nonumber\\
&& \qquad\quad
  \left[ G(n-m)(\nabla^2)^2 G(m-n)-\nabla^2 G(n-m)\nabla^2 G(m-n) \right] ,
\nonumber\\
\\
&& \frac{1}{2!}
\left\langle
\left(
-\sum_{n\mu}
\frac{1}{2} \nabla^2 \pi_n^a f^{abc}
 \left(\nabla_\mu \pi_n^b \hat A_{n\mu}^c
      +\bar \nabla_\mu \pi_n^b \hat A_{n-\hat\mu,\mu}^c \right)
\right)^2
\right\rangle_0
\nonumber\\
&&=
\frac{N}{8}\sum_{nm} \hat A_{n\mu}^a \hat A_{m\nu}^a \, \,
     (\nabla^2)^2 G(n-m) \times \nonumber\\
&& \quad
           \left[ G(m+\hat\nu-n-\hat\mu)  + G(m-n)
                 -G(m+\hat\nu-n)-G(m-n-\hat\mu) \right]
\nonumber\\
&&+
\frac{N}{8}\sum_{nm} \hat A_{n-\hat\mu,\mu}^a
                     \hat A_{m-\hat\nu,\nu}^a \, \,
     (\nabla^2)^2 G(n-m) \times \nonumber\\
&& \quad
           \left[ G(m-n) + G(m-\hat\nu-n+\hat\mu)
                 -G(m-\hat\nu-n)-G(m-n+\hat\mu) \right]
\nonumber\\
&&+
\frac{N}{8}\sum_{nm} \hat A_{n-\hat\mu,\mu}^a \hat A_{m\nu}^a \, \,
     (\nabla^2)^2 G(n-m) \times \nonumber\\
&& \quad
           \left[ G(m+\hat\nu-n) + G(m-n+\hat\mu)
                 -G(m-n)-G(m+\hat\nu-n+\hat\mu) \right]
\nonumber\\
&&+
\frac{N}{8}\sum_{nm} \hat A_{n\mu}^a \hat A_{m-\hat\nu,\nu}^a \, \,
     (\nabla^2)^2 G(n-m) \times \nonumber\\
&& \quad
           \left[ G(m-n-\hat\mu) + G(m-\hat\nu-n)
                 -G(m-\hat\nu-n-\hat\mu)-G(m-n) \right]
\nonumber\\
&&-
\frac{N}{8}\sum_{nm} \hat A_{n\mu}^a \hat A_{m\nu}^a \, \times
\nonumber\\
&& \quad
       \left[ \nabla^2 G(n-m-\hat\nu) - \nabla^2 G(n-m) \right] \,
       \left[ \nabla^2 G(m-n-\hat\mu) - \nabla^2 G(m-n) \right]
\nonumber\\
&&-
\frac{N}{8}\sum_{nm} \hat A_{n-\hat\mu,\mu}^a
                     \hat A_{m-\hat\nu,\nu}^a \, \times
\nonumber\\
&& \quad
       \left[ \nabla^2 G(n-m) - \nabla^2 G(n-m+\hat\nu) \right] \,
       \left[ \nabla^2 G(m-n) - \nabla^2 G(m-n+\hat\mu) \right]
\nonumber\\
&&-
\frac{N}{8}\sum_{nm} \hat A_{n-\hat\mu,\mu}^a \hat A_{m\nu}^a \, \times
\nonumber\\
&& \quad
       \left[ \nabla^2 G(n-m-\hat\nu) - \nabla^2 G(n-m) \right] \,
       \left[ \nabla^2 G(m-n) - \nabla^2 G(m-n+\hat\mu) \right]
\nonumber\\
&&-
\frac{N}{8}\sum_{nm} \hat A_{n\mu}^a \hat A_{m-\hat\nu,\nu}^a \, \times
\nonumber\\
&& \quad
       \left[ \nabla^2 G(n-m) - \nabla^2 G(n-m+\hat\nu) \right] \,
       \left[ \nabla^2 G(m-n-\hat\mu) - \nabla^2 G(m-n) \right] ,
\nonumber\\
\\
&&
\left\langle
\left(
\sum_{n\mu}
\frac{1}{2} \bar \nabla_\mu \hat A_{n\mu}^a f^{abc} \pi_n^b \nabla^2 \pi_n^c
\right)  \times
\right. \nonumber\\
&& \qquad \left.
\left(
-\sum_{n\mu}
\frac{1}{2} \nabla^2 \pi_n^a f^{abc}
 \left(\nabla_\mu \pi_n^b \hat A_{n\mu}^c
      +\bar \nabla_\mu \pi_n^b \hat A_{n-\hat\mu,\mu}^c \right)
\right)
\right\rangle_0
\nonumber\\
&& =
-\frac{N}{4} \sum_{nm}
\bar \nabla_\mu \hat A_{n\mu}^a \hat A_{m\nu}^a \, \times
\nonumber\\
&& \qquad
       \nabla^2 G(n-m)
       \left[ \nabla^2 G(m+\hat\nu-n)-\nabla^2 G(m-n)  \right]
\nonumber\\
&&-
\frac{N}{4} \sum_{nm}
\bar \nabla_\mu \hat A_{n\mu}^a \hat A_{m-\hat\nu,\nu}^a \, \times
\nonumber\\
&& \qquad
       \nabla^2 G(n-m)
       \left[ \nabla^2 G(m-n)-\nabla^2 G(m-\hat\nu-n)  \right]
\nonumber\\
&&+
\frac{N}{4} \sum_{nm}
\bar \nabla_\mu \hat A_{n\mu}^a \hat A_{m\nu}^a \, \times
\nonumber\\
&& \qquad
       \left[ G(n-m-\hat\nu)-G(n-m)  \right]
       (\nabla^2)^2 G(m-n)
\nonumber\\
&&+
\frac{N}{4} \sum_{nm}
\bar \nabla_\mu \hat A_{n\mu}^a \hat A_{m-\hat\nu,\nu}^a \, \times
\nonumber\\
&& \qquad
       \left[ G(n-m)- G(n-m+\hat\nu)  \right]
       (\nabla^2)^2 G(m-n) ,
\nonumber\\
\\
&& \frac{1}{2!}
\left\langle
\left(
\sum_n \frac{1}{2} f^{abc}
   \bar \xi_n^a \bar \nabla_\mu
  \left\{
   \left( \xi_n^b + \xi_{n+\hat\mu}^b \right) \hat A_{n\mu}^c
  \right\}
\right)^2
\right\rangle_0
\nonumber\\
&&=
\frac{N}{8} \sum_{nm}
\hat A_{n\mu}^a \hat A_{m,\nu}^a \, \times
\nonumber\\
&& \qquad
       \left[ G_c(n-m-\hat\nu) - G_c(n-m) \right] \,
       \left[ G_c(m-n-\hat\mu) - G_c(m-n) \right]
\nonumber\\
&&+
\frac{N}{8} \sum_{nm}
\hat A_{n\mu}^a \hat A_{m,\nu}^a \, \times
\nonumber\\
&& \qquad
    \left[ G_c(n+\hat\mu -m-\hat\nu) - G_c(n+\hat\mu-m) \right]
\times \nonumber\\
&& \qquad\qquad
    \left[ G_c(m+\hat\nu-n-\hat\mu) - G_c(m+\hat\nu-n) \right]
\nonumber\\
&&+
\frac{N}{8} \sum_{nm}
\hat A_{n\mu}^a \hat A_{m,\nu}^a \, \times
\nonumber\\
&&\qquad
       \left[ G_c(n-m-\hat\nu) - G_c(n-m) \right] \,
       \left[ G_c(m+\hat\nu-n-\hat\mu) - G_c(m+\hat\nu-n) \right]
\nonumber\\
&&+
\frac{N}{8} \sum_{nm}
\hat A_{n\mu}^a \hat A_{m,\nu}^a \, \times
\nonumber\\
&& \qquad
       \left[ G_c(n+\hat\mu -m-\hat\nu) - G_c(n+\hat\mu-m) \right] \,
       \left[ G_c(m-n-\hat\mu) - G_c(m-n) \right] .
\nonumber\\
\end{eqnarray}

As for the second contribution, we can see by noting the identity,
\begin{equation}
  \nabla^2 G(n-m) = - G_c(n-m) + {\cal O}(\mu^4) ,
\end{equation}
that the last four terms are canceled by the forth contribution
of the ghost fields.

Under the assumption that the classical solution is sufficiently
slowly varying,
\begin{equation}
  \nabla_\nu A_{n\mu} << A_{n\mu} ,
\end{equation}
we extract the first order term in the local operator limit,
\begin{equation}
  A_{m\nu}^a \longrightarrow A_{n\nu}^a
                           + (m-n) \bar \nabla_\nu A_{n\nu}^a
                           + \cdots .
\end{equation}
Then the first contribution turns out to vanishes.
In the second contribution, the remaining four terms are
already local and turn out to be
\begin{eqnarray}
&&  \frac{N}{8}\sum_{n} \hat A_{n\mu}^a \hat A_{n\nu}^a \,
      \left[ G(\hat\nu-\hat\mu) + G(0)
                 -G(\hat\nu)-G(-\hat\mu) \right]
\nonumber\\
&&+
\frac{N}{8}\sum_{n} \hat A_{n-\hat\mu,\mu}^a
                     \hat A_{n-\hat\nu,\nu}^a \,
           \left[ G(0) + G(-\hat\nu+\hat\mu)
                 -G(-\hat\nu)-G(\hat\mu) \right]
\nonumber\\
&&+
\frac{N}{8}\sum_{n} \hat A_{n-\hat\mu,\mu}^a \hat A_{n\nu}^a \,
           \left[ G(\hat\nu) + G(\hat\mu)
                 -G(0)-G(\hat\nu+\hat\mu) \right]
\nonumber\\
&&+
\frac{N}{8}\sum_{nm} \hat A_{n\mu}^a \hat A_{n-\hat\nu,\nu}^a \,
           \left[ G(-\hat\mu) + G(-\hat\nu)
                 -G(-\hat\nu-\hat\mu)-G(0) \right] .
\nonumber\\
\end{eqnarray}
And we can see that they cancel the last four terms in
Eq.~(\ref{eq:reduced-formula-of-first-term-in-A2-appendix}).
As to the third contribution,
the nonlocal product of the correlation functions has the following
local expansion:
\begin{eqnarray}
&& \nabla^2 G(n-m)
  \left[ \nabla^2 G(m+\hat\nu-n)-\nabla^2 G(m-n) \right]
\nonumber\\
&& \quad = \delta_{nm} \left[ G(1)-G(0) \right]
           -\nabla_\nu \delta_{nm} \, \bar G  \cdots,
\end{eqnarray}
where
\begin{eqnarray}
\label{eq:one-loop-correction-to-classical-action-divergent-part-appendix}
  \bar G &=& \frac{1}{2} \int \frac{d^4 k}{(2\pi)^4}
           \frac{ \left( \sum_\mu 4 \sin^2 \frac{k_\mu}{2} \right)
                   \left( \sum_\mu \sin^2 k_\mu \right) }
           {\left[
              \left(\sum_\mu 4 \sin^2 \frac{k_\mu}{2}\right)^2+\mu_0^4
            \right]^2 } \nonumber\\
&\simeq& - \frac{1}{16\pi^2} \, \ln (a\mu_0) + C .
\end{eqnarray}
This is because in momentum space, the product of the correlation functions
has the expression
\begin{eqnarray}
&& \nabla^2 G(n-m)
  \left[ \nabla^2 G(m+\hat\nu-n)-\nabla^2 G(m-n) \right]
\nonumber\\
&& \qquad \equiv
\int \frac{d^4 p}{(2\pi)^4} e^{i p(n-m)} \, \Gamma_\nu(p) , \nonumber\\
\end{eqnarray}
where
\begin{eqnarray}
\Gamma_\nu(p) &=&
\int \frac{d^4 k}{(2\pi)^4} \,
\frac{\left( \sum_\mu 4 \sin^2 \frac{(k+p)_\mu}{2} \right)}
     {\left( \sum_\mu 4 \sin^2 \frac{(k+p)_\mu}{2} \right)^2 +\mu^4}
\frac{\left( \sum_\mu 4 \sin^2 \frac{k_\mu}{2} \right)
      \left(e^{i k_\nu} - 1 \right)}
     {\left( \sum_\mu 4 \sin^2 \frac{k_\mu}{2} \right)^2 +\mu^4}  ,
\nonumber\\
\end{eqnarray}
and $\Gamma_\nu(p)$ can be expanded in terms of the momentum $p$ as
\begin{equation}
\Gamma_\nu(p) =   \left[ G(1)-G(0) \right]
            - i p_\nu \, \bar G  + {\cal O}(\mu^4,p^2) .
\end{equation}
Then the local operator limit of the third contribution
can be evaluated as
\begin{equation}
\frac{1}{2} \sum_{n}
\bar \nabla_\mu \hat A_{n\mu}^a  \bar \nabla_\nu \hat A_{n\nu}^a \,
 N \bar G .
\end{equation}
Combining these results, we have
\begin{eqnarray}
\Delta S_2 [A^2] &= &
\frac{1}{2} \sum_{n}
\bar \nabla_\mu \hat A_{n\mu}^a  \bar \nabla_\nu \hat A_{n\nu}^a \,
 \left( N \bar G
       +\left(\frac{N^2-1}{N}\right)\, \left[ G(0)-G(1) \right]
 \right)  . \nonumber\\
\end{eqnarray}

The contributions of ghost fields
Eqs.~(\ref{eq:one-loop-effective-action-c2-appendix})
and (\ref{eq:one-loop-effective-action-Ac2-appendix})
can be calculated in a similar manner and we finally obtain
Eq.~(\ref{eq:one-loop-correction-to-classical-action})
\begin{eqnarray}
\label{eq:one-loop-correction-to-classical-action-appendix}
\Delta S_1 + \Delta S_2 + \Delta S_{1c} + \Delta S_{2c}
&=& - \lambda^2 \left[ N \bar G
                      + \left( \frac{N^2-1}{N} \right) [G(0)-G(1)] \right]
    \, {\cal O}_0 . \nonumber\\
\end{eqnarray}

\section{Effect of Gauge symmetry breaking term}
\label{appendix:effect-gauge-breaking-term}
\reseteqnum

In this appendix, we describe in some detail the calculation of
the one-loop contributions in the case with the explicit gauge
symmetry breaking term.
We first give the explicit formula of the explicit gauge symmetry
breaking term expanded in terms of the fluctuation of the gauge
freedom,
Eq.~(\ref{eq:explicit-gauge-break-in-gaussian-gauge-fluctuation}):
\begin{eqnarray}
\label{eq:explicit-gauge-break-in-gaussian-gauge-fluctuation-appendix}
  S_B[ \exp(i\lambda \pi) g ]
&=& S_B[ g ] + S_{B\, 0}[\pi^2] \nonumber\\
&+& S_{B\, 1}[\pi^2, \sin A ] + S_{B\, 2}[\pi^2, \cos A -1 ]
+ {\cal O}(\pi^3) .
\end{eqnarray}
\begin{eqnarray}
S_{B\, 0}[\pi^2] &=&  -\frac{1}{2} K \lambda^2  \sum_{n\mu}
   \nabla_{\mu}\pi_n^a \nabla_{\mu}\pi_n^a ,  \\
S_{B\, 1}[\pi^2, \sin A ]
&=& + \frac{1}{2} K \lambda^2
    f^{abc} \pi^a_n \pi^b_{n+\hat\mu} \hat A_{n\mu}^c ,
\end{eqnarray}
and
\begin{eqnarray}
&& S_{B\, 2}[\pi^2, \cos A -1 ]  \nonumber\\
&&= - K \lambda^2
    {\rm Tr}\left\{
            \sum_\mu \left( \pi_n^2 + \pi_{n+\hat\mu}^2
                   - \{ \pi_n, \pi_{n+\hat\mu} \} \right) \,
             \left(\cos A_{n\mu} -1 \right) \right\} . \nonumber\\
\end{eqnarray}

The propagator of the quantum fluctuation of the gauge freedom is
given by
\begin{eqnarray}
\label{eq:massless-dipole-propagator-of-pi-with-gauge-breaking-appendix}
  \langle \pi_n^a \pi_m^b \rangle
&=& \delta^{ab} \int \frac{d^4 p}{(2\pi)^4} e^{ip(n-m)}
    \frac{1}{\left( \sum_\mu 4 \sin^2 \frac{p_\mu}{2} \right)^2
             + M_0^2
             \left( \sum_\mu 4 \sin^2 \frac{p_\mu}{2} \right) }
\nonumber\\
&\equiv& \delta^{ab} G_B(n-m) .
\end{eqnarray}
Here we have set
\begin{equation}
  M_0^2 = K \lambda^2 .
\end{equation}

The quantum correction to the classical action can be divided into
two classes: the first class consists of the contributions which does not
have the gauge-breaking vertexes from $S_B$ and are given by the same
expressions as
Eqs.~(\ref{eq:one-loop-effective-action-A1}),
     (\ref{eq:one-loop-effective-action-A2}),
     (\ref{eq:one-loop-effective-action-c2}),
     (\ref{eq:one-loop-effective-action-Ac2}).
Corresponding equations are in the appendix
\ref{appendix:detail-one-loop-calculation} as
Eqs.~(\ref{eq:one-loop-effective-action-A1-appendix}),
     (\ref{eq:one-loop-effective-action-A2-appendix}),
     (\ref{eq:one-loop-effective-action-c2-appendix}),
     (\ref{eq:one-loop-effective-action-Ac2-appendix}).
Here we have taken account of the fact that $M_0^2$ should be
counted to have mass dimension two and omit
Eqs.~(\ref{eq:one-loop-effective-action-A3})
and (\ref{eq:one-loop-effective-action-A4}).

The second class consists of the additional contribution due to the
gauge-breaking vertexes.
They are given up to the second order in $A_\mu$ as follows:
\begin{eqnarray}
\label{eq:one-loop-effective-action-A1-gauge-breaking-appendix}
\Delta S_{B\, 1}[A]   &=&  \langle S_{B\, 1}[A] \rangle_0 , \\
\label{eq:one-loop-effective-action-A2-gauge-breaking-appendix}
\Delta S_{B\, 2}[A^2] &=&  \langle S_{B\, 2}[A^2] \rangle_0
                          + \frac{1}{2} \langle S_{B\, 1}[A]^2 \rangle_0
                          + \langle S_{B\, 1}[A] S_1[A] \rangle_0 .
\end{eqnarray}

The contributions of the first class can be evaluated in a similar
manner described in the
appendix \ref{appendix:detail-one-loop-calculation}, by just noting
the differences in the relations
\begin{eqnarray}
(\nabla^2)^2 G_B(n-m) &=& \delta_{nm} + M_0^2  \nabla^2 G_B(n-m)  ,  \\
  \nabla^2   G_B(n-m) &=& - G_c(n-m)  + M_0^2  G_B(n-m)  ,
\end{eqnarray}
and by counting $M_0^2$ as mass dimension two.
They are given as
\begin{eqnarray}
\label{eq:quantum-correction-first-class-appendix}
&& \langle S_2[A^2] \rangle_0
+ \frac{1}{2!} \langle S_1[A]^2 \rangle_0  \nonumber\\
&&=
\frac{1}{2}
\sum_n \bar \nabla_\mu \hat A_{n\mu}^a \bar \nabla_\nu \hat A_{n\nu}^a
\left( N\bar G_B
      + \left(\frac{N^2-1}{N}\right) \, \left[G_B(0)-G_B(1)\right]
\right)
\nonumber\\
&&\quad
- M_0^2 \nabla^2 G_B(0) \left(\frac{N^2-1}{N}\right)
          \sum_n {\rm Tr}\left(\cos A_{n\mu}-1\right)
\nonumber\\
&&\quad
+\frac{N}{8} M_0^2 \sum_{nm} \hat A_{n\mu}^a \hat A_{m\nu}^a \, \,
\times \nonumber\\
&& \qquad
 \left[ \nabla^2 G_B(n-m) + \nabla^2 G_B(n+\hat\mu-m-\hat\nu)
 \right.
 \nonumber\\
&& \qquad\qquad\qquad
  \left.
      +\nabla^2 G_B(n+\hat\mu-m) + \nabla^2 G_B(n-m-\hat\nu)
  \right]
\times \nonumber\\
&& \qquad\qquad
   \left[ G_B(m+\hat\nu-n-\hat\mu)  + G_B(m-n)
   \right. \nonumber\\
&& \qquad\qquad\qquad
   \left.
         -G_B(m+\hat\nu-n)-G_B(m-n-\hat\mu) \right]
\nonumber\\
&&\quad
-\frac{N}{8} \sum_{nm} \hat A_{n\mu}^a \hat A_{m\nu}^a \, \,
\times \nonumber\\
&& \qquad
 \left[ \nabla^2 G_B(n-m) - \nabla^2 G_B(n+\hat\mu-m-\hat\nu)
 \right.
 \nonumber\\
&& \qquad\qquad\qquad
  \left.
      +\nabla^2 G_B(n+\hat\mu-m) - \nabla^2 G_B(n-m-\hat\nu)
  \right]
\times \nonumber\\
&& \qquad
 \left[ \nabla^2 G_B(m-n) - \nabla^2 G_B(m+\hat\nu-n-\hat\mu)
 \right.
 \nonumber\\
&& \qquad\qquad\qquad
  \left.
      -\nabla^2 G_B(m-n-\hat\mu) + \nabla^2 G_B(m+\hat\nu-n)
  \right]
\nonumber\\
&&\quad
+\frac{N}{8} \sum_{nm} \hat A_{n\mu}^a \hat A_{m\nu}^a \, \,
\times \nonumber\\
&& \qquad
 \left[ G_c(n-m) + G_c(n+\hat\mu-m-\hat\nu)
 \right.
 \nonumber\\
&& \qquad\qquad\qquad
  \left.
      - G_c(n+\hat\mu-m) - G_c(n-m-\hat\nu)
  \right]
\times \nonumber\\
&& \qquad
 \left[ G_c(m-n) + G_c(m+\hat\nu-n-\hat\mu)
 \right.
 \nonumber\\
&& \qquad\qquad\qquad
  \left.
      - G_c(m-n-\hat\mu) - G_c(m+\hat\nu-n)
  \right] ,
\nonumber\\
\end{eqnarray}
where
\begin{eqnarray}
\label{eq:one-loop-correction-to-classical-action-divergent-part
-with-breaking-appendix}
  \bar G_B &=& \frac{1}{2} \int \frac{d^4 k}{(2\pi)^4}
           \frac{ \left( \sum_\mu \sin^2 k_\mu \right) }
           {\left[
             \sum_\mu 4 \sin^2 \frac{k_\mu}{2} + M_0^2
            \right]^3 } \nonumber\\
&\simeq& - \frac{1}{16\pi^2} \, \ln (M_0) + \bar C_B  \qquad (M_0 << 1) .
\end{eqnarray}

The last three terms in the above are evaluated further as
\begin{eqnarray}
&&-\frac{N}{8} M_0^2 \sum_{nm} \hat A_{n\mu}^a \hat A_{m\nu}^a \, \,
\times \nonumber\\
&& \quad
 \left\{
G_c(n-m)
\right.
\times \nonumber\\
&& \qquad
   \left[-G_B(m-n) + 3 G_B(m+\hat\nu-n-\hat\mu)
   \right. \nonumber\\
&& \qquad
   \left.
         -G_B(m+\hat\nu-n)-G_B(m-n-\hat\mu) \right]
\nonumber\\
&& \quad
+G_c(n+\hat\mu-m-\hat\nu)
\times \nonumber\\
&& \qquad
   \left[3G_B(m-n)- G_B(m+\hat\nu-n-\hat\mu)
   \right. \nonumber\\
&& \qquad
   \left.
         -G_B(m+\hat\nu-n)-G_B(m-n-\hat\mu) \right]
\nonumber\\
&& \quad
-G_c(n+\hat\mu-m)
\times \nonumber\\
&& \qquad
   \left[-G_B(m-n)- G_B(m+\hat\nu-n-\hat\mu)
   \right. \nonumber\\
&& \qquad
   \left.
         +3G_B(m+\hat\nu-n)-G_B(m-n-\hat\mu) \right]
\nonumber\\
&& \quad
-G_c(n-m-\hat\nu)
\times \nonumber\\
&& \qquad
   \left[-G_B(m-n)- G_B(m+\hat\nu-n-\hat\mu)
   \right. \nonumber\\
&& \qquad
\left.
   \left.
         -G_B(m+\hat\nu-n)+3G_B(m-n-\hat\mu) \right]
\right\} + {\cal O}(M_0^4)
\nonumber\\
&=& -\frac{N}{8} M_0^2 \sum_{nm} \hat A_{n\mu}^a \hat A_{m\nu}^a \, \,
\times \nonumber\\
&& \quad
 \left[ \, 4 G_c(n-m) \, G_B(m+\hat\nu-n-\hat\mu)
       +4 G_c(n+\hat\mu-m-\hat\nu) \, G_B(m-n)
 \right.
 \nonumber\\
&&\quad
-4 G_c(n+\hat\mu-m) \, G_B(m+\hat\nu-n)
-4 G_c(n-m-\hat\nu) \, G_B(m-n-\hat\mu)
 \nonumber\\
&&\qquad
-\left( G_c(n-m) + G_c(n+\hat\mu-m-\hat\nu)
 \right.
 \nonumber\\
&& \qquad\qquad\qquad
  \left.
      - G_c(n+\hat\mu-m) - G_c(n-m-\hat\nu)
  \right)
\times \nonumber\\
&& \qquad\qquad
 \left( G_B(m-n) + G_B(m+\hat\nu-n-\hat\mu)
 \right.
 \nonumber\\
&& \qquad\qquad\qquad
\left.\left.
      + G_B(m-n-\hat\mu) + G_B(m+\hat\nu-n)
  \right)
  \right]  + {\cal O}(M_0^4)
\nonumber\\
&=& -N \tilde G_B \, M_0^2 \, \,
\frac{1}{2} \sum_{n} \hat A_{n\mu}^a \hat A_{n\nu}^a \,
+ {\cal O}(M_0^4) , \nonumber\\
\end{eqnarray}
where
\begin{eqnarray}
\label{eq:one-loop-correction-to-explicit-breaking-divergent-part-appendix}
  \tilde G_B &=& \frac{3}{4} \int \frac{d^4 k}{(2\pi)^4}
           \frac{
                   \left( \sum_\mu \sin^2 k_\mu \right) }
           {\left[
          \left( \sum_\mu 4 \sin^2 \frac{k_\mu}{2} \right)^2
          \left(\sum_\mu 4 \sin^2 \frac{k_\mu}{2}+ M_0^2 \right)
            \right] } \nonumber\\
&\simeq& - \frac{3}{32\pi^2} \, \ln (M_0) + \tilde C_B
\qquad (M_0 << 1) .
\end{eqnarray}

As to the contributions of the second class,
it is easily seen that
Eq.~(\ref{eq:one-loop-effective-action-A1-gauge-breaking-appendix})
vanishes.
In Eq.~(\ref{eq:one-loop-effective-action-A2-gauge-breaking-appendix}),
only the first term contributes to the local operators of dimension
less than five. It is finite and is evaluated as follows:
\begin{eqnarray}
&& \Delta S_{B\, 2}[A^2] \nonumber\\
&&=  M_0^2 \left(\frac{N^2-1}{2N}\right)
\left[ G_B(0)-G_B(1) \right]
\left( \sum_{n\mu} \frac{1}{2} \, A_{n\mu}^a A_{n\mu}^a  \right) .
\end{eqnarray}

Combining these result, we obtain

\begin{eqnarray}
\label{eq:one-loop-correction-to-classical-action-with
-explicit-breaking-all-appendix}
&&\Delta S_1 + \Delta S_2 + \Delta S_{1c} + \Delta S_{2c}
\nonumber\\
&&\qquad
= - \lambda^2 \left[ N \bar G_B
                   + \left( \frac{N^2-1}{N} \right) [G_B(0)-G_B(1)] \right]
    \, {\cal O}_0  \nonumber\\
&&\qquad \qquad
 + M_0^2
   \left[ N \tilde G_B
            - \left( \frac{N^2-1}{2N}\right)  \nabla^2 G_B(0) \right] \,
  {\cal O}_1 , \nonumber\\
&&\Delta S_{B\, 1} + \Delta S_{B\, 2} \nonumber\\
&&\qquad
= - M_0^2
\left(\frac{N^2-1}{2N}\right)
\left[ G_B(0)-G_B(1) \right] \, {\cal O}_1 . \nonumber\\
\end{eqnarray}

\end{document}